\documentclass[superscriptaddress,amsmath,prd, amssymb,aps]{revtex4-2}
\usepackage{graphicx}
\usepackage{dcolumn}
\usepackage{hyperref}
\hypersetup{colorlinks=true, citecolor=blue, urlcolor=blue, linkcolor=blue}
\usepackage{amsmath,amssymb}
\usepackage{mathrsfs}
\usepackage{esint}
\usepackage{comment}
\usepackage{tcolorbox}
\usepackage{booktabs}
\usepackage{caption}
\usepackage{ulem}

\newcommand{\myDelta}{\Delta\hspace{-2.3ex}\raisebox{0.3ex}{\Large$\sim$}}
\newcommand{\mydelta}{\Delta\hspace{-1.78ex}\raisebox{0.5ex}{\normalsize$\rule{0.25cm}{1pt}$}}

\begin{document}

\title{Generalized relativistic second-order spin hydrodynamics from Zubarev's non-equilibrium statistical operator}
\medskip

\author{Duan She}
\email{sheduan@hnas.ac.cn}
\affiliation{Institute of Physics, Henan Academy of Sciences, Zhengzhou 450046, China}

\author{Yi-Wei Qiu}
\email{qiuyiwei1998@foxmail.com}
\affiliation{Institute of Particle Physics and Key Laboratory of Quark and Lepton Physics (MOE),
	Central China Normal University, Wuhan 430079, China}

\author{Ze-Fang Jiang}
\email{jiangzf@mails.ccnu.edu.cn}
\affiliation{Institute of Particle Physics and Key Laboratory of Quark and Lepton Physics (MOE),
	Central China Normal University, Wuhan 430079, China}
\affiliation{Department of Physics and Electronic-Information Engineering,
	Hubei Engineering University, Xiaogan, Hubei, 432000, China}

\author{Defu Hou}
\email{houdf@mail.ccnu.edu.cn}
\affiliation{Institute of Particle Physics and Key Laboratory of Quark and Lepton Physics (MOE),
	Central China Normal University, Wuhan 430079, China}

\date{\today}

\begin{abstract}

Inspired by the work in Ref.~\cite{Harutyunyan:2025fgs}, which considers the additional second-order contributions arising from nonlocal corrections due to two-point correlation functions of tensors of different ranks at distinct spacetime points, we similarly employ the nonequilibrium statistical operator method to extend this framework to include spin degrees of freedom. In addition to obtaining analogous extra second-order terms in the shear stress tensor, bulk viscous pressure, and charge diffusion currents resulting from such contributions, we further derive additional second-order terms originating from the same mechanism in the charge diffusion currents, rotational stress tensor and the boost heat vector. Furthermore, we express all transport coefficients represented by two-point or three-point correlations in terms of retarded Green's functions.

\end{abstract}


 \maketitle

\section{Introduction}
\label{section1}

In non-central relativistic heavy-ion collisions, substantial orbital angular momentum is generated in the reaction zone, manifesting as fluid vortices within the quark-gluon plasma (QGP). Due to spin-orbit coupling, quarks and gluons polarize along the direction of the system's orbital angular momentum, giving rise to the global polarization of the QGP.

A pivotal theoretical breakthrough came in 2005 when Liang and Wang first proposed the concept of global polarization in quark matter, establishing a connection between the orbital angular momentum in non-central collisions and hadronic spin polarization. Their work predicted hyperon spin polarization and vector meson spin alignment, marking the inception of spin-related studies in the QGP~\cite{Liang:2004ph,Liang:2004xn}. Later, in 2013, Becattini et al. used the equilibrium density matrix formalism to predict that vortices induced by angular momentum would lead to the spin polarization of $\Lambda$ hyperons~\cite{Becattini:2013fla}, significantly advancing our theoretical understanding of spin effects in the QGP.

Experimental confirmation followed in 2017 when the STAR Collaboration at the Relativistic Heavy-Ion Collider (RHIC) reported the first measurement of $\Lambda$ hyperon global polarization in Au+Au collisions, published as a cover article in Nature~\cite{STAR:2017ckg}. Subsequent measurements extended these observations to multi-strange hyperons ($\Xi^-$,$\Omega^-$)~\cite{STAR:2020xbm}, providing definitive evidence for global polarization in quark matter. Further validation came with the detection of spin alignment in
$\phi$ meson and $K^{*0}$ mesons, with results featured in Nature in 2023~\cite{STAR:2022fan}. These landmark findings have solidified spin physics as a leading research frontier in high-energy nuclear physics.

Despite these advances, key questions remain unresolved. Notably, discrepancies persist between $\Lambda$ and $\bar{\Lambda}$ hyperon polarization~\cite{STAR:2023nvo,Peng:2022cya}, and the rapidity dependence of global polarization requires further clarification~\cite{Liang:2019pst}. Additionally, the mechanisms underlying hyperon local polarization~\cite{Karpenko:2016jyx} and differences in spin alignment among vector mesons with varying quark flavors~\cite{ALICE:2022dyy} demand deeper investigation.

The experimental progress in hyperon polarization and vector meson spin alignment has spurred theoretical efforts to model spin transport in the QGP. Over the past decade, relativistic hydrodynamics has emerged as a powerful tool for describing the hot, dense matter produced in heavy-ion collisions at RHIC and the LHC, successfully reproducing observables such as elliptic flow and particle spectra. This success has motivated the development of relativistic spin hydrodynamics, which incorporates angular momentum dynamics into the hydrodynamic framework.

Relativistic spin hydrodynamics must simultaneously satisfy energy-momentum conservation and angular momentum conservation, introducing additional evolution equations for the spin tensor. When internal symmetries (e.g., $U(1)$ baryon number conservation) are present, the corresponding charge conservation laws must also be considered.

Recent theoretical approaches to spin hydrodynamics include: entropy current analysis~\cite{Hattori:2019lfp,Fukushima:2020ucl,Hongo:2021ona,Li:2020eon,She:2021lhe,Daher:2022xon,Cao:2022aku,Biswas:2023qsw}, quantum kinetic theories~\cite{Peng:2021ago,Florkowski:2017ruc,Florkowski:2018fap,Li:2019qkf,Bhadury:2020puc,Shi:2020htn,Hongo:2022izs,Weickgenannt:2022zxs,Bhadury:2022ulr,Weickgenannt:2022qvh}, holographic duality~\cite{Gallegos:2021bzp,Gallegos:2022jow}, effective Lagrangian approaches~\cite{Montenegro:2020paq}, and Zubarev's non-equilibrium statistical operator methods~\cite{Hu:2021lnx,She:2024rnx}. While these frameworks differ in methodology, they collectively enhance our understanding of spin dynamics in relativistic fluids.

Despite progress, several theoretical issues remain unresolved: the physical meaning of pseudo-gauge transformations, the stability and causality of spin hydrodynamic equations, and the calculation of the new transport coefficient. Additionally, the numerical solution of relativistic spin hydrodynamics equations and their application to high-energy heavy-ion collisions represent critical frontier for future research efforts~\cite{Singh:2024cub,Sapna:2025yss}.

First-order spin hydrodynamics suffers from acausal modes and numerical instabilities at high momenta~\cite{Daher:2022wzf,Sarwar:2022yzs,Xie:2023gbo}, necessitating second-order formulations with relaxation terms. Previous works have derived second-order spin hydrodynamics via: spin kinetic theory (incorporating collision terms)~\cite{Weickgenannt:2022zxs,Weickgenannt:2022qvh}, entropy current methods (phenomenological approach)~\cite{Biswas:2023qsw}, and Zubarev's formalism (under different power-counting schemes for the spin chemical potential $\omega_{\mu\nu}$)~\cite{She:2024rnx}. These developments provide a robust foundation for future studies on spin polarization and transport in relativistic systems.

Relativistic spin hydrodynamics offers a promising framework for interpreting polarization phenomena in heavy-ion collisions. Future numerical simulations must account for spin chemical potential and spin density effects, which modify the polarization vector~\cite{Liu:2021nyg}. Overcoming current challenges—particularly in numerical stability and transport coefficient determination—will be crucial for advancing this rapidly evolving field.

Recent work~\cite{Harutyunyan:2025fgs} demonstrates that two-point correlation functions involving tensors of different ranks at distinct spacetime points introduce additional second-order nonlocal corrections to hydrodynamic constitutive relations. Building on this insight, we extend our earlier formulation of relativistic second-order spin hydrodynamics~\cite{She:2024rnx}—derived via the nonequilibrium statistical operator method—by incorporating these contributions. Specifically, we generalize the second-order viscous hydrodynamic framework of Ref.~\cite{Harutyunyan:2025fgs} to include spin degrees of freedom. Beyond recovering the previously identified second-order corrections to the shear stress tensor, bulk viscous pressure, and charge diffusion currents~\cite{Harutyunyan:2025fgs}, we derive novel second-order terms in the rotational stress tensor and boost heat vector arising from the same nonlocal correlations. As shown in Ref.~\cite{Harutyunyan:2025fgs}, all such corrections are proportional to the comoving derivatives of the flow velocity. Furthermore, we express the associated transport coefficients, governed by two-point or three-point correlation functions, explicitly in terms of retarded Green’s functions.

The structure of this paper is organized as follows. Section~\ref{section2} presents a concise overview of relativistic spin hydrodynamics and the Zubarev formalism for non-equilibrium statistical operators. In Section~\ref{section3}, we derive the complete set of second-order constitutive equations, which incorporate novel terms describing the shear stress tensor, bulk viscous pressure, charge-diffusion currents, rotational stress tensor, boost heat vector, and dissipative spin tensor fluxes. Finally, Section~\ref{section4} summarizes and discusses our results. Throughout this work, we consider a flat spacetime with metric signature $g_{\mu\nu}=\text{diag}\left(+,-,-,-\right)$ and adopt natural units where $\hbar=k_{B}=c=1$. For any tensor $X^{\mu\nu}$, we define its symmetric and antisymmetric components respectively as: $X^{\left(\mu\nu\right)}=\frac{1}{2}\left(X^{\mu\nu}+X^{\nu\mu}\right)$ and $X^{[\mu\nu]}=\frac{1}{2}\left(X^{\mu\nu}-X^{\nu\mu}\right)$. The fluid four-velocity $u^\mu$ satisfies the normalization condition $u^\mu u_\mu=1$. The orthogonal projector is given by $\Delta^{\mu\nu}= g^{\mu\nu}-u^{\mu}u^{\nu}$, which satisfies $\Delta^{\mu\nu}u_\mu=0$. The orthogonal projection of a four-vector $X^\mu$ is denoted by $X^{\left\langle \mu\right\rangle }=\Delta^{\mu\nu}X_{\nu}$. We further define the following projection operators: the traceless  symmetric projector $\Delta_{\alpha\beta}^{\mu\nu}=\frac{1}{2}\left(\Delta_{\alpha}^{\mu}\Delta_{\beta}^{\nu}+\Delta_{\beta}^{\mu}\Delta_{\alpha}^{\nu}-\frac{2}{3}\Delta^{\mu\nu}\Delta_{\alpha\beta}\right)$, the antisymmetric projector $\mydelta_{\alpha\beta}^{\mu\nu}=\frac{1}{2}\left(\Delta_{\alpha}^{\mu}\Delta_{\beta}^{\nu}-\Delta_{\beta}^{\mu}\Delta_{\alpha}^{\nu}\right)$, and the totally antisymmetric projector $\myDelta_{\alpha\beta\gamma}^{\lambda\mu\nu}=\frac{1}{6}\left(\Delta_{\alpha}^{\lambda}\Delta_{\beta}^{\mu}\Delta_{\gamma}^{\nu}+\Delta_{\alpha}^{\mu}\Delta_{\beta}^{\nu}\Delta_{\gamma}^{\lambda}+\Delta_{\alpha}^{\nu}\Delta_{\beta}^{\lambda}\Delta_{\gamma}^{\mu}-\Delta_{\alpha}^{\mu}\Delta_{\beta}^{\lambda}\Delta_{\gamma}^{\nu}-\Delta_{\alpha}^{\nu}\Delta_{\beta}^{\mu}\Delta_{\gamma}^{\lambda}-\Delta_{\alpha}^{\lambda}\Delta_{\beta}^{\nu}\Delta_{\gamma}^{\mu}\right)$.

\section{The nonequilibrium statistical-operator formalism}
\label{section2}

For strongly interacting systems, a rigorous quantum-statistical approach based on the Liouville equation for the non-equilibrium statistical operator becomes essential. This approach provides the foundation for our subsequent analysis. Within this theoretical framework, Zubarev's formalism—alternatively termed the non-equilibrium statistical operator method—enables the systematic derivation of hydrodynamic equations for strongly correlated systems. The formalism extends the Gibbs canonical ensemble to non-equilibrium conditions by constructing the statistical operator as a non-local functional of thermodynamic parameters and their spatial gradients.

Under the assumption that thermodynamic parameters vary sufficiently slowly compared to the characteristic correlation lengths of the microscopic theory, the statistical operator can be expanded in a Taylor series with respect to hydrodynamic gradients, truncated at the desired order. The resulting dissipative hydrodynamic equations are then obtained through complete statistical averaging of the relevant quantum-mechanical operators.

We employ Zubarev's non-equilibrium statistical operator formalism to study a general quantum system in the spin hydrodynamic regime. The analysis begins with the operator-valued conservation laws for the energy-momentum tensor, charge currents, and total angular momentum tensor
\begin{align}
\partial_{\mu}\hat{N}_{a}^{\mu}=&0, \label{1}\\
\partial_{\mu}\hat{T}^{\mu\nu}=&0, \label{2}\\
\partial_{\lambda}\hat{J}^{\lambda\mu\nu}=&\partial_{\lambda}\hat{S}^{\lambda\mu\nu}+2\hat{T}^{[\mu\nu]}=0, \label{3}
\end{align}
where the index $a=1,2,\cdots,l$ labels the conserved charges, with $l$ denoting the total number of charge conservation laws. In Eq.~\eqref{3}, the total angular momentum tensor $\hat{J}^{\lambda\mu\nu}$ decomposes into orbital and intrinsic spin components:  $\hat{L}^{\lambda\mu\nu}=x^{\mu}\hat{T}^{\lambda\nu}-x^{\nu}\hat{T}^{\lambda\mu}$ for orbital angular momentum and $\hat{S}^{\lambda\mu\nu}$ for spin angular momentum. Compared to the viscous fluid in Ref.~\cite{Harutyunyan:2025fgs}, the spin fluid requires additional consideration of the angular momentum conservation law, where the antisymmetric part of the energy-momentum tensor serves as the source term for spin current conservation.

The total angular momentum current can be decomposed into energy-momentum and spin tensor components in multiple ways. As discussed in Refs.~\cite{Daher:2022xon,Speranza:2020ilk}, these tensor pairs are related through pseudogauge transformations. A prominent choice is the Belinfante pseudogauge, which employs the symmetric Belinfante energy-momentum tensor~\cite{BELINFANTE1939887,BELINFANTE1940449}. Alternatively, the phenomenological pseudogauge utilizes a general energy-momentum tensor with both symmetric and antisymmetric parts, coupled to a spin tensor that is antisymmetric in its last two indices. This approach has proven effective for constructing hydrodynamic frameworks, as demonstrated in Refs.~\cite{Hattori:2019lfp,Fukushima:2020ucl,Daher:2022xon,Weyssenhoff:1947iua}. At the microscopic level, the canonical conserved currents consist of an asymmetric energy-momentum tensor and a fully antisymmetric spin tensor. This choice naturally arises from applying the Noether theorem to Dirac fermions in quantum field theory, as shown in Ref.~\cite{Hongo:2021ona}. Unlike the phenomenological approach, we adopt these canonical-like conserved currents, which provide a more fundamental microscopic description of the system. 
Further support for the canonical pseudogauge comes from Ref.~\cite{Buzzegoli:2024mra}, which demonstrates that the thermodynamics of a field theory with axial current interactions coincides with the Zubarev local equilibrium operator when the canonical pseudogauge is selected. This consistency reinforces the canonical pseudogauge as a robust foundation for developing spin hydrodynamics.

In the framework of spin hydrodynamics, the conservation equations can be derived by computing the statistical averages of the operators $\hat{T}^{\mu\nu}$, $\hat{N}_{a}^{\mu}$, and $\hat{S}^{\lambda\mu\nu}$ using the complete nonequilibrium statistical operator. For a multicomponent system, this operator is defined as~\cite{She:2024rnx}
\begin{eqnarray}
	\hat{\rho}(t)=Q^{-1} e^{-\hat{A}+\hat{B}}, \quad Q=\operatorname{Tr} e^{-\hat{A}+\hat{B}},
	\label{4}
\end{eqnarray}
where the operators $\hat{A}$ and $\hat{B}$ are given by
\begin{align}
	\hat{A}(t)= & \int d^{3}x\Bigl[\beta_{\nu}(\boldsymbol{x},t)\hat{T}^{0\nu}(\boldsymbol{x},t)-\sum_{a}\alpha_{a}(\boldsymbol{x},t)\hat{N}_{a}^{0}(\boldsymbol{x},t)-\frac{1}{2}\Omega_{\alpha\beta}(\boldsymbol{x},t)\hat{S}^{0\alpha\beta}(\boldsymbol{x},t)\Bigr],\label{5}\\
	\hat{B}(t)= & \int d^{3}x\int_{-\infty}^{t}dt_{1}e^{\varepsilon\left(t_{1}-t\right)}\hat{C}\left(\boldsymbol{x},t_{1}\right),\label{6}\\
	\hat{C}(\boldsymbol{x},t)= & \hat{T}^{\mu\nu}(\boldsymbol{x},t)\partial_{\mu}\beta_{\nu}(\boldsymbol{x},t)-\sum_{a}\hat{N}_{a}^{\mu}(\boldsymbol{x},t)\partial_{\mu}\alpha_{a}(\boldsymbol{x},t)-\frac{1}{2}\hat{S}^{\lambda\alpha\beta}\left(\boldsymbol{x},t\right)\partial_{\lambda}\Omega_{\alpha\beta}(\boldsymbol{x},t)+\Omega_{\alpha\beta}(\boldsymbol{x},t)\hat{T}^{[\alpha\beta]}(\boldsymbol{x},t).\label{7}
\end{align}
Here, the parameters $\beta^\mu,\alpha_a$, and $\Omega_{\alpha\beta}$ are defined as
\begin{eqnarray} \beta^{\nu}(x)=\beta(x)u^{\nu}(x),\quad\alpha_{a}(x)=\beta(x)\mu_{a}(x),\quad\Omega_{\alpha\beta}\left(x\right)=\beta\left(x\right)\omega_{\alpha\beta}\left(x\right),\label{8}
\end{eqnarray}
where $\beta^{-1},\mu_a,\omega_{\alpha\beta}$, and $u^\nu$ represent the local temperature, chemical potentials, spin chemical potential, and fluid four-velocity, respectively. We assume that these quantities vary slowly in space and time. To account for the irreversibility of thermodynamic processes, we introduce an additional source term of order $\mathcal{O}\left(\varepsilon\right)$ in the Liouville equation for the statistical operator. This term breaks time-reversal symmetry due to its retarded solution. To preserve irreversibility in the calculations, the thermodynamic limit must be taken before the limit $\varepsilon \to 0^+$.

The statistical operator defined in Eq.~\eqref{4} serves as the foundation for deriving transport equations that describe dissipative currents. In this framework, the operators $\hat{A}$ and $\hat{B}$ represent the equilibrium and non-equilibrium components of the statistical operator, respectively, while the operator $\hat{C}$ is constructed from thermodynamic parameters or their space-time derivatives. To develop the second-order spin hydrodynamics theory, we expand the statistical operator up to second order in $\hat{B}$~\cite{Harutyunyan:2021rmb}
\begin{eqnarray}
	\hat{\rho}=\hat{\rho}_{l}+\hat{\rho}_{1}+\hat{\rho}_{2}.
	\label{9}
\end{eqnarray}
Here, $\hat{\rho}_{l}=e^{-\hat{A}}/\text{Tr}e^{-\hat{A}}$ represents the equilibrium component, with the first-order and second-order corrections given by
\begin{eqnarray}
	\hat{\rho}_{1}(t)=\int d^{4}x_{1}\int_{0}^{1}d\tau\left[\hat{C}_{\tau}\left(x_{1}\right)-\left\langle \hat{C}_{\tau}\left(x_{1}\right)\right\rangle _{l}\right]\hat{\rho}_{l},
	\label{10}
\end{eqnarray}
and
\begin{eqnarray}
\begin{aligned}
\hat{\rho}_{2}(t) =&\frac{1}{2}\int d^{4}x_{1}d^{4}x_{2}\int_{0}^{1}d\tau\int_{0}^{1}d\lambda\biggl[\widetilde{T}\left\{ \hat{C}_{\lambda}\left(x_{1}\right)\hat{C}_{\tau}\left(x_{2}\right)\right\} -\left\langle \widetilde{T}\left\{ \hat{C}_{\lambda}\left(x_{1}\right)\hat{C}_{\tau}\left(x_{2}\right)\right\} \right\rangle _{l}\\
&-\left\langle \hat{C}_{\lambda}\left(x_{1}\right)\right\rangle _{l}\hat{C}_{\tau}\left(x_{2}\right)-\hat{C}_{\lambda}\left(x_{1}\right)\left\langle \hat{C}_{\tau}\left(x_{2}\right)\right\rangle _{l}+2\left\langle \hat{C}_{\lambda}\left(x_{1}\right)\right\rangle _{l}\left\langle \hat{C}_{\tau}\left(x_{2}\right)\right\rangle _{l}\biggr]\hat{\rho}_{l}.
\end{aligned}
\label{11}
\end{eqnarray}
We employ the notation $\hat{X}_{\tau}=e^{-\tau\hat{A}}\hat{X}e^{\tau\hat{A}}$ for any operator $\hat{X}$. The integration $\int d^{4}x_{1}$ is defined as $\int d^{3}x_{1}\int_{-\infty}^{t}dt_{1}e^{\varepsilon\left(t_{1}-t\right)}$, and $\widetilde{T}$ denotes the antichronological time-ordering operator with respect to the parameters $\tau$ and $\lambda$.

From Eqs.~\eqref{9}-\eqref{11}, the statistical average of an arbitrary operator $\hat{X}(x)$ can be expressed as
\begin{eqnarray} \langle\hat{X}(x)\rangle=\text{Tr}\bigl[\hat{\rho}\left(t\right)\hat{X}\left(x\right)\bigr]=\langle\hat{X}(x)\rangle_{l}+\int d^{4}x_{1}\left(\hat{X}\left(x\right),\hat{C}\left(x_{1}\right)\right)+\int d^{4}x_{1}\int d^{4}x_{2}\left(\hat{X}\left(x\right),\hat{C}\left(x_{1}\right),\hat{C}\left(x_{2}\right)\right),
\label{12}
\end{eqnarray} 
where $\langle\hat{X}(x)\rangle_{l}=\text{Tr}\left[\hat{\rho}_{l}\left(t\right)\hat{X}\left(x\right)\right]$ denotes the local-equilibrium average. The two-point correlation function is defined by
\begin{equation}
	\left(\hat{X}(x),\hat{Y}\left(x_{1}\right)\right)=\int_{0}^{1}d\tau\left\langle \hat{X}(x)\Bigl[\hat{Y}_{\tau}\left(x_{1}\right)-\bigl\langle\hat{Y}_{\tau}\left(x_{1}\right)\bigr\rangle_{l}\Bigr]\right\rangle _{l},
	\label{13}
\end{equation}
and the three-point correlation function is given by
\begin{eqnarray}
	\begin{aligned}
		\left(\hat{X}(x),\hat{Y}\left(x_{1}\right),\hat{Z}\left(x_{2}\right)\right) \equiv &\frac{1}{2}\int_{0}^{1}d\tau\int_{0}^{1}d\lambda\biggl\langle\widetilde{T}\Bigl\{\hat{X}(x)\Bigl[\hat{Y}_{\lambda}\left(x_{1}\right)\hat{Z}_{\tau}\left(x_{2}\right)-\bigl\langle\widetilde{T}\hat{Y}_{\lambda}\left(x_{1}\right)\hat{Z}_{\tau}\left(x_{2}\right)\bigr\rangle_{l}\\ &-\bigl\langle\hat{Y}_{\lambda}\left(x_{1}\right)\bigr\rangle_{l}\hat{Z}_{\tau}\left(x_{2}\right)-\hat{Y}_{\lambda}\left(x_{1}\right)\bigl\langle\hat{Z}_{\tau}\left(x_{2}\right)\bigr\rangle_{l}+2\bigl\langle\hat{Y}_{\lambda}\left(x_{1}\right)\bigr\rangle_{l}\bigl\langle\hat{Z}_{\tau}\left(x_{2}\right)\bigr\rangle_{l}\Bigr]\Bigr\}\biggr\rangle_{l}.
	\end{aligned}
	\label{14}
\end{eqnarray}
From Eq.~\eqref{14}, we derive the symmetric property of the three-point correlation function 
\begin{eqnarray}
	\int d^{4} x_{1} d^{4} x_{2}\left(\hat{X}(x), \hat{Y}\left(x_{1}\right), \hat{Z}\left(x_{2}\right)\right)=\int d^{4} x_{1} d^{4} x_{2}\left(\hat{X}(x), \hat{Z}\left(x_{1}\right), \hat{Y}\left(x_{2}\right)\right).
	\label{15}
\end{eqnarray}
This property will be frequently used in subsequent derivations.

\subsection{Spin hydrodynamic equations: canonical-like framework}

To derive the fluid equations, we decompose the energy-momentum tensor, charge currents, and spin tensor into equilibrium and dissipative components, expressed in terms of the fluid four-velocity~\cite{She:2024rnx}
\begin{align} &\hat{T}^{\mu\nu}=\hat{\epsilon}u^{\mu}u^{\nu}-\hat{p}\Delta^{\mu\nu}+\hat{h}^{\mu}u^{\nu}+\hat{h}^{\nu}u^{\mu}+\hat{\pi}^{\mu\nu}+\hat{q}^{\mu}u^{\nu}-\hat{q}^{\nu}u^{\mu}+\hat{\phi}^{\mu\nu},\label{16}\\
&\hat{N}_{a}^{\mu}=\hat{n}_{a}u^{\mu}+\hat{j}_{a}^{\mu},\label{17}\\	&\hat{S}^{\lambda\mu\nu}=u^{\lambda}\hat{S}^{\mu\nu}+u^{\mu}\hat{S}^{\nu\lambda}+u^{\nu}\hat{S}^{\lambda\mu}+\hat{\varpi}^{\lambda\mu\nu}.\label{18}
\end{align}
Here, $\Delta^{\mu\nu}=g^{\mu\nu}-u^{\mu}u^{\nu}$ is the projection tensor orthogonal to the fluid four-velocity $u^\mu$. Within the hydrodynamic gradient expansion, the energy density $\hat{\epsilon}$, particle number densities $\hat{n}_{a}$, spin density $S^{\mu\nu}$, and four-velocity $u^{\mu}$ are zeroth-order ($\mathcal{O}\left(\partial^{0}\right)$) quantities. 
In contrast, the dissipative components--diffusion currents $\hat{j}_{a}^{\mu}$, shear stress tensor $\hat{\pi}^{\mu\nu}$, heat flux $\hat{h}^{\mu}$, boost heat vector $\hat{q}^{\mu}$, rotational stress tensor $\hat{\phi}^{\mu\nu}$, and spin-related dissipative flux $\hat{\varpi}^{\lambda\mu\nu}$--are first order ($\mathcal{O}\left(\partial\right)$) terms. 
These dissipative quantities satisfy the following orthogonality and symmetry conditions
\begin{eqnarray}
	\begin{aligned}
		&u_{\nu}\hat{h}^{\nu}=0,\,\,u_{\nu}\hat{j}_{a}^{\nu}=0,\,\,u_{\nu}\hat{\pi}^{\mu\nu}=0,\,\,u_{\mu}\hat{\pi}^{\mu\nu}=0,\,\,u_{\nu}\hat{q}^{\nu}=0,\,\,u_{\nu}\hat{\phi}^{\mu\nu}=0,\,\,u_{\mu}\hat{\phi}^{\mu\nu}=0,\\
		&\hat{\pi}^{\mu\nu}=\hat{\pi}^{\nu\mu},\,\,\hat{\phi}^{\mu\nu}=-\hat{\phi}^{\nu\mu},\,\,\hat{\pi}_{\,\,\,\mu}^{\mu}=0,\,\,u_{\lambda}\hat{\varpi}^{\lambda\mu\nu}=0,\,\,u_{\mu}\hat{\varpi}^{\lambda\mu\nu}=0,\,\,u_{\nu}\hat{\varpi}^{\lambda\mu\nu}=0.
	\end{aligned}
	\label{19}
\end{eqnarray}

The spin density $\hat{S}^{\mu\nu}$ is antisymmetric ($\hat{S}^{\mu\nu}=-\hat{S}^{\nu\mu}$) and satisfies the Frenkel condition, $\hat{S}^{\mu\nu}u_{\mu}=0$, due to the full antisymmetric nature of the spin current. Consequently, the spin chemical potential $\omega_{\mu\nu}$ is also transverse to the fluid velocity ($\omega_{\mu\nu}u^\mu=0$). In this work, we restrict our analysis to cases where the spin chemical potential is a first-order quantity, as it reduces to the thermal vorticity tensor $\varpi=-\partial_{[\mu}\beta_{\nu]}$ in global thermal equilibrium. Notably, the first-order dissipative current $\hat{\varpi}^{\lambda\mu\nu}$ is fully antisymmetric.

It is important to note that the equilibrium and bulk-viscous contributions to the pressure are not explicitly separated in this analysis. The statistical average of the operator $\hat{p}$ yields the true isotropic pressure under non-equilibrium conditions, which generally differs from the equilibrium pressure, $p\bigl(\langle\hat{\epsilon}\rangle,\langle\hat{n}_{a}\rangle,\langle\hat{S}^{\alpha\beta}\rangle\bigr)$, defined via averaging $\hat{p}$ over the local equilibrium distribution (evaluated at fixed thermodynamic parameters). The bulk viscous pressure is then defined as the difference between these two quantities.

The inclusion of spin degrees of freedom modifies the standard thermodynamic relations as follows
\begin{align}
Tds &= d\epsilon-\sum_{a}\mu_{a}dn_{a}-\frac{1}{2}\omega_{\alpha\beta}dS^{\alpha\beta} \qquad \text{First law of thermodynamics}, \label{20} \\
h = \epsilon + p &= Ts+\sum_{a}\mu_{a}n_{a}+\frac{1}{2}\omega_{\alpha\beta}S^{\alpha\beta} \qquad\,\,\,\, \text{Euler's relation}, \label{21} \\
dp &= sdT+\sum_{a}n_{a}d\mu_{a}+\frac{1}{2}S^{\alpha\beta}d\omega_{\alpha\beta} \quad\, \text{Gibbs-Duhem relation}, \label{22}
\end{align}
where $h$ represents the enthalpy density.

The operators appearing on the right-hand sides of Eqs.~\eqref{16}-\eqref{18} are defined through projections of the energy-momentum tensor $\hat{T}^{\mu\nu}$, charge currents $\hat{N}_{a}^{\mu}$, and spin tensor $\hat{S}^{\lambda\mu\nu}$
\begin{align} &\hat{\epsilon}=u_{\mu}u_{\nu}\hat{T}^{\mu\nu},\quad\hat{p}=-\frac{1}{3}\Delta_{\mu\nu}\hat{T}^{\mu\nu},\quad\hat{h}^{\mu}=\Delta_{(\alpha}^{\mu}u_{\beta)}\hat{T}^{\alpha\beta},\quad\hat{\pi}^{\mu\nu}=\Delta_{\alpha\beta}^{\mu\nu}\hat{T}^{\alpha\beta},\quad\hat{q}^{\mu}=\Delta_{[\alpha}^{\mu}u_{\beta]}\hat{T}^{\alpha\beta},\label{23}\\ &\hat{\phi}^{\mu\nu}=\mydelta_{\alpha\beta}^{\mu\nu}\hat{T}^{\alpha\beta},\quad\hat{n}_{a}=u_{\mu}\hat{N}_{a}^{\mu},\quad\hat{j}_{a}^{\nu}=\Delta_{\mu}^{\nu}\hat{N}_{a}^{\mu},\quad\hat{S}^{\mu\nu}=u_{\lambda}\hat{S}^{\lambda\mu\nu},\quad\hat{\varpi}^{\lambda\mu\nu}=\myDelta_{\rho\sigma\delta}^{\lambda\mu\nu}\hat{S}^{\rho\sigma\delta}.\label{24}
\end{align}
Here, the projectors orthogonal to $u^\mu$ are defined as
\begin{align} \Delta_{\rho\sigma}^{\mu\nu}&=\frac{1}{2}\left(\Delta_{\rho}^{\mu}\Delta_{\sigma}^{\nu}+\Delta_{\sigma}^{\mu}\Delta_{\rho}^{\nu}\right)-\frac{1}{3}\Delta^{\mu\nu}\Delta_{\rho\sigma},\label{25}\\ \mydelta_{\rho\sigma}^{\mu\nu}&=\frac{1}{2}\left(\Delta_{\rho}^{\mu}\Delta_{\sigma}^{\nu}-\Delta_{\sigma}^{\mu}\Delta_{\rho}^{\nu}\right),\label{26}\\ \myDelta_{\alpha\beta\gamma}^{\lambda\mu\nu}&=\frac{1}{6}\left(\Delta_{\alpha}^{\lambda}\Delta_{\beta}^{\mu}\Delta_{\gamma}^{\nu}+\Delta_{\alpha}^{\mu}\Delta_{\beta}^{\nu}\Delta_{\gamma}^{\lambda}+\Delta_{\alpha}^{\nu}\Delta_{\beta}^{\lambda}\Delta_{\gamma}^{\mu}-\Delta_{\alpha}^{\mu}\Delta_{\beta}^{\lambda}\Delta_{\gamma}^{\nu}-\Delta_{\alpha}^{\nu}\Delta_{\beta}^{\mu}\Delta_{\gamma}^{\lambda}-\Delta_{\alpha}^{\lambda}\Delta_{\beta}^{\nu}\Delta_{\gamma}^{\mu}\right).\label{27}
\end{align}

By substituting Eqs.~\eqref{16} to Eqs.~\eqref{18} into the conservation laws~\eqref{1}-~\eqref{3} and taking the statistical average with respect to the non-equilibrium operator, we derive the equations of relativistic spin hydrodynamics
\begin{align}
&Dn_{a}+n_{a}\theta+\partial_{\mu}j_{a}^{\mu}=0,\label{28}\\
&D\epsilon+(\epsilon+p+\Pi)\theta+\partial_{\mu}h^{\mu}-h^{\mu}Du_{\mu}-\pi^{\mu\nu}\sigma_{\mu\nu}+\partial_{\mu}q^{\mu}+q^{\mu}Du_{\mu}-\phi^{\mu\nu}\partial_{\mu}u_{\nu}=0,\label{29}\\
&(\epsilon+p+\Pi)Du_{\alpha}-\nabla_{\alpha}(p+\Pi)+\Delta_{\alpha\mu}Dh^{\mu}+h^{\mu}\partial_{\mu}u_{\alpha}+h_{\alpha}\theta+\Delta_{\alpha\nu}\partial_{\mu}\pi^{\mu\nu}+q^{\mu}\partial_{\mu}u_{\alpha}-q_{\alpha}\theta-\Delta_{\alpha\nu}Dq^{\nu}+\Delta_{\alpha\nu}\partial_{\mu}\phi^{\mu\nu}=0,\label{30}\\
&DS^{\mu\nu}+\theta S^{\mu\nu}+u^{\mu}\partial_{\lambda}S^{\nu\lambda}+S^{\nu\lambda}\partial_{\lambda}u^{\mu}+u^{\nu}\partial_{\lambda}S^{\lambda\mu}+S^{\lambda\mu}\partial_{\lambda}u^{\nu}+\partial_{\lambda}\varpi^{\lambda\mu\nu}+2q^{\mu}u^{\nu}-2q^{\nu}u^{\mu}+2\phi^{\mu\nu}=0.\label{31}
\end{align}
Here, the averaged quantities are defined as $n_{a}\equiv\langle\hat{n}_{a}\rangle$, $j_{a}^{\mu}\equiv\langle\hat{j}_{a}^{\mu}\rangle$, $\epsilon\equiv\langle\hat{\epsilon}\rangle$, $h^{\mu}\equiv\langle\hat{h}^{\mu}\rangle$, $\pi^{\mu\nu}\equiv\langle\hat{\pi}^{\mu\nu}\rangle$, $q^{\mu}\equiv\langle\hat{q}^{\mu}\rangle$, $\phi^{\mu\nu}\equiv\langle\hat{\phi}^{\mu\nu}\rangle$, $S^{\mu\nu}\equiv\langle\hat{S}^{\mu\nu}\rangle$, and $\varpi^{\lambda\mu\nu}\equiv\langle\hat{\varpi}^{\lambda\mu\nu}\rangle$. The local equilibrium pressure $p\equiv p\left(\epsilon,n_{a},S^{\alpha\beta}\right)$ is determined by the equation of state (EoS), with $\Pi$ denoting the non-equilibrium correction. We employ the comoving derivative $D\equiv u^{\mu}\partial_{\mu}$, covariant spatial derivative $\nabla_\alpha \equiv \Delta_{\alpha \beta} \partial^\beta$, shear tensor $\sigma_{\mu \nu} \equiv \Delta_{\mu \nu}^{\alpha \beta} \partial_\alpha u_\beta$, and expansion scalar $\theta\equiv\partial_{\mu}u^{\mu}$, which measures the fluid's expansion ($\theta>0$) or contraction ($\theta<0$). The velocity gradient decomposes as $\partial_{\mu}u_{\nu}=u_{\mu}Du_{\nu}+\frac{1}{3}\theta\Delta_{\mu\nu}+\sigma_{\mu\nu}+\nabla_{[\mu}u_{\nu]}$. Eqs.~\eqref{29} and \eqref{30} are derived by contracting Eq.~\eqref{2} with $u_\nu$ and $\Delta_{\nu\alpha}$, respectively. This study emphasizes generality by not assuming a specific frame unless explicitly required.

By substituting Eqs.~\eqref{16}-\eqref{18} into Eqs~\eqref{1}-\eqref{3} and performing and average over the local equilibrium distribution, we obtain the equations of ideal spin hydrodynamics, 
\begin{align}
	&Dn_{a}+n_{a}\theta=0,\label{32}\\
	&D\epsilon+\left(\epsilon+p\right)\theta=0,\label{33}\\
	&DS^{\mu\nu}+\theta S^{\mu\nu}+u^{\mu}\partial_{\lambda}S^{\nu\lambda}+S^{\nu\lambda}\partial_{\lambda}u^{\mu}+u^{\nu}\partial_{\lambda}S^{\lambda\mu}+S^{\lambda\mu}\partial_{\lambda}u^{\nu}=0,\label{34}\\
	&\left(\epsilon+p\right)Du_{\alpha}=\nabla_{\alpha}p.\label{35}
\end{align}
Here, we have used the fact that the statistical average of dissipative operators vanishes in local equilibrium
\begin{equation}	\langle\hat{h}^{\mu}\rangle_{l}=0,\quad\langle\hat{\pi}^{\mu\nu}\rangle_{l}=0,\quad\langle\hat{q}^{\mu}\rangle_{l}=0,\quad\langle\hat{\phi}^{\mu\nu}\rangle_{l}=0,\quad\langle\hat{j}_{a}^{\mu}\rangle_{l}=0,\quad\langle\hat{\varpi}^{\lambda\mu\nu}\rangle_{l}=0.
\label{36}
\end{equation}

\subsection{Decomposition into different dissipative processes}

To compute the statistical averages of dissipative fluxes, we decompose the operator $\hat{C}$ in Eq.~\eqref{7} into contributions from distinct dissipative processes using Eqs.~\eqref{16}-\eqref{18}. By employing the orthogonality conditions in Eq.~\eqref{19}, we express $\hat{C}$ as~\cite{She:2024rnx},
\begin{eqnarray}
	\begin{aligned}
		\hat{C}= & \hat{\epsilon}D\beta-\hat{p}\beta\theta-\sum_{a}\hat{n}_{a}D\alpha_{a}-\frac{1}{2}\hat{S}^{\alpha\beta}D\Omega_{\alpha\beta}+\hat{h}^{\mu}\left(\beta Du_{\mu}+\partial_{\mu}\beta\right)-\sum_{a}\hat{j}_{a}^{\mu}\partial_{\mu}\alpha_{a}\\
		& +\hat{q}^{\mu}\left(\partial_{\mu}\beta-\beta Du_{\mu}\right)+\beta\hat{\pi}^{\mu\nu}\partial_{\mu}u_{\nu}-u^{\alpha}\hat{S}^{\beta\lambda}\partial_{\lambda}\Omega_{\alpha\beta}+\hat{\phi}^{\mu\nu}\left(\beta\partial_{\mu}u_{\nu}+\Omega_{\mu\nu}\right)-\frac{1}{2}\hat{\varpi}^{\lambda\alpha\beta}\partial_{\lambda}\Omega_{\alpha\beta}.
	\end{aligned}
	\label{37}
\end{eqnarray}
For conciseness, we reformulate the first law of thermodynamics and the Gibbs-Duhem relation as
\begin{eqnarray}
	ds=\beta d\epsilon-\sum_{a}\alpha_{a}dn_{a}-\frac{1}{2}\Omega_{\alpha\beta}dS^{\alpha\beta},\quad\beta dp=-h d\beta+\sum_{a}n_{a}d\alpha_{a}+\frac{1}{2}S^{\alpha\beta}d\Omega_{\alpha\beta}.
	\label{38}
\end{eqnarray}
From the first equation, we derive the following Maxwell relations
\begin{eqnarray}
	\begin{aligned}
		& \frac{\partial\beta}{\partial n_{a}}\bigg|_{\epsilon,n_{b}\neq n_{a},S^{\alpha\beta}}=-\frac{\partial\alpha_{a}}{\partial\epsilon}\bigg|_{n_{b},S^{\alpha\beta}},\quad\frac{\partial\alpha_{c}}{\partial n_{a}}\bigg|_{\epsilon,n_{b}\neq n_{a},S^{\alpha\beta}}=\frac{\partial\alpha_{a}}{\partial n_{c}}\bigg|_{\epsilon,n_{b}\neq n_{c},S^{\alpha\beta}},\\
		& \frac{\partial\beta}{\partial S^{\alpha\beta}}\bigg|_{\epsilon,n_{b}}=-\frac{1}{2}\frac{\partial\Omega_{\alpha\beta}}{\partial\epsilon}\bigg|_{n_{b},S^{\alpha\beta}},\quad\frac{1}{2}\frac{\partial\Omega_{\alpha\beta}}{\partial n_{a}}\bigg|_{\epsilon,n_{b}\neq n_{a},S^{\alpha\beta}}=\frac{\partial\alpha_{a}}{\partial S^{\alpha\beta}}\bigg|_{\epsilon,n_{b}},\\
		& \frac{\partial\Omega_{\alpha\beta}}{\partial S^{\lambda\rho}}\bigg|_{\epsilon,n_{b},S^{\alpha\beta}\neq S^{\lambda\rho}}=\frac{\partial\Omega_{\lambda\rho}}{\partial S^{\alpha\beta}}\bigg|_{\epsilon,n_{b},S^{\lambda\rho}\neq S^{\alpha\beta}},
	\end{aligned}
	\label{39}
\end{eqnarray}
while the Gibbs-Duhem relation yields
\begin{eqnarray}
	h=-\beta\frac{\partial p}{\partial\beta}\bigg|_{\alpha_{a},\Omega_{\alpha\beta}},\quad n_{a}=\beta\frac{\partial p}{\partial\alpha_{a}}\bigg|_{\beta,\alpha_{b}\neq\alpha_{a},\Omega_{\alpha\beta}},\quad S^{\alpha\beta}=2\beta\frac{\partial p}{\partial\Omega_{\alpha\beta}}\bigg|_{\beta,\alpha_{a}}.
	\label{40}
\end{eqnarray}

We now employ the full set of hydrodynamic equations~\eqref{28}-\eqref{31} to eliminate the terms $D\beta,D\alpha_{a}$, and $D\Omega_{\alpha\beta}$ in Eq.~\eqref{37}. Treating $\epsilon$, $n_a$, and $S^{\alpha\beta}$ as independent thermodynamic variables, we obtain the following expressions
\begin{eqnarray}
	\begin{aligned}
		D\beta= & \beta\theta\Gamma-\biggl(\Pi\theta+\partial_{\mu}h^{\mu}-h^{\mu}Du_{\mu}-\pi^{\mu\nu}\sigma_{\mu\nu}+\partial_{\mu}q^{\mu}+q^{\mu}Du_{\mu}-\phi^{\mu\nu}\partial_{\mu}u_{\nu}\biggr)\frac{\partial\beta}{\partial\epsilon}\bigg|_{n_{a},S^{\alpha\beta}}-\sum_{a}\partial_{\mu}j_{a}^{\mu}\frac{\partial\beta}{\partial n_{a}}\bigg|_{\epsilon,n_{b}\neq n_{a},S^{\alpha\beta}}\\ &-\left(u^{\alpha}\partial_{\lambda}S^{\beta\lambda}+S^{\beta\lambda}\partial_{\lambda}u^{\alpha}+u^{\beta}\partial_{\lambda}S^{\lambda\alpha}+S^{\lambda\alpha}\partial_{\lambda}u^{\beta}+\partial_{\lambda}\varpi^{\lambda\alpha\beta}+2q^{\alpha}u^{\beta}-2q^{\beta}u^{\alpha}+2\phi^{\alpha\beta}\right)\frac{\partial\beta}{\partial S^{\alpha\beta}}\bigg|_{\epsilon,n_{a}},
	\end{aligned}
	\label{41}
\end{eqnarray}
\begin{eqnarray}
	\begin{aligned}
		D\alpha_{c}= & -\beta\theta\delta_{c}-\biggl(\Pi\theta+\partial_{\mu}h^{\mu}-h^{\mu}Du_{\mu}-\pi^{\mu\nu}\sigma_{\mu\nu}+\partial_{\mu}q^{\mu}+q^{\mu}Du_{\mu}-\phi^{\mu\nu}\partial_{\mu}u_{\nu}\biggr)\frac{\partial\alpha_{c}}{\partial\epsilon}\bigg|_{n_{a},S^{\alpha\beta}}-\sum_{a}\partial_{\mu}j_{a}^{\mu}\frac{\partial\alpha_{c}}{\partial n_{a}}\bigg|_{\epsilon,n_{b}\neq n_{a},S^{\alpha\beta}}\\
		&-\left(u^{\alpha}\partial_{\lambda}S^{\beta\lambda}+S^{\beta\lambda}\partial_{\lambda}u^{\alpha}+u^{\beta}\partial_{\lambda}S^{\lambda\alpha}+S^{\lambda\alpha}\partial_{\lambda}u^{\beta}+\partial_{\lambda}\varpi^{\lambda\alpha\beta}+2q^{\alpha}u^{\beta}-2q^{\beta}u^{\alpha}+2\phi^{\alpha\beta}\right)\frac{\partial\alpha_{c}}{\partial S^{\alpha\beta}}\bigg|_{\epsilon,n_{a}},
	\end{aligned}
	\label{42}
\end{eqnarray}
and
\begin{eqnarray}
	\begin{aligned}
		D\Omega_{\alpha\beta}= & -2\beta\theta\mathcal{K}_{\alpha\beta}-\biggl(\Pi\theta+\partial_{\mu}h^{\mu}-h^{\mu}Du_{\mu}-\pi^{\mu\nu}\sigma_{\mu\nu}+\partial_{\mu}q^{\mu}+q^{\mu}Du_{\mu}-\phi^{\mu\nu}\partial_{\mu}u_{\nu}\biggr)\frac{\partial\Omega_{\alpha\beta}}{\partial\epsilon}\bigg|_{n_{a},S^{\alpha\beta}}-\sum_{a}\partial_{\mu}j_{a}^{\mu}\frac{\partial\Omega_{\alpha\beta}}{\partial n_{a}}\bigg|_{\epsilon,n_{b}\neq n_{a},S^{\alpha\beta}}\\
		&-\left(u^{\lambda}\partial_{\delta}S^{\rho\delta}+S^{\rho\delta}\partial_{\delta}u^{\lambda}+u^{\rho}\partial_{\delta}S^{\delta\lambda}+S^{\delta\lambda}\partial_{\delta}u^{\rho}+\partial_{\delta}\varpi^{\delta\lambda\rho}+2q^{\lambda}u^{\rho}-2q^{\rho}u^{\lambda}+2\phi^{\lambda\rho}\right)\frac{\partial\Omega_{\alpha\beta}}{\partial S^{\lambda\rho}}\bigg|_{\epsilon,n_{a}},
	\end{aligned}
	\label{43}
\end{eqnarray}
where
\begin{eqnarray}
	\Gamma\equiv\frac{\partial p}{\partial\epsilon}\bigg|_{n_{a},S^{\alpha\beta}},\quad\delta_{c}\equiv\frac{\partial p}{\partial n_{c}}\bigg|_{\epsilon,n_{b}\neq n_{c},S^{\alpha\beta}},\quad\mathcal{K}_{\alpha\beta}\equiv\frac{\partial p}{\partial S^{\alpha\beta}}\bigg|_{\epsilon,n_{a}}.
	\label{44}
\end{eqnarray}

The first four terms in Eq.~\eqref{37} can be combined into the following form
\begin{eqnarray}
	\begin{aligned} &\hat{\epsilon}D\beta-\hat{p}\beta\theta-\sum_{c}\hat{n}_{c}D\alpha_{c}-\frac{1}{2}\hat{S}^{\alpha\beta}D\Omega_{\alpha\beta}\\ =&-\beta\theta\hat{P}^{*}-\hat{\mathcal{A}}^{*}\left(\Pi\theta+\partial_{\mu}h^{\mu}-h^{\mu}Du_{\mu}-\pi^{\mu\nu}\sigma_{\mu\nu}+\partial_{\mu}q^{\mu}+q^{\mu}Du_{\mu}-\phi^{\mu\nu}\partial_{\mu}u_{\nu}\right)+\sum_{a}\hat{\mathcal{B}}_{a}^{*}\partial_{\mu}j_{a}^{\mu}\\ &+\frac{1}{2}\left(u^{\alpha}\partial_{\lambda}S^{\beta\lambda}+S^{\beta\lambda}\partial_{\lambda}u^{\alpha}+u^{\beta}\partial_{\lambda}S^{\lambda\alpha}+S^{\lambda\alpha}\partial_{\lambda}u^{\beta}+\partial_{\lambda}\varpi^{\lambda\alpha\beta}+2q^{\alpha}u^{\beta}-2q^{\beta}u^{\alpha}+2\phi^{\alpha\beta}\right)\hat{\mathcal{C}}_{\alpha\beta}^{*},
	\end{aligned}
	\label{45}
\end{eqnarray}
where we define the following composite operators
\begin{align}  \hat{P}^{*}=&\left(\hat{p}-\Gamma\hat{\epsilon}-\sum_{a}\delta_{a}\hat{n}_{a}\right)-\mathcal{K}_{\alpha\beta}\hat{S}^{\alpha\beta}=\hat{p}^{*}-\mathcal{K}_{\alpha\beta}\hat{S}^{\alpha\beta},\label{46}\\
	\hat{\mathcal{A}}^{*}=&\hat{\epsilon}\frac{\partial\beta}{\partial\epsilon}\bigg|_{n_{a},S^{\alpha\beta}}+\sum_{a}\hat{n}_{a}\frac{\partial\beta}{\partial n_{a}}\bigg|_{\epsilon,n_{b}\neq n_{a},S^{\alpha\beta}}+\hat{S}^{\alpha\beta}\frac{\partial\beta}{\partial S^{\alpha\beta}}\bigg|_{\epsilon,n_{b}}=\sum_{i}\hat{\mathfrak{D}}_{i}\partial_{\epsilon n}^{i}\beta+\hat{S}^{\alpha\beta}\mathcal{D}_{\alpha\beta},\label{47}\\
	\hat{\mathcal{B}}_{a}^{*}=&\hat{\epsilon}\frac{\partial\alpha_{a}}{\partial\epsilon}\bigg|_{n_{b},S^{\alpha\beta}}+\sum_{c}\hat{n}_{c}\frac{\partial\alpha_{a}}{\partial n_{c}}\bigg|_{\epsilon,n_{b}\neq n_{c},S^{\alpha\beta}}+\hat{S}^{\alpha\beta}\frac{\partial\alpha_{a}}{\partial S^{\alpha\beta}}\bigg|_{\epsilon,n_{b}}=\sum_{i}\hat{\mathfrak{D}}_{i}\partial_{\epsilon n}^{i}\alpha_{a}+\hat{S}^{\alpha\beta}\mathcal{E}_{\alpha\beta}^{a},\label{48}\\
	\hat{\mathcal{C}}_{\alpha\beta}^{*}=&\hat{\epsilon}\frac{\partial\Omega_{\alpha\beta}}{\partial\epsilon}\bigg|_{n_{b},S^{\alpha\beta}}+\sum_{c}\hat{n}_{c}\frac{\partial\Omega_{\alpha\beta}}{\partial n_{c}}\bigg|_{\epsilon,n_{b}\neq n_{c},S^{\alpha\beta}}+\hat{S}^{\delta\rho}\frac{\partial\Omega_{\alpha\beta}}{\partial S^{\delta\rho}}\bigg|_{\epsilon,n_{b},S^{\alpha\beta}\neq S^{\delta\rho}}=\sum_{i}\hat{\mathfrak{D}}_{i}\partial_{\epsilon n}^{i}\Omega_{\alpha\beta}+\hat{S}^{\delta\rho}\mathcal{F}_{\alpha\beta\delta\rho},\label{49}\\
	\mathcal{D}_{\alpha\beta}= & \frac{\partial\beta}{\partial S^{\alpha\beta}}\bigg|_{\epsilon,n_{b}},\mathcal{E}_{\alpha\beta}^{a}=\frac{\partial\alpha_{a}}{\partial S^{\alpha\beta}}\bigg|_{\epsilon,n_{b}},\mathcal{F}_{\alpha\beta\delta\rho}=\frac{\partial\Omega_{\alpha\beta}}{\partial S^{\delta\rho}}\bigg|_{\epsilon,n_{b},S^{\alpha\beta}\neq S^{\delta\rho}},\label{50}\\
	\hat{\mathfrak{D}}_{i}= & \left(\hat{\epsilon},\hat{n}_{1},\hat{n}_{2},\cdots,\hat{n}_{l}\right),\partial_{\epsilon n}^{i}=\left(\frac{\partial}{\partial\epsilon},\frac{\partial}{\partial n_{1}},\frac{\partial}{\partial n_{2}},\cdots,\frac{\partial}{\partial n_{l}}\right),\sum_{i}=\sum_{i=0}^{l}=\sum_{i=0}+\sum_{a=1}^{l}.\label{51}
\end{align}
Here, the index $i=0$ corresponds to the first element, with $\hat{\mathfrak{D}}_{0}=\hat{\epsilon}$ and $\partial_{\epsilon n}^{0}=\frac{\partial}{\partial\epsilon}$, while indices $i\geq1$ denote the subsequent elements in the sequence.

We now combine Eq.~\eqref{30} with the modified pressure gradient obtained from the second relation in Eq.~\eqref{38}, yielding
\begin{eqnarray}
\begin{aligned}
hDu_{\sigma}= & -\Pi Du_{\sigma}-hT\nabla_{\sigma}\beta+T\sum_{a}n_{a}\nabla_{\sigma}\alpha_{a}+\frac{1}{2}TS^{\alpha\beta}\nabla_{\sigma}\Omega_{\alpha\beta}+\nabla_{\sigma}\Pi-\Delta_{\sigma\mu}Dh^{\mu}\\
&-h^{\mu}\partial_{\mu}u_{\sigma}-h_{\sigma}\theta-\Delta_{\sigma\nu}\partial_{\mu}\pi^{\mu\nu}-q^{\mu}\partial_{\mu}u_{\sigma}+q_{\sigma}\theta+\Delta_{\sigma\nu}Dq^{\nu}-\Delta_{\sigma\nu}\partial_{\mu}\phi^{\mu\nu}.
\end{aligned}
\label{52}
\end{eqnarray}
This expression is then used to modify the vector terms containing $\hat{h}^\mu$ and $\hat{q}^\mu$ in Eq.~\eqref{37}. Utilizing the identities $\hat{h}^{\sigma}\partial_{\sigma}=\hat{h}^{\sigma}\nabla_{\sigma}$ and $\hat{q}^{\sigma}\partial_{\sigma}=\hat{q}^{\sigma}\nabla_{\sigma}$, we obtain
\begin{eqnarray}
\begin{aligned}\hat{h}^{\sigma}\left(\beta Du_{\sigma}+\partial_{\sigma}\beta\right)= & \sum_{a}n_{a}h^{-1}\hat{h}^{\sigma}\nabla_{\sigma}\alpha_{a}-\hat{h}^{\sigma}\beta h^{-1}\biggl(-\frac{1}{2}\beta^{-1}S^{\alpha\beta}\nabla_{\sigma}\Omega_{\alpha\beta}-\nabla_{\sigma}\Pi+\Pi Du_{\sigma}+\Delta_{\sigma\nu}Dh^{\nu}\\
&+h^{\mu}\partial_{\mu}u_{\sigma}+h_{\sigma}\theta+\Delta_{\sigma\nu}\partial_{\mu}\pi^{\mu\nu}+q^{\mu}\partial_{\mu}u_{\sigma}-q_{\sigma}\theta-\Delta_{\sigma\nu}Dq^{\nu}+\Delta_{\sigma\nu}\partial_{\mu}\phi^{\mu\nu}\biggr),
\end{aligned}
\label{53}
\end{eqnarray}
and
\begin{eqnarray}
\begin{aligned}
\hat{q}^{\mu}\left(\partial_{\mu}\beta-\beta Du_{\mu}\right)= & -\beta\hat{q}^{\mu}\left(h^{-1}\nabla_{\mu}p+\beta\nabla_{\mu}T\right)+\hat{q}^{\mu}\beta h^{-1}\Bigl(\Pi Du_{\mu}-\nabla_{\mu}\Pi+\Delta_{\mu\nu}Dh^{\nu}\\
&+h^{\nu}\partial_{\nu}u_{\mu}+h_{\mu}\theta+\Delta_{\mu\nu}\partial_{\rho}\pi^{\rho\nu}+q^{\nu}\partial_{\nu}u^{\mu}-q_{\mu}\theta-\Delta_{\mu\nu}Dq^{\nu}+\Delta_{\mu\nu}\partial_{\rho}\phi^{\rho\nu}\Bigr).
\end{aligned}
\label{54}
\end{eqnarray}

By combining Eqs.~\eqref{37}, \eqref{45}, \eqref{53}, \eqref{54} with the velocity gradient decomposition $\partial_{\mu}u_{\nu}=u_{\mu}Du_{\nu}+\frac{1}{3}\theta\Delta_{\mu\nu}+\sigma_{\mu\nu}+\nabla_{[\mu}u_{\nu]}$, we can express the operator $\hat{C}$ as a sum of first-order and second-order contributions
\begin{eqnarray}
	\hat{C}\left(x\right)=\hat{C}_{1}\left(x\right)+\hat{C}_{2}\left(x\right),
	\label{55}
\end{eqnarray}
where the components are given by
\begin{align}
	\hat{C}_{1}\left(x\right)=&-\beta\theta\hat{p}^{*}-\sum_{a}\hat{\mathscr{J}}_{a}^{\sigma}\nabla_{\sigma}\alpha_{a}+\beta\hat{q}^{\mu}M_{\mu}+\beta\hat{\pi}^{\mu\nu}\sigma_{\mu\nu}+\beta\hat{\phi}^{\mu\nu}\xi_{\mu\nu},\label{56}\\
	\hat{C}_{2}\left(x\right)=&\beta\hat{S}^{\alpha\beta}\mathcal{R}_{\alpha\beta}+\beta\sum_{i}\left[\left(\hat{\mathfrak{D}}_{i}\partial_{\epsilon n}^{i}\beta\right)\mathcal{X}+\sum_{a}\left(\hat{\mathfrak{D}_{i}}\partial_{\epsilon n}^{i}\alpha_{a}\right)\mathcal{Y}_{a}+\left(\hat{\mathfrak{D}}_{i}\partial_{\epsilon n}^{i}\Omega_{\mu\nu}\right)\mathcal{Z}^{\mu\nu}\right]+\beta\hat{h}^{\sigma}\mathcal{H}_{\sigma}+\beta\hat{q}^{\mu}\mathcal{Q}_{\mu}+\hat{\varpi}^{\lambda\alpha\beta}\varXi_{\lambda\alpha\beta}.
	\label{57}
\end{align}
The various quantities appearing in these expressions are defined as follows
\begin{eqnarray}
\begin{aligned}
\hat{\mathscr{J}}_{a}^{\sigma}= & \hat{j}_{a}^{\sigma}-\frac{n_{a}}{\epsilon+p}\hat{h}^{\sigma},M_{\mu}=-\left(h^{-1}\nabla_{\mu}p+\beta\nabla_{\mu}T\right),\xi_{\mu\nu}=\mydelta_{\mu\nu}^{\lambda\delta}\left(\partial_{\lambda}u_{\delta}+\beta^{-1}\Omega_{\lambda\delta}\right),\\
\mathcal{Z}^{\alpha\beta} & =\beta^{-1}\left(u^{[\alpha}\partial_{\lambda}S^{\beta]\lambda}+S^{[\beta\lambda}\partial_{\lambda}u^{\alpha]}+q^{\alpha}u^{\beta}-q^{\beta}u^{\alpha}+\phi^{\alpha\beta}\right),\mathcal{W}_{\beta\lambda}=-\beta^{-1}\mydelta_{\beta\lambda}^{\rho\sigma}u^{\alpha}\partial_{\sigma}\Omega_{\alpha\rho}\\
\varXi_{\lambda\alpha\beta}= & -\frac{1}{2}\myDelta_{\lambda\alpha\beta\rho\sigma\delta}\partial^{\rho}\Omega^{\sigma\delta},\mathcal{Y}_{a}=\beta^{-1}\partial_{\mu}j_{a}^{\mu},\mathcal{G}^{\alpha\beta}=\frac{1}{2}\beta^{-1}\partial_{\lambda}\varpi^{\lambda\alpha\beta},\sigma_{\mu\nu}=\Delta_{\mu\nu}^{\lambda\delta}\partial_{\lambda}u_{\delta},\\
\mathcal{R}_{\alpha\beta}= & \theta\mathcal{K}_{\alpha\beta}+\mathcal{W}_{\alpha\beta},\mathcal{X}=-\beta^{-1}\left(\Pi\theta+\partial_{\mu}h^{\mu}-h^{\mu}Du_{\mu}-\pi^{\mu\nu}\sigma_{\mu\nu}+\partial_{\mu}q^{\mu}+q^{\mu}Du_{\mu}-\phi^{\mu\nu}\partial_{\mu}u_{\nu}\right),\\
\mathcal{H}_{\sigma}= & -h^{-1}\biggl(-\frac{1}{2}\beta^{-1}S^{\alpha\beta}\nabla_{\sigma}\Omega_{\alpha\beta}-\nabla_{\sigma}\Pi+\Pi Du_{\sigma}+\Delta_{\sigma\nu}Dh^{\nu}+h^{\mu}\partial_{\mu}u_{\sigma}+h_{\sigma}\theta\\
&\qquad\qquad+\Delta_{\sigma\nu}\partial_{\mu}\pi^{\mu\nu}+q^{\mu}\partial_{\mu}u_{\sigma}-q_{\sigma}\theta-\Delta_{\sigma\nu}Dq^{\nu}+\Delta_{\sigma\nu}\partial_{\mu}\phi^{\mu\nu}\biggr),\\
\mathcal{Q}_{\mu}= & h^{-1}\left(\Pi Du_{\mu}-\nabla_{\mu}\Pi+\Delta_{\mu\nu}Dh^{\nu}+h^{\nu}\partial_{\nu}u_{\mu}+h_{\mu}\theta+\Delta_{\mu\nu}\partial_{\rho}\pi^{\rho\nu}+q^{\nu}\partial_{\nu}u^{\mu}-q_{\mu}\theta-\Delta_{\mu\nu}Dq^{\nu}+\Delta_{\mu\nu}\partial_{\rho}\phi^{\rho\nu}\right).
\end{aligned}
\label{58}
\end{eqnarray}
An important consistency requirement emerges from the Maxwell relations: when the spin chemical potential scales as $\omega_{\mu\nu}\sim\mathcal{O}\left(\partial\right)$, we must take $\mathcal{K}_{\alpha\beta},\mathcal{D}_{\alpha\beta}$, and $\mathcal{E}_{\alpha\beta}^{a}$ to be first-order in gradients, while $\mathcal{F}_{\alpha\beta\delta\rho}$ must be treated as a second-order gradient expansion.

The expressions in Eq.~\eqref{57} can be naturally interpreted as generalized thermodynamic forces. These second-order quantities involve either spacetime derivatives of dissipative fluxes or products of two first-order terms. Of particular significance is the thermodynamic force $\hat{C}$, which appears in the correlators of Eq.~\eqref{12} with the thermodynamic parameters evaluated at the point $x_1$. This introduces non-local corrections to dissipative current averages through second-order two-point correlators.

As demonstrated in Ref.~\cite{Harutyunyan:2025fgs}, the two-point correlation functions between tensors of different ranks at distinct spacetime points introduce additional second-order nonlocal corrections. To compute these nonlocal corrections in spin fluids, we expand the thermodynamic forces $\partial^{\mu}\beta^{\nu},\partial_{\mu}\alpha_{a},\partial_{\lambda}\Omega_{\alpha\beta}$ and $\Omega_{\alpha\beta}$ in the operator $\hat{C}$ around the point $x$ using a first-order Taylor series as $\partial^{\mu}\beta^{\nu}\left(x_{1}\right)=\partial^{\mu}\beta^{\nu}\left(x\right)+\partial_{\tau}\partial^{\mu}\beta^{\nu}\left(x\right)\left(x_{1}-x\right)^{\tau},\partial_{\mu}\alpha_{a}\left(x_{1}\right)=\partial_{\mu}\alpha_{a}\left(x\right)+\partial_{\tau}\partial_{\mu}\alpha_{a}\left(x\right)\left(x_{1}-x\right)^{\tau},\partial_{\lambda}\Omega_{\alpha\beta}\left(x_{1}\right)=\partial_{\mu}\Omega_{\alpha\beta}\left(x\right)+\partial_{\tau}\partial_{\mu}\Omega_{\alpha\beta}\left(x\right)\left(x_{1}-x\right)^{\tau},\Omega_{\alpha\beta}\left(x_{1}\right)=\Omega_{\alpha\beta}\left(x\right)+\partial_{\tau}\Omega_{\alpha\beta}\left(x\right)\left(x_{1}-x\right)^{\tau}$, and
\begin{eqnarray} \hat{C}\left(x_{1}\right)=\hat{C}\left(x_{1}\right)_{x}+\left(x_{1}-x\right)^{\tau}\partial_{\tau}\hat{C}.
	\label{59}
\end{eqnarray}
Here, $\hat{C}\left(x_{1}\right)_{x}$ is obtained by replacing all hydrodynamic quantities in $\hat{C}\left(x_{1}\right)$ with their values at $x$ while keeping the microscopic quantum operators ($\hat{T}^{\mu\nu},\hat{N}_{a}^{\mu}$ and $\hat{S}^{\lambda\alpha\beta}$) unchanged. 

The derivative operator $\partial_{\tau}\hat{C}$ can be computed using the relation $\partial_{\tau}\Delta_{\gamma\delta}=-u_{\gamma}\partial_{\tau}u_{\delta}-u_{\delta}\partial_{\tau}u_{\gamma}$ from Eqs.~\eqref{25}-\eqref{26}, yielding
\begin{eqnarray}
	\begin{aligned}
		\partial_{\tau}\Delta_{\gamma\delta\rho\sigma}= & -\frac{1}{2}\biggl[\left(u_{\gamma}\partial_{\tau}u_{\rho}+u_{\rho}\partial_{\tau}u_{\gamma}\right)\Delta_{\delta\sigma}+\Delta_{\gamma\rho}\left(u_{\sigma}\partial_{\tau}u_{\delta}+u_{\delta}\partial_{\tau}u_{\sigma}\right)\\
		&+\left(u_{\gamma}\partial_{\tau}u_{\sigma}+u_{\sigma}\partial_{\tau}u_{\gamma}\right)\Delta_{\delta\rho}+\Delta_{\gamma\sigma}\left(u_{\delta}\partial_{\tau}u_{\rho}+u_{\rho}\partial_{\tau}u_{\delta}\right)\biggr]\\
		&+\frac{1}{3}\biggl[\left(u_{\gamma}\partial_{\tau}u_{\delta}+u_{\delta}\partial_{\tau}u_{\gamma}\right)\Delta_{\rho\sigma}+\Delta_{\gamma\delta}\left(u_{\rho}\partial_{\tau}u_{\sigma}+u_{\sigma}\partial_{\tau}u_{\rho}\right)\biggr],
	\end{aligned}
	\label{60}
\end{eqnarray}
\begin{eqnarray}
	\begin{aligned}
		\partial_{\tau}\mydelta_{\gamma\delta\rho\sigma}= & -\frac{1}{2}\biggl[\left(u_{\gamma}\partial_{\tau}u_{\rho}+u_{\rho}\partial_{\tau}u_{\gamma}\right)\Delta_{\delta\sigma}+\Delta_{\gamma\rho}\left(u_{\sigma}\partial_{\tau}u_{\delta}+u_{\delta}\partial_{\tau}u_{\sigma}\right)\\
		&-\left(u_{\gamma}\partial_{\tau}u_{\sigma}+u_{\sigma}\partial_{\tau}u_{\gamma}\right)\Delta_{\delta\rho}-\Delta_{\gamma\sigma}\left(u_{\delta}\partial_{\tau}u_{\rho}+u_{\rho}\partial_{\tau}u_{\delta}\right)\biggr],
	\end{aligned}
	\label{61}
\end{eqnarray}
these results will be essential for our subsequent analysis. 

By substituting Eqs.~\eqref{16}-\eqref{18} into Eq.~\eqref{7}, we derive the following expression for $\partial_{\tau}\hat{C}$
\begin{eqnarray}
\begin{aligned}
\partial_{\tau}\hat{C}=&\hat{T}_{\rho\sigma}\partial_{\tau}\partial^{\rho}\left(\beta u^{\sigma}\right)-\sum_{a}\hat{N}_{a}^{\rho}\partial_{\tau}\partial_{\rho}\alpha_{a}-\frac{1}{2}\hat{S}^{\lambda\alpha\beta}\partial_{\tau}\partial_{\lambda}\Omega_{\alpha\beta}+\partial_{\tau}\Omega_{\alpha\beta}\hat{T}^{[\alpha\beta]}\\		=&\left(\hat{\epsilon}u_{\rho}u_{\sigma}-\hat{p}\Delta_{\rho\sigma}+\hat{h}_{\rho}u_{\sigma}+\hat{h}_{\sigma}u_{\rho}+\hat{\pi}_{\rho\sigma}+\hat{q}_{\rho}u_{\sigma}-\hat{q}_{\sigma}u_{\rho}+\hat{\phi}_{\rho\sigma}\right)\\ &\times\left[\beta\partial_{\tau}\partial^{\rho}u^{\sigma}+u^{\sigma}\partial_{\tau}\partial^{\rho}\beta+\left(\partial_{\tau}\beta\right)\left(\partial^{\rho}u^{\sigma}\right)+\left(\partial_{\tau}u^{\sigma}\right)\left(\partial^{\rho}\beta\right)\right]\\ &-\sum_{a}\left(\hat{n}_{a}u^{\rho}+\hat{j}_{a}^{\rho}\right)\partial_{\tau}\partial_{\rho}\alpha_{a}+\partial_{\tau}\Omega_{\alpha\beta}\left(\hat{q}^{\alpha}u^{\beta}-\hat{q}^{\beta}u^{\alpha}+\hat{\phi}^{\alpha\beta}\right)\\ =&\hat{\epsilon}X_{\tau}^{\left(\epsilon\right)}-\hat{p}X_{\tau}^{\left(p\right)}+\hat{h}_{\rho}X_{\tau}^{\left(h\right)}+\hat{\pi}_{\rho\sigma}X_{\tau}^{\left(\pi\right)}+\hat{q}_{\rho}X_{\tau}^{\left(q\right)}+\hat{\phi}_{\rho\sigma}X_{\tau}^{\left(\phi\right)}-\sum_{a}\left(\hat{n}_{a}X_{\tau}^{\left(n_{a}\right)}+\hat{j}_{a}^{\rho}X_{\tau}^{\left(j_{a}\right)}\right),
\end{aligned}
\label{62}
\end{eqnarray}
where we have neglected the higher-order term $\frac{1}{2}\hat{S}^{\lambda\alpha\beta}\partial_{\tau}\partial_{\lambda}\Omega_{\alpha\beta}$. The corresponding thermodynamic forces are given by
\begin{align}
	X_{\tau}^{\left(\epsilon\right)}= & \beta u^{\rho}u^{\sigma}\left(\partial_{\tau}\partial_{\rho}u_{\sigma}\right)+u^{\rho}\left(\partial_{\tau}\partial_{\rho}\beta\right)=\Gamma\partial_{\tau}\left(\beta\theta\right)+\beta\theta\left(\partial_{\tau}\Gamma\right)-\left(\partial_{\tau}u^{\sigma}\right)\sum_{a}\frac{n_{a}}{h}\left(\nabla_{\sigma}\alpha_{a}\right),\label{63}\\
	X_{\tau}^{\left(p\right)}= & \beta\partial_{\tau}\theta-\beta u^{\rho}u^{\sigma}\left(\partial_{\tau}\partial_{\rho}u_{\sigma}\right)+\theta\partial_{\tau}\beta+\left(\partial_{\tau}u^{\sigma}\right)\left(\nabla_{\sigma}\beta\right)=\partial_{\tau}\left(\beta\theta\right)+\left(\partial_{\tau}u^{\sigma}\right)\sum_{a}\frac{n_{a}}{h}\left(\nabla_{\sigma}\alpha_{a}\right),\label{64}\\
	X_{\tau}^{\left(h\right)}= & \beta u_{\sigma}\partial_{\tau}\partial^{\rho}u^{\sigma}+\partial_{\tau}\partial^{\rho}\beta+\beta u_{\sigma}\partial_{\tau}\partial^{\sigma}u^{\rho}+\left(\partial_{\tau}\beta\right)\left(u_{\sigma}\partial^{\sigma}u^{\rho}\right)+\left(\partial_{\tau}u^{\rho}\right)\left(u_{\sigma}\partial^{\sigma}\beta\right)\nonumber\\
	= & -2\beta\sigma^{\rho\sigma}\left(\partial_{\tau}u_{\sigma}\right)-2\left(\frac{1}{3}-\Gamma\right)\beta\theta\left(\partial_{\tau}u^{\rho}\right)+\sum_{a}\partial_{\tau}\left(n_{a}h^{-1}\right)\nabla^{\rho}\alpha_{a}+\sum_{a}\frac{n_{a}}{h}\partial_{\tau}\left(\nabla^{\rho}\alpha_{a}\right),\label{65}\\
	X_{\tau}^{\left(\pi\right)}= & \beta\partial_{\tau}\partial^{\rho}u^{\sigma}+\left(\partial_{\tau}\beta\right)\left(\partial^{\rho}u^{\sigma}\right)+\left(\partial_{\tau}u^{\sigma}\right)\left(\partial^{\rho}\beta\right)=\partial_{\tau}\left(\beta\sigma^{\rho\sigma}\right)+\left(\partial_{\tau}u^{\rho}\right)\sum_{a}\frac{n_{a}}{h}\nabla^{\sigma}\alpha_{a},\label{66}\\
	X_{\tau}^{\left(q\right)}= & \beta u_{\sigma}\partial_{\tau}\partial^{\rho}u^{\sigma}+\partial_{\tau}\partial^{\rho}\beta-\beta u_{\sigma}\partial_{\tau}\partial^{\sigma}u^{\rho}-\left(\partial_{\tau}\beta\right)\left(u_{\sigma}\partial^{\sigma}u^{\rho}\right)-\left(\partial_{\tau}u^{\rho}\right)\left(u_{\sigma}\partial^{\sigma}\beta\right)+2u_{\sigma}\partial_{\tau}\Omega^{\rho\sigma}\nonumber\\
	= & \partial_{\tau}\left(\beta M^{\rho}\right)-2\beta\xi^{\rho\sigma}\left(\partial_{\tau}u_{\sigma}\right),\label{67}\\
	X_{\tau}^{\left(\phi\right)}= & \beta\partial_{\tau}\partial^{\rho}u^{\sigma}+\left(\partial_{\tau}\beta\right)\left(\partial^{\rho}u^{\sigma}\right)+\left(\partial_{\tau}u^{\sigma}\right)\left(\partial^{\rho}\beta\right)+\partial_{\tau}\Omega^{\rho\sigma}=\partial_{\tau}\left(\beta\xi^{\rho\sigma}\right)+\left(\partial_{\tau}u^{\rho}\right)\sum_{a}\frac{n_{a}}{h}\nabla^{\sigma}\alpha_{a},\label{68}\\
	X_{\tau}^{\left(n_{a}\right)}= & u^{\rho}\partial_{\tau}\partial_{\rho}\alpha_{a}=-\delta_{a}\partial_{\tau}\left(\beta\theta\right)-\beta\theta\left(\partial_{\tau}\delta_{a}\right)-\left(\partial_{\tau}u^{\rho}\right)\left(\nabla_{\rho}\alpha_{a}\right),\label{69}\\
	X_{\tau}^{\left(j_{a}\right)}= & \partial_{\tau}\left(\partial_{\rho}\alpha_{a}\right)=-\left(\beta\theta\delta_{a}\right)\left(\partial_{\tau}u_{\rho}\right)+\partial_{\tau}\left(\nabla_{\rho}\alpha_{a}\right).\label{70}
\end{align}
In deriving these expressions, we have employed Eqs.~\eqref{41}-\eqref{42} and \eqref{52}. The orthogonality relations specified in Eq.~\eqref{19} were systematically applied to Eqs.~\eqref{65}-\eqref{68} and \eqref{70}. Furthermore, Eqs.~\eqref{60} and \eqref{61} were utilized in Eqs.~\eqref{66} and \eqref{68}, with the exclusion of terms proportional to $u^{\rho},u^{\sigma},\Delta^{\rho\sigma}$ that are orthogonal to $\hat{\pi}_{\rho\sigma}$, as well as terms proportional to $u^{\rho},u^{\sigma}$ that are orthogonal to $\hat{\phi}_{\rho\sigma}$.

Substituting these equations into Eq.~\eqref{62} yields
\begin{eqnarray}
\begin{aligned}
\partial_{\tau}\hat{C}= &-\hat{p}^{*}\partial_{\tau}\left(\beta\theta\right)+\beta\theta\left[\hat{\epsilon}\left(\partial_{\tau}\Gamma\right)+\sum_{a}\hat{n}_{a}\left(\partial_{\tau}\delta_{a}\right)\right]-\left(\partial_{\tau}u_{\rho}\right)\sum_{a}\left(\nabla^{\rho}\alpha_{a}\right)\left[\frac{n_{a}}{h}\left(\hat{\epsilon}+\hat{p}\right)-\hat{n}_{a}\right]\\		&+\hat{h}_{\rho}\left[-2\beta\sigma^{\rho\sigma}\left(\partial_{\tau}u_{\sigma}\right)-2\left(\frac{1}{3}-\Gamma\right)\beta\theta\left(\partial_{\tau}u^{\rho}\right)+\sum_{a}\partial_{\tau}\left(n_{a}h^{-1}\right)\nabla^{\rho}\alpha_{a}+\beta\theta\left(\partial_{\tau}u^{\rho}\right)h^{-1}\sum_{a}n_{a}\delta_{a}\right]\\
&+\hat{q}_{\rho}\left[\partial_{\tau}\left(\beta M^{\rho}\right)-2\beta\xi^{\rho\sigma}\left(\partial_{\tau}u_{\sigma}\right)\right]+\sum_{a}\hat{\mathscr{J}}_{a}^{\rho}\left[\beta\theta\left(\partial_{\tau}u_{\rho}\right)\delta_{a}-\partial_{\tau}\left(\nabla_{\rho}\alpha_{a}\right)\right]\\		&+\hat{\pi}_{\rho\sigma}\left[\partial_{\tau}\left(\beta\sigma^{\rho\sigma}\right)+\left(\partial_{\tau}u^{\rho}\right)\sum_{a}\frac{n_{a}}{h}\nabla^{\sigma}\alpha_{a}\right]+\hat{\phi}_{\rho\sigma}\left[\partial_{\tau}\left(\beta\xi^{\rho\sigma}\right)+\left(\partial_{\tau}u^{\rho}\right)\sum_{a}\frac{n_{a}}{h}\nabla^{\sigma}\alpha_{a}\right].
\end{aligned}
\label{71}
\end{eqnarray}
It is noted that in the operator $\partial_{\tau}\hat{C}$, the thermodynamic forces are evaluated at the point $x$, while the operators are evaluated at the point $x_1$.

According to Eqs.~\eqref{55} and \eqref{59}, the operator $\hat{C}(x_1)$ can be reformulated as~\cite{Harutyunyan:2025fgs}
\begin{eqnarray} \hat{C}\left(x_{1}\right)=\hat{C}_{1}\left(x_{1}\right)_{x}+\hat{C}_{2}\left(x_{1}\right)_{x}+\partial_{\tau}\hat{C}\left(x_{1}\right)_{x}\left(x_{1}-x\right)^{\tau}.
\label{72}
\end{eqnarray}
Combining this result with Eq.~\eqref{12}, the second-order statistical average of an arbitrary operator $\hat{X}\left(x\right)$ decomposes into three components~\cite{Harutyunyan:2025fgs}
\begin{eqnarray} \langle\hat{X}\left(x\right)\rangle=\langle\hat{X}\left(x\right)\rangle_{l}+\langle\hat{X}\left(x\right)\rangle_{1}+\langle\hat{X}\left(x\right)\rangle_{2}.
\label{73}
\end{eqnarray}
The first-order correction is expressed as
\begin{eqnarray}
	\langle\hat{X}\left(x\right)\rangle_{1}=\int d^{4}x_{1}\left(\hat{X}\left(x\right),\hat{C}_{1}\left(x_{1}\right)_{x}\right),
	\label{74}
\end{eqnarray}
while the second-order correction $\langle\hat{X}\left(x\right)\rangle_{2}$ comprises three distinct contributions
\begin{eqnarray} \langle\hat{X}\left(x\right)\rangle_{2}=\langle\hat{X}\left(x\right)\rangle_{2}^{1}+\langle\hat{X}\left(x\right)\rangle_{2}^{2}+\langle\hat{X}\left(x\right)\rangle_{2}^{3}
\label{75}
\end{eqnarray}
where
\begin{align} 
	& \langle\hat{X}\left(x\right)\rangle_{2}^{1}=\int d^{4}x_{1}\left(\hat{X}\left(x\right),\partial_{\tau}\hat{C}\left(x_{1}\right)_{x}\right)\left(x_{1}-x\right)^{\tau},\label{76}\\
	& \langle\hat{X}\left(x\right)\rangle_{2}^{2}=\int d^{4}x_{1}\left(\hat{X}\left(x\right),\hat{C}_{2}\left(x_{1}\right)_{x}\right),\label{77}\\
	& \langle\hat{X}\left(x\right)\rangle_{2}^{3}=\int d^{4}x_{1}d^{4}x_{2}\left(\hat{X}\left(x\right),\hat{C}_{1}\left(x_{1}\right)_{x},\hat{C}_{1}\left(x_{2}\right)_{x}\right).\label{78}
\end{align}
Notably, the second and third terms in Eq.~\eqref{72} have been omitted from Eq.~\eqref{78} as they contribute only to third-order and higher corrections. Each term in Eq.~\eqref{75} has distinct physical origins: (i) The first term $\langle\hat{X}\rangle_{2}^{1}$ represents non-local corrections arising from the operator $\hat{C}\left(x_{1}\right)$; (ii) The second term $\langle\hat{X}\rangle_{2}^{2}$ contains corrections from generalized thermodynamic forces; (iii) The third term $\langle\hat{X}\rangle_{2}^{3}$ corresponds to quadratic corrections involving five thermodynamic forces: $\theta,\nabla_\sigma \alpha_a,M_\mu,\sigma_{\rho\sigma}$, and $\xi_{\mu\nu}$. This systematic decomposition enables clear identification of the physical mechanisms contributing to each order of correction.

\section{Computing the dissiptive quantities}
\label{section3}

\subsection{First-order spin hydrodynamics}

By substituting Eq.~\eqref{56} into Eq.~\eqref{74} and applying Curie's theorem, we compute the first-order corrections to the shear stress tensor, bulk viscous pressure, diffusion currents, rotational stress tensor, and boost heat vector~\cite{She:2024rnx}
\begin{align}
	\langle\hat{\pi}_{\mu\nu}\left(x\right)\rangle_{1}= & \beta\left(x\right)\sigma^{\rho\sigma}\left(x\right)\int d^{4}x_{1}\left(\hat{\pi}_{\mu\nu}\left(x\right),\hat{\pi}_{\rho\sigma}\left(x_{1}\right)_{x}\right),\label{79}\\
	\langle\hat{p}^{*}\left(x\right)\rangle_{1}= & -\beta\left(x\right)\theta\left(x\right)\int d^{4}x_{1}\left(\hat{p}^{*}\left(x\right),\hat{p}^{*}\left(x_{1}\right)_{x}\right),\label{80}\\
	\langle\hat{\mathscr{J}}_{a}^{\mu}\left(x\right)\rangle_{1}= & -\sum_{b}\nabla_{\sigma}\alpha_{b}\left(x\right)\int d^{4}x_{1}\left(\hat{\mathscr{J}}_{a}^{\mu}\left(x\right),\hat{\mathscr{J}}_{b}^{\sigma}\left(x_{1}\right)_{x}\right)+\beta\left(x\right)M_{\alpha}\left(x\right)\int d^{4}x_{1}\left(\hat{\mathscr{J}}_{a}^{\mu}\left(x\right),\hat{q}^{\alpha}\left(x_{1}\right)_{x}\right),\label{81}\\
	\langle\hat{\phi}^{\mu\nu}\left(x\right)\rangle_{1}= & \beta\left(x\right)\xi_{\rho\sigma}\left(x\right)\int d^{4}x_{1}\left(\hat{\phi}^{\mu\nu}\left(x\right),\hat{\phi}^{\rho\sigma}\left(x_{1}\right)_{x}\right),\label{82}\\
	\langle\hat{q}^{\mu}\left(x\right)\rangle_{1}= & -\sum_{a}\nabla_{\sigma}\alpha_{a}\left(x\right)\int d^{4}x_{1}\left(\hat{q}^{\mu}\left(x\right),\hat{\mathscr{J}}_{a}^{\sigma}\left(x_{1}\right)_{x}\right)+\beta\left(x\right)M_{\alpha}\left(x\right)\int d^{4}x_{1}\left(\hat{q}^{\mu}\left(x\right),\hat{q}^{\alpha}\left(x_{1}\right)_{x}\right).\label{83}
\end{align}
Equations~\eqref{79}-\eqref{83} establish linear constitutive relations between dissipative currents and thermodynamic forces. Notably, the cross-correlation terms in Eqs.~\eqref{81} and \eqref{83} characterize nontrivial couplings between distinct dynamical variables. Medium isotropy and the constraints in Eq.~\eqref{19} enforce the following tensor decompositions
\begin{align} \biggl(\hat{\pi}_{\mu\nu}\left(x\right),\hat{\pi}_{\rho\sigma}\left(x_{1}\right)_{x}\biggr) & =\frac{1}{5}\Delta_{\mu\nu\rho\sigma}\left(x\right)\biggl(\hat{\pi}^{\lambda\eta}\left(x\right),\hat{\pi}_{\lambda\eta}\left(x_{1}\right)_{x}\biggr),\label{84}\\ \biggl(\hat{\mathcal{\mathscr{J}}}_{a}^{\mu}\left(x\right),\hat{\mathscr{J}}_{b}^{\nu}\left(x_{1}\right)_{x}\biggr) & =\frac{1}{3}\Delta^{\mu\nu}\left(x\right)\biggl(\hat{\mathscr{J}}_{a}^{\lambda}\left(x\right),\hat{\mathscr{J}}_{b\lambda}\left(x_{1}\right)_{x}\biggr),\label{85}\\ \left(\hat{\mathscr{J}}_{a}^{\mu}\left(x\right),\hat{q}^{\alpha}\left(x_{1}\right)_{x}\right) & =\frac{1}{3}\Delta^{\mu\alpha}\left(x\right)\left(\hat{\mathscr{J}}_{a}^{\lambda}\left(x\right),\hat{q}_{\lambda}\left(x_{1}\right)_{x}\right),\label{86}\\ \biggl(\hat{\phi}^{\mu\nu}\left(x\right),\hat{\phi}^{\rho\sigma}\left(x_{1}\right)_{x}\biggr) & =\frac{1}{3}\mydelta^{\mu\nu\rho\sigma}\left(x\right)\biggl(\hat{\phi}^{\lambda\eta}\left(x\right),\hat{\phi}_{\lambda\eta}\left(x_{1}\right)_{x}\biggr),\label{87}\\ \left(\hat{q}^{\mu}\left(x\right),\hat{\mathscr{J}}_{a}^{\sigma}\left(x_{1}\right)_{x}\right) & =\frac{1}{3}\Delta^{\mu\sigma}\left(x\right)\left(\hat{q}^{\lambda}\left(x\right),\hat{\mathscr{J}}_{a\lambda}\left(x_{1}\right)_{x}\right),\label{88}\\
\biggl(\hat{q}^{\mu}\left(x\right),\hat{q}^{\alpha}\left(x_{1}\right)_{x}\biggr) & =\frac{1}{3}\Delta^{\mu\alpha}\left(x\right)\biggl(\hat{q}^{\lambda}\left(x\right),\hat{q}_{\lambda}\left(x_{1}\right)_{x}\biggr).\label{89}
\end{align}
Substituting Eqs.~\eqref{84}-\eqref{89} into Eqs.~\eqref{79}-\eqref{83} yields the first-order transport laws
\begin{align}
	\pi_{\mu\nu}= & 2\eta\sigma_{\mu\nu},\label{90}\\
	\Pi= & -\zeta\theta,\label{91}\\
	\phi^{\mu\nu}= & 2\gamma\xi^{\mu\nu},\label{92}\\
	\mathscr{J}_{a}^{\mu}= & \sum_{b}\chi_{ab}\nabla^{\mu}\alpha_{b}+\chi_{\mathscr{J}_{a}q}M^{\mu},\label{93}\\
	q^{\mu}= & -\lambda M^{\mu}+\sum_{a}\lambda_{q\mathscr{J}_{a}}\nabla^{\mu}\alpha_{a}.\label{94}
\end{align}
The transport coefficients are defined through Kubo formulas (see Appendix B of Ref.~\cite{Harutyunyan:2025fgs})
\begin{align}
	\eta= & \frac{\beta}{10}\int d^{4}x_{1}\biggl(\hat{\pi}_{\mu\nu}\left(x\right),\hat{\pi}^{\mu\nu}\left(x_{1}\right)_{x}\biggr)=-\frac{1}{10}\frac{d}{d\omega}\text{Im}G_{\hat{\pi}_{\mu\nu}\hat{\pi}^{\mu\nu}}^{R}\left(\omega\right)\bigg|_{\omega=0},\label{95}\\
	\zeta= & \beta\int d^{4}x_{1}\left(\hat{p}^{*}\left(x\right),\hat{p}^{*}\left(x_{1}\right)_{x}\right)=-\frac{d}{d\omega}\text{Im}G_{\hat{p}^{*}\hat{p}^{*}}^{R}\left(\omega\right)\bigg|_{\omega=0},\label{96}\\
	\gamma= & \frac{1}{6}\beta\int d^{4}x_{1}\left(\hat{\phi}^{\lambda\eta}\left(x\right),\hat{\phi}_{\lambda\eta}\left(x_{1}\right)_{x}\right)=-\frac{1}{6}\frac{d}{d\omega}\text{Im}G_{\hat{\phi}^{\lambda\eta}\hat{\phi}_{\lambda\eta}}^{R}\left(\omega\right)\bigg|_{\omega=0},\label{97}\\
	\chi_{ab}= & -\frac{1}{3}\int d^{4}x_{1}\left(\hat{\mathscr{J}}_{a}^{\lambda}\left(x\right),\hat{\mathscr{J}}_{b\lambda}\left(x_{1}\right)_{x}\right)=\frac{T}{3}\frac{d}{d\omega}\text{Im}G_{\hat{\mathscr{J}}_{a}^{\lambda}\hat{\mathscr{J}}_{b\lambda}}^{R}\left(\omega\right)\bigg|_{\omega=0},\label{98}\\
	\chi_{\mathscr{Ja}_{a}q}= & \frac{1}{3}\beta\int d^{4}x_{1}\left(\hat{\mathscr{J}}_{a}^{\lambda}\left(x\right),\hat{q}_{\lambda}\left(x_{1}\right)_{x}\right)=-\frac{1}{3}\frac{d}{d\omega}\text{Im}G_{\hat{\mathscr{J}}_{a}^{\lambda}\hat{q}_{\lambda}}^{R}\left(\omega\right)\bigg|_{\omega=0},\label{99}\\
	\lambda= & -\frac{1}{3}\beta\int d^{4}x_{1}\left(\hat{q}^{\lambda}\left(x\right),\hat{q}_{\lambda}\left(x_{1}\right)_{x}\right)=\frac{1}{3}\frac{d}{d\omega}\text{Im}G_{\hat{q}^{\lambda}\hat{q}_{\lambda}}^{R}\left(\omega\right)\bigg|_{\omega=0},\label{100}\\
	\lambda_{q\mathscr{J}_{a}}= & -\frac{1}{3}\int d^{4}x_{1}\left(\hat{q}^{\lambda}\left(x\right),\hat{\mathscr{J}}_{a\lambda}\left(x_{1}\right)_{x}\right)=\frac{T}{3}\frac{d}{d\omega}\text{Im}G_{\hat{q}^{\lambda}\hat{\mathscr{J}}_{a\lambda}}^{R}\left(\omega\right)\bigg|_{\omega=0},\label{101}
\end{align}
where
\begin{eqnarray}
	G_{\hat{X}\hat{Y}}^{R}\left(\omega\right)=-i\int_{0}^{\infty}dte^{i\omega t}\int d^{3}x\left\langle\left[\hat{X}\left(\boldsymbol{x},t\right),\hat{Y}\left(\boldsymbol{0},0\right)\right]\right\rangle_{l}
	\label{140}
\end{eqnarray}
denotes the zero-wavenumber limit of the Fourier transform of the retarded two-point commutator, with square brackets representing the commutator. The first terms on the right-hand-sides of Eqs.~\eqref{90}-\eqref{94} represent the Navier-Stokes' contributions to the dissipative quantities with the first-order coefficient of the shear viscosity $\eta$, bulk viscosity $\zeta$, rotational viscosity $\gamma$, matrix of diffusion coefficients $\chi_{ab}$, boost heat conductivity $\lambda$, respectively; these coefficients are expressed via two-point retarded correlation functions via the Kubo formulas \eqref{95}-\eqref{98} and \eqref{100}. Note that there are cross terms for both the charge-diffusion currents and the boost heat vector. The transport coefficients of the cross terms can be expressed via two-point retarded correlation functions via the Kubo formulas~\eqref{99} and \eqref{101}.

\subsection{Second-order spin hydrodynamics}

\subsubsection{Second-order corrections to the shear stress tensor}

Through substitution of Eqs.~\eqref{71} and~\eqref{84} into Eq.~\eqref{76} under application of Curie's theorem, we obtain the second-order correction for the shear stress tensor
\begin{eqnarray}
	\left\langle \hat{\pi}_{\mu\nu}\left(x\right)\right\rangle _{2}^{1}=\frac{1}{5}\Delta_{\mu\nu\rho\sigma}\left(x\right)\left[\partial_{\tau}\left(\beta\sigma^{\rho\sigma}\right)+\left(\partial_{\tau}u^{\rho}\right)\sum_{a}\frac{n_{a}}{h}\nabla^{\sigma}\alpha_{a}\right]_{x}\int d^{4}x_{1}\left(\hat{\pi}^{\lambda\eta}\left(x\right),\hat{\pi}_{\lambda\eta}\left(x_{1}\right)_{x}\right)\left(x_{1}-x\right)^{\tau}.
	\label{103}
\end{eqnarray}
The second term inside the square brackets, which represents the coupling between shear stresses and diffusion currents, was not considered in our earlier analysis~\cite{She:2024rnx}.

The following equation is obtained from the appendix of Ref.~\cite{Harutyunyan:2025fgs}
\begin{eqnarray}
	\frac{\beta\left(x\right)}{10}\int d^{4}x_{1}\Delta_{\mu\nu\rho\sigma}\left(x\right)\left(\hat{\pi}^{\lambda\eta}\left(x\right),\hat{\pi}_{\lambda\eta}\left(x_{1}\right)_{x}\right)\left(x_{1}-x\right)^{\tau}=\widetilde{\eta}u^{\tau}.
	\label{104}
\end{eqnarray}
The parameter $\widetilde{\eta}$ is defined as
\begin{align} \widetilde{\eta}=i\frac{d}{d\omega}\eta\left(\omega\right)\bigg|_{\omega=0}=-\frac{1}{20}\frac{d^{2}}{d\omega^{2}}\text{Re}G_{\hat{\pi}_{\mu\nu}\hat{\pi}^{\mu\nu}}^{R}\left(\omega\right)\bigg|_{\omega=0},\label{105}
\end{align}	
where $\eta(\omega)$ is frequency-dependent transport coefficients given by the integral
\begin{align}	
	\eta\left(\omega\right)=\frac{\beta}{10}\int d^{4}x_{1}\int_{-\infty}^{t}e^{i\omega\left(t-t_{1}\right)}\left(\hat{\pi}_{\mu\nu}\left(\boldsymbol{x},t\right),\hat{\pi}^{\mu\nu}\left(\boldsymbol{x}_{1},t_{1}\right)\right).\label{106}
\end{align}

By combining Eqs.~\eqref{103} and \eqref{104} and applying the approximation $D\beta\simeq\beta\theta\Gamma$, we derive the non-local corrections to the shear stress tensor from the two-point correlation function. The resulting expression is given by
\begin{eqnarray}
	\begin{aligned}
		\left\langle \hat{\pi}_{\mu\nu}\left(x\right)\right\rangle _{2}^{1} & =2\widetilde{\eta}\beta^{-1}\Delta_{\mu\nu\rho\sigma}\left[D\left(\beta\sigma^{\rho\sigma}\right)+\left(Du^{\rho}\right)\sum_{a}\frac{n_{a}}{h}\nabla^{\sigma}\alpha_{a}\right]\\
		&=2\widetilde{\eta}\left(\Delta_{\mu\nu\rho\sigma}D\sigma^{\rho\sigma}+\Gamma\theta\sigma_{\mu\nu}\right)+2\widetilde{\eta}Th^{-1}\sum_{a}n_{a}\dot{u}_{\langle\mu}\nabla_{\nu\rangle}\alpha_{a}.
	\end{aligned}
\label{107}
\end{eqnarray}
Here, $\dot{u}^{\mu}$ denotes the convective derivative, defined as $\dot{u}^{\mu}=Du^{\mu}$. The final terms on the right-hand side of both lines in Eq.~\eqref{107} represent novel contributions. This term signify new physical effects pertinent to the non-local behavior of the shear stress tensor, highlighting complexities in the system's dynamics not captured by previous analyses.

As established in our prior work~\cite{She:2024rnx}, the second-order corrections to the shear-stress tensor expectation value, $\langle\hat{\pi}_{\mu\nu}\rangle$, arising from generalized thermodynamic forces and three-point correlation functions are given by
\begin{align}
\left\langle \hat{\pi}_{\mu\nu}\right\rangle _{2}^{2}= & 0,\label{108}\\
\left\langle \hat{\pi}_{\mu\nu}\right\rangle _{2}^{3}= & 2\eta_{\pi p\pi}\theta\sigma_{\mu\nu}+\sum_{ab}\eta_{\pi\mathscr{J}_{a}\mathscr{J}_{b}}\nabla_{\langle\mu}\alpha_{a}\nabla_{\nu\rangle}\alpha_{b}+2\sum_{a}\eta_{\pi\mathscr{J}_{a}q}\nabla_{\langle\mu}\alpha_{a}M_{\nu\rangle}\nonumber\\
 & +\eta_{\pi qq}M_{\langle\mu}M_{\nu\rangle}+\eta_{\pi\pi\pi}\sigma_{\alpha\langle\mu}\sigma_{\nu\rangle}^{\,\,\,\,\alpha}+2\eta_{\pi\pi\phi}\sigma_{\alpha\langle\mu}\xi_{\nu\rangle}^{\,\,\,\,\alpha}+\eta_{\pi\phi\phi}\xi_{\alpha\langle\mu}\xi_{\nu\rangle}^{\,\,\,\,\alpha},\label{109}
\end{align}
where we have introduced the following transport coefficients (see Appendix B of Ref.~\cite{Harutyunyan:2025fgs})
\begin{align}
\eta_{\pi p\pi} & =-\frac{1}{5}\beta^{2}\int d^{4}x_{1}d^{4}x_{2}\left(\hat{\pi}_{\gamma\delta}(x),\hat{p}^{*}\left(x_{1}\right),\hat{\pi}^{\gamma\delta}\left(x_{2}\right)\right)=\frac{1}{10}\frac{\partial}{\partial\omega_{1}}\frac{\partial}{\partial\omega_{2}}\text{Re}G_{\hat{\pi}_{\gamma\delta}\hat{p}^{*}\hat{\pi}^{\gamma\delta}}^{R}\left(\omega_{1},\omega_{2}\right)\bigg|_{\omega_{1,2}=0},\label{110}\\
\eta_{\pi\mathscr{J}_{a}\mathscr{J}_{b}} & =\frac{1}{5}\int d^{4}x_{1}d^{4}x_{2}\left(\hat{\pi}_{\gamma\delta}\left(x\right),\hat{\mathscr{J}}_{a}^{\gamma}\left(x_{1}\right),\hat{\mathscr{J}}_{b}^{\delta}\left(x_{2}\right)\right)=-\frac{T^{2}}{10}\frac{\partial}{\partial\omega_{1}}\frac{\partial}{\partial\omega_{2}}\text{Re}G_{\hat{\pi}_{\gamma\delta}\hat{\mathscr{J}}_{a}^{\gamma}\hat{\mathscr{J}}_{b}^{\delta}}^{R}\left(\omega_{1},\omega_{2}\right)\bigg|_{\omega_{1,2}=0},\label{111}\\
\eta_{\pi\mathscr{J}_{a}q} & =-\frac{1}{5}\beta\int d^{4}x_{1}d^{4}x_{2}\left(\hat{\pi}_{\gamma\delta}\left(x\right),\hat{\mathscr{J}}_{a}^{\gamma}\left(x_{1}\right),\hat{q}^{\delta}\left(x_{2}\right)\right)=\frac{T}{10}\frac{\partial}{\partial\omega_{1}}\frac{\partial}{\partial\omega_{2}}\text{Re}G_{\hat{\pi}_{\gamma\delta}\hat{\mathscr{J}}_{a}^{\gamma}\hat{q}^{\delta}}^{R}\left(\omega_{1},\omega_{2}\right)\bigg|_{\omega_{1,2}=0},\label{112}\\
\eta_{\pi qq} & =\frac{1}{5}\beta^{2}\int d^{4}x_{1}d^{4}x_{2}\left(\hat{\pi}_{\gamma\delta}\left(x\right),\hat{q}^{\gamma}\left(x_{1}\right),\hat{q}^{\delta}\left(x_{2}\right)\right)=-\frac{1}{10}\frac{\partial}{\partial\omega_{1}}\frac{\partial}{\partial\omega_{2}}\text{Re}G_{\hat{\pi}_{\gamma\delta}\hat{q}^{\gamma}\hat{q}^{\delta}}^{R}\left(\omega_{1},\omega_{2}\right)\bigg|_{\omega_{1,2}=0},\label{113}\\
\eta_{\pi\pi\pi} & =\frac{12}{35}\beta^{2}\int d^{4}x_{1}d^{4}x_{2}\left(\hat{\pi}_{\gamma}^{\,\,\,\delta}\left(x\right),\hat{\pi}_{\delta}^{\,\,\,\lambda}\left(x_{1}\right),\hat{\pi}_{\lambda}^{\,\,\,\gamma}\left(x_{2}\right)\right)=-\frac{6}{35}\frac{\partial}{\partial\omega_{1}}\frac{\partial}{\partial\omega_{2}}\text{Re}G_{\hat{\pi}_{\gamma}^{\,\,\,\delta}\hat{\pi}_{\delta}^{\,\,\,\lambda}\hat{\pi}_{\lambda}^{\,\,\,\gamma}}^{R}\left(\omega_{1},\omega_{2}\right)\bigg|_{\omega_{1,2}=0},\label{114}\\
\eta_{\pi\pi\phi} & =-\frac{4}{15}\beta^{2}\int d^{4}x_{1}d^{4}x_{2}\left(\hat{\pi}_{\gamma}^{\,\,\,\delta}\left(x\right),\hat{\pi}_{\delta}^{\,\,\,\lambda}\left(x_{1}\right),\hat{\phi}_{\lambda}^{\,\,\,\gamma}\left(x_{2}\right)\right)=\frac{2}{15}\frac{\partial}{\partial\omega_{1}}\frac{\partial}{\partial\omega_{2}}\text{Re}G_{\hat{\pi}_{\gamma}^{\,\,\,\delta}\hat{\pi}_{\delta}^{\,\,\,\lambda}\hat{\phi}_{\lambda}^{\,\,\,\gamma}}^{R}\left(\omega_{1},\omega_{2}\right)\bigg|_{\omega_{1,2}=0},\label{115}\\
\eta_{\pi\phi\phi} & =\frac{4}{5}\beta^{2}\int d^{4}x_{1}d^{4}x_{2}\left(\hat{\pi}_{\gamma}^{\,\,\,\delta}\left(x\right),\hat{\phi}_{\delta}^{\,\,\,\lambda}\left(x_{1}\right),\hat{\phi}_{\lambda}^{\,\,\,\gamma}\left(x_{2}\right)\right)=-\frac{2}{5}\frac{\partial}{\partial\omega_{1}}\frac{\partial}{\partial\omega_{2}}\text{Re}G_{\hat{\pi}_{\gamma}^{\,\,\,\delta}\hat{\phi}_{\delta}^{\,\,\,\lambda}\hat{\phi}_{\lambda}^{\,\,\,\gamma}}^{R}\left(\omega_{1},\omega_{2}\right)\bigg|_{\omega_{1,2}=0}.\label{116}
\end{align}
These coefficients are expressed in terms of integrals of three-point correlation functions and, equivalently, as derivatives of the real part of specific retarded Green's functions evaluated at zero frequencies.

The complete second-order expression for the shear stress tensor is obtained by combining the corrections presented in Eqs.~\eqref{36}, \eqref{90}, \eqref{107}, \eqref{108}, and \eqref{109} with the results from Eqs.~\eqref{73} and \eqref{75}
\begin{equation}
\begin{aligned}
\pi_{\mu\nu}= & 2\eta\sigma_{\mu\nu}+2\widetilde{\eta}\left(\Delta_{\mu\nu\rho\sigma}D\sigma^{\rho\sigma}+\theta\Gamma\sigma_{\mu\nu}\right)+2\widetilde{\eta}Th^{-1}\sum_{a}n_{a}\dot{u}_{\langle\mu}\nabla_{\nu\rangle}\alpha_{a}\\
 & +2\eta_{\pi p\pi}\theta\sigma_{\mu\nu}+\sum_{ab}\eta_{\pi\mathscr{J}_{a}\mathscr{J}_{b}}\nabla_{\langle\mu}\alpha_{a}\nabla_{\nu\rangle}\alpha_{b}+2\sum_{a}\eta_{\pi\mathscr{J}_{a}q}\nabla_{\langle\mu}\alpha_{a}M_{\nu\rangle}\\
 & +\eta_{\pi qq}M_{\langle\mu}M_{\nu\rangle}+\eta_{\pi\pi\pi}\sigma_{\alpha\langle\mu}\sigma_{\nu\rangle}^{\,\,\,\,\alpha}+2\eta_{\pi\pi\phi}\sigma_{\alpha\langle\mu}\xi_{\nu\rangle}^{\,\,\,\,\alpha}+\eta_{\pi\phi\phi}\xi_{\alpha\langle\mu}\xi_{\nu\rangle}^{\,\,\,\,\alpha}.
\end{aligned}
\label{117}
\end{equation}
Here, the second-order terms on the first line represent non-local corrections, while the remaining second-order terms account for nonlinear effects originating from three-point correlations.

To derive a relaxation-type equation for the shear stress tensor $\pi_{\mu\nu}$ from Eq.~\eqref{117}, we approximate the term $2\sigma^{\rho\sigma}$ in the second term on the right-hand side by its Navier-Stokes expression, $\eta^{-1}\pi^{\rho\sigma}$. This substitution is justified because the term is already of second order in spacetime gradients, rendering higher-order corrections negligible within this approximation scheme. Applying this substitution and utilizing Eqs.~\eqref{41}-\eqref{42}, we obtain the approximation
\begin{eqnarray}
2\widetilde{\eta}\Delta_{\mu\nu\rho\sigma}D\sigma^{\rho\sigma}&\simeq-\tau_{\pi}\Delta_{\mu\nu\rho\sigma}D\pi^{\rho\sigma}+\tau_{\pi}\eta^{-1}\beta\left(\frac{\partial\eta}{\partial\beta}\Gamma-\sum_{a}\frac{\partial\eta}{\partial\alpha_{a}}\delta_{a}\right)\theta\pi_{\mu\nu}.
\label{118}
\end{eqnarray}
We then define the coefficients
\begin{align} 
&\tau_{\pi}=-\widetilde{\eta}\eta^{-1},\label{119}\\
&\widetilde{\eta}_{\pi}=\tau_{\pi}\eta^{-1}\beta\left(\frac{\partial\eta}{\partial\beta}\Gamma-\sum_{a}\frac{\partial\eta}{\partial\alpha_{a}}\delta_{a}\right).\label{120}
\end{align}
Combining Eq.~\eqref{117} with the approximated Eq.~\eqref{118} leads to the following relaxation equation for $\pi_{\mu\nu}$
\begin{eqnarray}
\begin{aligned}\tau_{\pi}\dot{\pi}_{\mu\nu}+\pi_{\mu\nu}= & 2\eta\sigma_{\mu\nu}+\widetilde{\eta}_{\pi}\theta\pi_{\mu\nu}+2\widetilde{\eta}\theta\Gamma\sigma_{\mu\nu}+2\widetilde{\eta}Th^{-1}\sum_{a}n_{a}\dot{u}_{\langle\mu}\nabla_{\nu\rangle}\alpha_{a}\\
 & +2\eta_{\pi p\pi}\theta\sigma_{\mu\nu}+\sum_{ab}\eta_{\pi\mathscr{J}_{a}\mathscr{J}_{b}}\nabla_{\langle\mu}\alpha_{a}\nabla_{\nu\rangle}\alpha_{b}+2\sum_{a}\eta_{\pi\mathscr{J}_{a}q}\nabla_{\langle\mu}\alpha_{a}M_{\nu\rangle}\\
 & +\eta_{\pi qq}M_{\langle\mu}M_{\nu\rangle}+\eta_{\pi\pi\pi}\sigma_{\alpha\langle\mu}\sigma_{\nu\rangle}^{\,\,\,\,\alpha}+2\eta_{\pi\pi\phi}\sigma_{\alpha\langle\mu}\xi_{\nu\rangle}^{\,\,\,\,\alpha}+\eta_{\pi\phi\phi}\xi_{\alpha\langle\mu}\xi_{\nu\rangle}^{\,\,\,\,\alpha},
\end{aligned}
\label{121}
\end{eqnarray}
where $\dot{\pi}_{\mu\nu}$ is defined as $\dot{\pi}_{\mu\nu}=\Delta_{\mu\nu\rho\sigma}D\pi^{\rho\sigma}$. 
The term $\tau_{\pi}\dot{\pi}_{\mu\nu}$ on the left-hand side of Eqs.~\eqref{121} represents the relaxation of the shear stress towards its Navier-Stokes limit, characterized by the relaxation time $\tau_{\pi}$. This time scale is related to the relevant first-order transport coefficients as specified in Eqs.~\eqref{105}. Notably, the term $2\widetilde{\eta}Th^{-1}\sum_{a}n_{a}\dot{u}_{\langle\mu}\nabla_{\nu\rangle}\alpha_{a}$ on the first line of the right-hand side represents a new contribution that distinguishes this result from our previous work~\cite{She:2024rnx}, the contribution of this term is also present in viscous fluids without spin degrees of freedom~\cite{Harutyunyan:2025fgs}.

\subsubsection{Second-order corrections to the bulk viscous pressure}

It is well-established that the bulk viscous pressure serves as a quantitative measure of the deviation between the actual thermodynamic pressure $\left\langle \hat{p}\right\rangle $ and its equilibrium value $p\left(\epsilon,n_{a},S^{\alpha\beta}\right)$, which is determined by the equation of state (EoS). This discrepancy arises from fluid expansion or compression and is mathematically expressed as
\begin{eqnarray} \Pi=\langle\hat{p}\rangle-p\left(\epsilon,n_{a},S^{\alpha\beta}\right)=\langle\hat{p}\rangle_{l}+\langle\hat{p}\rangle_{1}+\langle\hat{p}\rangle_{2}-p\left(\epsilon,n_{a},S^{\alpha\beta}\right).
\label{122}
\end{eqnarray}
Considering potential deviations of the energy, charge, and spin densities from their equilibrium values, namely $\epsilon=\langle\hat{\epsilon}\rangle_{l}+\Delta\epsilon,n_{a}=\langle\hat{n}_{a}\rangle_{l}+\Delta n_{a}$, and $S^{\alpha\beta}=\langle\hat{S}^{\alpha\beta}\rangle_{l}+\Delta S^{\alpha\beta}$, we can derive the following expression for $\langle\hat{p}\rangle_{l}$ through a Taylor expansion around the equilibrium densities
\begin{eqnarray}
	\begin{aligned}
		\langle\hat{p}\rangle_{l}\equiv & p\left(\langle\hat{\epsilon}\rangle_{l},\langle\hat{n}_{a}\rangle_{l},\langle\hat{S}^{\alpha\beta}\rangle_{l}\right)\\
		= & p\left(\epsilon-\Delta\epsilon,n_{a}-\Delta n_{a},S^{\alpha\beta}-\Delta S^{\alpha\beta}\right)\\
		= & p\left(\epsilon,n_{a},S^{\alpha\beta}\right)-\Gamma\Delta\epsilon-\sum_{a}\delta_{a}\Delta n_{a}-\mathcal{K}_{\alpha\beta}\Delta S^{\alpha\beta}+\frac{1}{2}\frac{\partial^{2}p}{\partial\epsilon^{2}}\left(\Delta\epsilon\right)^{2}+\frac{1}{2}\times2\sum_{a}\frac{\partial^{2}p}{\partial\epsilon\partial n_{a}}\Delta\epsilon\Delta n_{a}\\
		& +\frac{1}{2}\sum_{ab}\frac{\partial^{2}p}{\partial n_{a}\partial n_{b}}\Delta n_{a}\Delta n_{b}+\frac{1}{2}\times2\sum_{a}\frac{\partial^{2}p}{\partial n_{a}\partial S^{\alpha\beta}}\Delta n_{a}\Delta S^{\alpha\beta}+\frac{1}{2}\frac{\partial^{2}p}{\partial S^{\alpha\beta}\partial S^{\rho\sigma}}\Delta S^{\alpha\beta}\Delta S^{\rho\sigma}+\frac{1}{2}\times2\frac{\partial^{2}p}{\partial\epsilon\partial S^{\alpha\beta}}\Delta\epsilon\Delta S^{\alpha\beta}.
	\end{aligned}
	\label{123}
\end{eqnarray}
Substituting this expansion into Eq.~\eqref{122}, the bulk-viscous pressure can be formulated as
\begin{eqnarray}
	\begin{aligned}
		\Pi= & \langle\hat{p}\rangle_{1}+\langle\hat{p}\rangle_{2}-\Gamma\Delta\epsilon-\sum_{a}\delta_{a}\Delta n_{a}-\mathcal{K}_{\alpha\beta}\Delta S^{\alpha\beta}+\frac{1}{2}\frac{\partial^{2}p}{\partial\epsilon^{2}}\left(\Delta\epsilon\right)^{2}+\sum_{a}\frac{\partial^{2}p}{\partial\epsilon\partial n_{a}}\Delta\epsilon\Delta n_{a}\\
		& +\frac{1}{2}\sum_{ab}\frac{\partial^{2}p}{\partial n_{a}\partial n_{b}}\Delta n_{a}\Delta n_{b}+\sum_{a}\frac{\partial^{2}p}{\partial n_{a}\partial S^{\alpha\beta}}\Delta n_{a}\Delta S^{\alpha\beta}+\frac{1}{2}\frac{\partial^{2}p}{\partial S^{\alpha\beta}\partial S^{\rho\sigma}}\Delta S^{\alpha\beta}\Delta S^{\rho\sigma}+\frac{\partial^{2}p}{\partial\epsilon\partial S^{\alpha\beta}}\Delta\epsilon\Delta S^{\alpha\beta}.
	\end{aligned}
	\label{124}
\end{eqnarray}
By substituting $\Delta\epsilon=\langle\hat{\epsilon}\rangle_{1}+\langle\hat{\epsilon}\rangle_{2}$, $\Delta n_{a}=\langle\hat{n}_{a}\rangle_{1}+\langle\hat{n}_{a}\rangle_{2}$, and $\Delta S^{\alpha\beta}=\langle\hat{S}^{\alpha\beta}\rangle_{1}+\langle\hat{S}^{\alpha\beta}\rangle_{2}$ and neglecting higher-order terms (specifically quadratic terms in $\langle\cdots\rangle_{2}$ and cross terms between $\langle\cdots\rangle_{1}$ and $\langle\cdots\rangle_{2}$), we arrive at the following expression, utilizing the definition of $\hat{p}^*$ from Eq.~\eqref{46}
\begin{eqnarray}
\begin{aligned}
	\Pi= \langle\hat{p}^{*}\rangle_{1}+\langle\hat{p}^{*}\rangle_{2}+\frac{1}{2}\frac{\partial^{2}p}{\partial\epsilon^{2}}\langle\hat{\epsilon}\rangle_{1}^{2}+\frac{1}{2}\sum_{ab}\frac{\partial^{2}p}{\partial n_{a}\partial n_{b}}\langle\hat{n}_{a}\rangle_{1}\langle\hat{n}_{b}\rangle_{1}+\sum_{a}\frac{\partial^{2}p}{\partial\epsilon\partial n_{a}}\langle\hat{\epsilon}\rangle_{1}\langle\hat{n}_{a}\rangle_{1}-\mathcal{K}_{\alpha\beta}\langle\hat{S}^{\alpha\beta}\rangle_{1}.
\end{aligned}
\label{125}
\end{eqnarray}
Introducing the coefficients defined by correlation functions (see Appendix B of Ref.~\cite{Harutyunyan:2025fgs})
\begin{align}
	\zeta_{\epsilon p} & =\beta\int d^{4}x_{1}\left(\hat{\epsilon}\left(x\right),\hat{p}^{*}\left(x_{1}\right)_{x}\right)=-\frac{d}{d\omega}\text{Im}G_{\hat{\epsilon}\hat{p}^{*}}^{R}\left(\omega\right)\bigg|_{\omega=0},\label{126}\\
	\zeta_{n_{a}p} & =\beta\int d^{4}x_{1}\left(\hat{n}_{a}\left(x\right),\hat{p}^{*}\left(x_{1}\right)_{x}\right)=-\frac{d}{d\omega}\text{Im}G_{\hat{n}_{a}\hat{p}^{*}}^{R}\left(\omega\right)\bigg|_{\omega=0},\label{127}\\
	\zeta_{S\phi} & =-\frac{1}{3}\beta\int d^{4}x_{1}\left(\hat{S}^{\lambda\eta}\left(x\right),\hat{\phi}_{\lambda\eta}\left(x_{1}\right)_{x}\right)=\frac{1}{3}\frac{d}{d\omega}\text{Im}G_{\hat{S}^{\lambda\eta}\hat{\phi}_{\lambda\eta}}^{R}\left(\omega\right)\bigg|_{\omega=0}.\label{128}
\end{align}
and leveraging Eqs.~\eqref{56} and \eqref{74}, the first-order averages $\langle\hat{\epsilon}\rangle_{1},\langle\hat{n}_{a}\rangle_{1}$, and $\langle\hat{S}^{\alpha\beta}\rangle_{1}$ can be expressed as
\begin{align}
	\langle\hat{\epsilon}\rangle_{1}= & -\zeta_{\epsilon p}\theta,\label{129}\\
	\langle\hat{n}_{a}\rangle_{1}= & -\zeta_{n_{a}p}\theta,\label{130}\\
	\langle\hat{S}^{\mu\nu}\rangle_{1}= & -\zeta_{S\phi}\xi^{\mu\nu}.\label{131}
\end{align}
Substituting Eqs.~\eqref{129}-\eqref{131} into Eq.~\eqref{125}, we obtain the bulk viscous pressure
\begin{eqnarray}
\begin{aligned}
	\Pi=-\zeta\theta+\langle\hat{p}^{*}\rangle_{2}+\frac{1}{2}\frac{\partial^{2}p}{\partial\epsilon^{2}}\zeta_{\epsilon p}^{2}\theta^{2}+\frac{1}{2}\sum_{ab}\frac{\partial^{2}p}{\partial n_{a}\partial n_{b}}\zeta_{n_{a}p}\zeta_{n_{b}p}\theta^{2}+\sum_{a}\frac{\partial^{2}p}{\partial\epsilon\partial n_{a}}\zeta_{\epsilon p}\zeta_{n_{a}p}\theta^{2}+\mathcal{K}_{\alpha\beta}\zeta_{S\phi}\xi^{\alpha\beta}.
\end{aligned}
\label{132}
\end{eqnarray}

We now proceed with the calculation of $\langle\hat{p}^{*}\rangle_{2}$. By substituting Eq.~\eqref{71} into Eq.~\eqref{76} and applying Curie's theorem, we obtain the following expression for $\langle\hat{p}^{*}\left(x\right)\rangle_{2}^{1}$
\begin{equation}
\begin{aligned}
	\langle\hat{p}^{*}\left(x\right)\rangle_{2}^{1} =&-\partial_{\tau}\left(\beta\theta\right)\int d^{4}x_{1}\left(\hat{p}^{*}\left(x\right),\hat{p}^{*}\left(x_{1}\right)_{x}\right)\left(x_{1}-x\right)^{\tau}\\
	& +\beta\theta\int d^{4}x_{1}\left(\hat{p}^{*}\left(x\right),\left[\hat{\epsilon}\left(\partial_{\tau}\Gamma\right)+\sum_{a}\hat{n}_{a}\left(\partial_{\tau}\delta_{a}\right)\right]\right)\left(x_{1}-x\right)^{\tau}\\
	& -\sum_{a}\left(\partial_{\tau}u_{\rho}\right)\nabla^{\rho}\alpha_{a}\int d^{4}x_{1}\left(\hat{p}^{*}\left(x\right),\left[n_{a}h^{-1}\left(\hat{\epsilon}+\hat{p}\right)-\hat{n}_{a}\right]\right)\left(x_{1}-x\right)^{\tau}.
\end{aligned}
\label{133}
\end{equation}
The final term in Eq.~\eqref{133} originates from the interaction between the bulk viscous pressure and the diffusion currents. This interaction was not considered in our previous analysis, underscoring its significance in the present context.
  
The following relations are obtained from the Appendix of Ref.~\cite{Harutyunyan:2025fgs}
\begin{align}
	\int d^{4}x_{1}\left(\hat{p}^{*}\left(x\right),\hat{p}^{*}\left(x_{1}\right)_{x}\right)\left(x_{1}-x\right)^{\tau}= & u^{\tau}\beta^{-1}\widetilde{\zeta},\label{134}\\
	\int d^{4}x_{1}\left(\hat{p}^{*}\left(x\right),\left[\hat{\epsilon}\left(\partial_{\tau}\Gamma\right)+\sum_{a}\hat{n}_{a}\left(\partial_{\tau}\delta_{a}\right)\right]\right)\left(x_{1}-x\right)^{\tau}= & \beta^{-1}\left(\widetilde{\zeta}_{p\epsilon}D\Gamma+\sum_{a}\widetilde{\zeta}_{pn_{a}}D\delta_{a}\right),\label{135}\\
	\int d^{4}x_{1}\left(\hat{p}^{*}\left(x\right),\left[n_{a}h^{-1}\left(\hat{\epsilon}+\hat{p}\right)-\hat{n}_{a}\right]\right)\left(x_{1}-x\right)^{\tau}= & \beta^{-1}u^{\tau}\left[n_{a}h^{-1}\widetilde{\zeta}+\left(1+\Gamma\right)n_{a}h^{-1}\widetilde{\zeta}_{p\epsilon}+\sum_{b}\widetilde{\zeta}_{pn_{b}}\left(n_{a}\delta_{b}h^{-1}-\delta_{ab}\right)\right].\label{136}
\end{align}
Here, the coefficients $\widetilde{\zeta},\widetilde{\zeta}_{p\epsilon}$, and $\widetilde{\zeta}_{pn_{a}}$ are defined as
\begin{align} 
	&\widetilde{\zeta}=i\frac{d}{d\omega}\zeta(\omega)\bigg|_{\omega=0}=-\frac{1}{2}\frac{d^{2}}{d\omega^{2}}\mathrm{Re}G_{\hat{p}^{*}\hat{p}^{*}}^{R}(\omega)\bigg|_{\omega=0},\label{137}\\
	&\widetilde{\zeta}_{p\epsilon}=i\frac{d}{d\omega}\zeta_{p\epsilon}(\omega)\bigg|_{\omega=0}=-\frac{1}{2}\frac{d^{2}}{d\omega^{2}}\mathrm{Re}G_{\hat{p}^{*}\hat{\epsilon}}^{R}(\omega)\bigg|_{\omega=0},\label{138}\\
	&\widetilde{\zeta}_{pn_{a}}=i\frac{d}{d\omega}\zeta_{pn_{a}}(\omega)\bigg|_{\omega=0}=-\frac{1}{2}\frac{d^{2}}{d\omega^{2}}\mathrm{Re}G_{\hat{p}^{*}\hat{n}_{a}}^{R}(\omega)\bigg|_{\omega=0},\label{139}
\end{align}
where $\zeta\left(\omega\right),\zeta_{p\epsilon}\left(\omega\right)$, and $\zeta_{pn_a}\left(\omega\right)$ are frequency-dependent transport coefficients given by
\begin{align}
	\zeta\left(\omega\right) & =\beta\int d^{4}x_{1}\int_{-\infty}^{t}e^{i\omega\left(t-t_{1}\right)}\biggl(\hat{p}^{*}\left(\boldsymbol{x},t\right),\hat{p}^{*}\left(\boldsymbol{x}_{1},t_{1}\right)\biggr)\text{,}\label{140}\\
	\zeta_{p\epsilon}\left(\omega\right) & =\beta\int d^{4}x_{1}\int_{-\infty}^{t}e^{i\omega\left(t-t_{1}\right)}\biggl(\hat{p}^{*}\left(\boldsymbol{x},t\right),\hat{\epsilon}\left(\boldsymbol{x}_{1},t_{1}\right)\biggr),\label{141}\\
	\zeta_{pn_{a}}\left(\omega\right) & =\beta\int d^{4}x_{1}\int_{-\infty}^{t}e^{i\omega\left(t-t_{1}\right)}\biggl(\hat{p}^{*}\left(\boldsymbol{x},t\right),\hat{n}_{a}\left(\boldsymbol{x}_{1},t_{1}\right)\biggr).\label{142}
\end{align}
Combining Eqs.~\eqref{133}-\eqref{136} and applying the approximation $D\beta\simeq\beta\theta\Gamma$, we derive the non-local corrections to the bulk viscous pressure arising from the two-point correlation function
\begin{eqnarray}
	\begin{aligned}
		\langle\hat{p}^{*}\left(x\right)\rangle_{2}^{1}= & -\widetilde{\zeta}\left(D\theta+\Gamma\theta^{2}\right)+\theta\left(\widetilde{\zeta}_{p\epsilon}D\Gamma+\sum_{a}\widetilde{\zeta}_{pn_{a}}D\delta_{a}\right)-\beta^{-1}\dot{u}_{\rho}\sum_{a}\left(\nabla^{\rho}\alpha_{a}\right)\\
		&\times\left[n_{a}h^{-1}\widetilde{\zeta}+\left(1+\Gamma\right)n_{a}h^{-1}\widetilde{\zeta}_{p\epsilon}+\sum_{b}\widetilde{\zeta}_{pn_{b}}\left(n_{a}\delta_{b}h^{-1}-\delta_{ab}\right)\right].
	\end{aligned}
	\label{143}
\end{eqnarray}
Here, $\delta_{ab}$ denotes the Kronecker delta. This expression quantifies the contributions of various physical parameters and their interactions to the non-local bulk viscous pressure.

We then derive the expressions for the material derivatives of  $\Gamma$ and $\delta_a$:
\begin{align}
	D\Gamma= & \frac{\partial^{2}p}{\partial\epsilon^{2}}D\epsilon+\sum_{a}\frac{\partial^{2}p}{\partial\epsilon\partial n_{a}}Dn_{a}+\frac{\partial^{2}p}{\partial\epsilon\partial S^{\alpha\beta}}DS^{\alpha\beta}\nonumber\\
	= & -\frac{\partial^{2}p}{\partial\epsilon^{2}}w\theta-\sum_{a}\frac{\partial^{2}p}{\partial\epsilon\partial n_{a}}n_{a}\theta-\frac{\partial^{2}p}{\partial\epsilon\partial S^{\alpha\beta}}\left[S^{\alpha\beta}\theta+u^{\alpha}\partial_{\lambda}S^{\beta\lambda}+S^{\beta\lambda}\partial_{\lambda}u^{\alpha}+u^{\beta}\partial_{\lambda}S^{\lambda\alpha}+S^{\lambda\alpha}\partial_{\lambda}u^{\beta}\right],\label{144}\\
	D\delta_{a}= & \frac{\partial^{2}p}{\partial\epsilon\partial n_{a}}D\epsilon+\sum_{b}\frac{\partial^{2}p}{\partial n_{a}\partial n_{b}}Dn_{b}+\frac{\partial^{2}p}{\partial n_{a}\partial S^{\alpha\beta}}DS^{\alpha\beta}\nonumber\\
	= & -\frac{\partial^{2}p}{\partial\epsilon\partial n_{a}}w\theta-\sum_{b}\frac{\partial^{2}p}{\partial n_{a}\partial n_{b}}n_{b}\theta-\frac{\partial^{2}p}{\partial n_{a}\partial S^{\alpha\beta}}\left[S^{\alpha\beta}\theta+u^{\alpha}\partial_{\lambda}S^{\beta\lambda}+S^{\beta\lambda}\partial_{\lambda}u^{\alpha}+u^{\beta}\partial_{\lambda}S^{\lambda\alpha}+S^{\lambda\alpha}\partial_{\lambda}u^{\beta}\right],\label{145}
\end{align}
where we have substituted the expressions for $D\epsilon,Dn_a$ and $DS^{\alpha\beta}$ from Eqs.~\eqref{33}-\eqref{36}. Subsequently, we define the following quantities
\begin{eqnarray}
\begin{aligned}
	\widetilde{\Gamma}= & \frac{\partial^{2}p}{\partial\epsilon^{2}}h+\sum_{a}\frac{\partial^{2}p}{\partial\epsilon\partial n_{a}}n_{a}+\frac{\partial^{2}p}{\partial\epsilon\partial S^{\alpha\beta}}\left[S^{\alpha\beta}+\theta^{-1}\left(u^{\alpha}\partial_{\lambda}S^{\beta\lambda}+S^{\beta\lambda}\partial_{\lambda}u^{\alpha}+u^{\beta}\partial_{\lambda}S^{\lambda\alpha}+S^{\lambda\alpha}\partial_{\lambda}u^{\beta}\right)\right],\\
	\widetilde{\delta}_{a}= & \frac{\partial^{2}p}{\partial\epsilon\partial n_{a}}h+\sum_{b}\frac{\partial^{2}p}{\partial n_{a}\partial n_{b}}n_{b}+\frac{\partial^{2}p}{\partial n_{a}\partial S^{\alpha\beta}}\left[S^{\alpha\beta}+\theta^{-1}\left(u^{\alpha}\partial_{\lambda}S^{\beta\lambda}+S^{\beta\lambda}\partial_{\lambda}u^{\alpha}+u^{\beta}\partial_{\lambda}S^{\lambda\alpha}+S^{\lambda\alpha}\partial_{\lambda}u^{\beta}\right)\right],\\
	\overline{\zeta}_{a}= & Tn_{a}h^{-1}\left[\widetilde{\zeta}+\left(1+\Gamma\right)\widetilde{\zeta}_{p\epsilon}\right]+T\sum_{b}\widetilde{\zeta}_{pn_{b}}\left(n_{a}\delta_{b}h^{-1}-\delta_{ab}\right).
\end{aligned}
\label{146}
\end{eqnarray} 
Using these definitions, the non-local corrections to the bulk viscous pressure stemming from the two-point correlation function are expressed as
\begin{eqnarray}
	\langle\hat{p}^{*}\rangle_{2}^{1}=-\widetilde{\zeta}D\theta-\widetilde{\zeta}\theta^{2}\Gamma-\widetilde{\zeta}_{p\epsilon}\theta^{2}\widetilde{\Gamma}-\sum_{a}\widetilde{\zeta}_{pn_{a}}\theta^{2}\widetilde{\delta}_{a}-\sum_{a}\overline{\zeta}_{a}\dot{u}_{\rho}\nabla^{\rho}\alpha_{a}.
	\label{147}
\end{eqnarray}
The final term in Eq.~\eqref{147} represents a novel contribution to the non-local bulk viscous pressure expression. This term is also present in viscous fluids lacking spin degrees of freedom~\cite{Harutyunyan:2025fgs}. 

Following the framework established in our prior research \cite{She:2024rnx}, the second-order corrections to the pressure deviation $\hat{p}^{*}$ arise from generalized thermodynamic forces and three-point correlation functions. These contributions to the second-order correction $\langle\hat{p}^{*}\rangle_{2}$ are given by
\begin{align}
\langle\hat{p}^{*}\rangle_{2}^{2}=&\sum_{i}\zeta_{p\mathfrak{D}_{i}}\left[\left(\partial_{\epsilon n}^{i}\beta\right)\mathcal{X}+\sum_{a}\left(\partial_{\epsilon n}^{i}\alpha_{a}\right)\mathcal{Y}_{a}+\left(\partial_{\epsilon n}^{i}\Omega_{\alpha\beta}\right)\mathcal{Z}^{\alpha\beta}\right],\label{148}\\
\langle\hat{p}^{*}\left(x\right)\rangle_{2}^{3}=&\zeta_{ppp}\theta^{2}+\sum_{ab}\zeta_{p\mathscr{J}_{a}\mathscr{J}_{b}}\nabla_{\alpha}\alpha_{a}\nabla^{\alpha}\alpha_{b}+2\sum_{a}\zeta_{p\mathscr{J}_{a}q}\nabla^{\sigma}\alpha_{a}M_{\sigma}+\zeta_{pqq}M^{\sigma}M_{\sigma}+\zeta_{p\pi\pi}\sigma^{\rho\sigma}\sigma_{\rho\sigma}+\zeta_{p\phi\phi}\xi^{\rho\sigma}\xi_{\rho\sigma}.\label{149}
\end{align}
The transport coefficients appearing in these expressions are defined by two-point or three-point correlation functions, related to derivatives of retarded Green's functions evaluated at zero frequency (see the appendix of Ref.~\cite{Harutyunyan:2025fgs})
\begin{align}
\zeta_{p\mathfrak{D}_{i}}= & \beta\int d^{4}x_{1}\left(\hat{p}^{*}\left(x\right),\hat{\mathfrak{D}}_{i}\left(x_{1}\right)\right)=-\frac{d}{d\omega}\text{Im}G_{\hat{p}^{*}\hat{\mathfrak{D}}_{i}}^{R}\left(\omega\right)\bigg|_{\omega=0},\label{150}\\
\zeta_{ppp} = &\beta^{2}\int d^{4}x_{1}d^{4}x_{2}\left(\hat{p}^{*}\left(x\right),\hat{p}^{*}\left(x_{1}\right),\hat{p}^{*}\left(x_{2}\right)\right)=-\frac{1}{2}\frac{\partial}{\partial\omega_{1}}\frac{\partial}{\partial\omega_{2}}\text{Re}G_{\hat{p}^{*}\hat{p}^{*}\hat{p}^{*}}^{R}\left(\omega_{1},\omega_{2}\right)\bigg|_{\omega_{1,2}=0},\label{151}\\
\zeta_{p\mathscr{J}_{a}\mathscr{J}_{b}} = &\frac{1}{3}\int d^{4}x_{1}d^{4}x_{2}\left(\hat{p}^{*}\left(x\right),\hat{\mathscr{J}}_{a\gamma}\left(x_{1}\right),\hat{\mathscr{J}}_{b}^{\gamma}\left(x_{2}\right)\right)=-\frac{T^{2}}{6}\frac{\partial}{\partial\omega_{1}}\frac{\partial}{\partial\omega_{2}}\text{Re}G_{\hat{p}^{*}\hat{\mathscr{J}}_{a\gamma}\hat{\mathscr{J}}_{b}^{\gamma}}^{R}\left(\omega_{1},\omega_{2}\right)\bigg|_{\omega_{1,2}=0},\label{152}\\
\zeta_{p\mathscr{J}_{a}q} = &-\frac{1}{3}\beta\int d^{4}x_{1}d^{4}x_{2}\left(\hat{p}^{*}\left(x\right),\hat{\mathscr{J}}_{a\gamma}\left(x_{1}\right),\hat{q}^{\gamma}\left(x_{2}\right)\right)=\frac{T}{6}\frac{\partial}{\partial\omega_{1}}\frac{\partial}{\partial\omega_{2}}\text{Re}G_{\hat{p}^{*}\hat{\mathscr{J}}_{a\gamma}\hat{q}^{\gamma}}^{R}\left(\omega_{1},\omega_{2}\right)\bigg|_{\omega_{1,2}=0},\label{153}\\
\zeta_{pqq} = &\frac{1}{3}\beta^{2}\int d^{4}x_{1}d^{4}x_{2}\left(\hat{p}^{*}(x),\hat{q}_{\gamma}(x_{1}),\hat{q}^{\gamma}(x_{2})\right)=-\frac{1}{6}\frac{\partial}{\partial\omega_{1}}\frac{\partial}{\partial\omega_{2}}\text{Re}G_{\hat{p}^{*}\hat{q}_{\gamma}\hat{q}^{\gamma}}^{R}\left(\omega_{1},\omega_{2}\right)\bigg|_{\omega_{1,2}=0},\label{154}\\
\zeta_{p\pi\pi} = &\frac{1}{5}\beta^{2}\int d^{4}x_{1}d^{4}x_{2}\left(\hat{p}^{*}\left(x\right),\hat{\pi}_{\gamma\delta}\left(x_{1}\right),\hat{\pi}^{\gamma\delta}\left(x_{2}\right)\right)=-\frac{1}{10}\frac{\partial}{\partial\omega_{1}}\frac{\partial}{\partial\omega_{2}}\text{Re}G_{\hat{p}^{*}\hat{\pi}_{\gamma\delta}\hat{\pi}^{\gamma\delta}}^{R}\left(\omega_{1},\omega_{2}\right)\bigg|_{\omega_{1,2}=0},\label{155}\\
\zeta_{p\phi\phi} = &\frac{1}{3}\beta^{2}\int d^{4}x_{1}d^{4}x_{2}\left(\hat{p}^{*}\left(x\right),\hat{\phi}_{\gamma\delta}\left(x_{1}\right),\hat{\phi}^{\gamma\delta}\left(x_{2}\right)\right)=-\frac{1}{6}\frac{\partial}{\partial\omega_{1}}\frac{\partial}{\partial\omega_{2}}\text{Re}G_{\hat{p}^{*}\hat{\phi}_{\gamma\delta}\hat{\phi}^{\gamma\delta}}^{R}\left(\omega_{1},\omega_{2}\right)\bigg|_{\omega_{1,2}=0}.\label{156}
\end{align}
Combining these second-order corrections to $\hat{p}^*$ with other contributions, the full expression for the bulk viscous pressure $\Pi$, accurate to second order in gradients, is given by
\begin{equation}
\begin{aligned}
\Pi= & -\zeta\theta+\frac{1}{2}\frac{\partial^{2}p}{\partial\epsilon^{2}}\zeta_{\epsilon p}^{2}\theta^{2}+\frac{1}{2}\sum_{ab}\frac{\partial^{2}p}{\partial n_{a}\partial n_{b}}\zeta_{n_{a}p}\zeta_{n_{b}p}\theta^{2}+\sum_{a}\frac{\partial^{2}p}{\partial\epsilon\partial n_{a}}\zeta_{\epsilon p}\zeta_{n_{a}p}\theta^{2}+\mathcal{K}_{\alpha\beta}\zeta_{S\phi}\xi^{\alpha\beta}\\
 &-\widetilde{\zeta}_{p\epsilon}\theta^{2}\widetilde{\Gamma}-\sum_{a}\widetilde{\zeta}_{pn_{a}}\theta^{2}\widetilde{\delta}_{a}-\widetilde{\zeta}\theta^{2}\Gamma-\widetilde{\zeta}D\theta-\sum_{a}\overline{\zeta}_{a}\dot{u}_{\rho}\nabla^{\rho}\alpha_{a}+\sum_{i}\zeta_{p\mathfrak{D}_{i}}\left[\left(\partial_{\epsilon n}^{i}\beta\right)\mathcal{X}+\sum_{a}\left(\partial_{\epsilon n}^{i}\alpha_{a}\right)\mathcal{Y}_{a}+\left(\partial_{\epsilon n}^{i}\Omega_{\alpha\beta}\right)\mathcal{Z}^{\alpha\beta}\right]\\
 &+\zeta_{ppp}\theta^{2}+\sum_{ab}\zeta_{p\mathscr{J}_{a}\mathscr{J}_{b}}\nabla_{\alpha}\alpha_{a}\nabla^{\alpha}\alpha_{b}+2\sum_{a}\zeta_{p\mathscr{J}_{a}q}\nabla^{\sigma}\alpha_{a}M_{\sigma}+\zeta_{pqq}M^{\sigma}M_{\sigma}+\zeta_{p\pi\pi}\sigma^{\rho\sigma}\sigma_{\rho\sigma}+\zeta_{p\phi\phi}\xi^{\rho\sigma}\xi_{\rho\sigma}.
\end{aligned}
\label{157}
\end{equation}

To derive the relaxation equation for the bulk viscous pressure, we substitute $\theta$ with $-\zeta^{-1}\Pi$ within the term $\widetilde{\zeta}D\theta$. By employing Eqs.~\eqref{41} and \eqref{42}, this substitution yields the following expression
\begin{eqnarray}
-\widetilde{\zeta}D\theta\simeq&\widetilde{\zeta}\zeta^{-1}D\Pi-\widetilde{\zeta}\zeta^{-2}\beta\left(\frac{\partial\zeta}{\partial\beta}\Gamma-\sum_{a}\frac{\partial\zeta}{\partial\alpha_{a}}\delta_{a}\right)\theta\Pi.
\label{158}
\end{eqnarray}
Combining Eqs.~\eqref{157} and \eqref{158} and introducing the coefficients
\begin{align}
\dot{\Pi}= & D\Pi,\label{159}\\
\tau_{\Pi}= & -\widetilde{\zeta}\zeta^{-1},\label{160}\\
\widetilde{\zeta}_{\Pi}= & \tau_{\Pi}\zeta^{-1}\beta\left(\frac{\partial\zeta}{\partial\beta}\Gamma-\sum_{a}\frac{\partial\zeta}{\partial\alpha_{a}}\delta_{a}\right),\label{161}
\end{align}
we obtain the relaxation equation for the bulk viscous pressure
\begin{equation}
\begin{aligned}
\Pi+\tau_{\Pi}\dot{\Pi}= & -\zeta\theta+\widetilde{\zeta}_{\Pi}\theta\Pi+\left[\frac{1}{2}\frac{\partial^{2}p}{\partial\epsilon^{2}}\zeta_{\epsilon p}^{2}+\frac{1}{2}\sum_{ab}\frac{\partial^{2}p}{\partial n_{a}\partial n_{b}}\zeta_{n_{a}p}\zeta_{n_{b}p}+\sum_{a}\frac{\partial^{2}p}{\partial\epsilon\partial n_{a}}\zeta_{\epsilon p}\zeta_{n_{a}p}\right]\theta^{2}+\zeta_{S\phi}\mathcal{K}_{\alpha\beta}\xi^{\alpha\beta}\\
&-\left[\Gamma\widetilde{\zeta}+\widetilde{\Gamma}\widetilde{\zeta}_{p\epsilon}+\sum_{a}\widetilde{\zeta}_{pn_{a}}\widetilde{\delta}_{a}\right]\theta^{2}-\sum_{a}\overline{\zeta}_{a}\dot{u}_{\rho}\nabla^{\rho}\alpha_{a}+\sum_{i}\zeta_{p\mathfrak{D}_{i}}\left[\left(\partial_{\epsilon n}^{i}\beta\right)\mathcal{X}+\sum_{a}\left(\partial_{\epsilon n}^{i}\alpha_{a}\right)\mathcal{Y}_{a}+\left(\partial_{\epsilon n}^{i}\Omega_{\alpha\beta}\right)\mathcal{Z}^{\alpha\beta}\right]\\
&+\zeta_{ppp}\theta^{2}+\sum_{ab}\zeta_{p\mathscr{J}_{a}\mathscr{J}_{b}}\nabla_{\alpha}\alpha_{a}\nabla^{\alpha}\alpha_{b}+2\sum_{a}\zeta_{p\mathscr{J}_{a}q}\nabla^{\sigma}\alpha_{a}M_{\sigma}+\zeta_{pqq}M^{\sigma}M_{\sigma}+\zeta_{p\pi\pi}\sigma^{\rho\sigma}\sigma_{\rho\sigma}+\zeta_{p\phi\phi}\xi^{\rho\sigma}\xi_{\rho\sigma}.
\end{aligned}
\label{162}
\end{equation}
A comparison with our previous work~\cite{She:2024rnx} reveals that the fourth term on the second line of the right-hand side constitutes a novel contribution to this equation. This term also exists in viscous fluids without spin degrees of freedom~\cite{Harutyunyan:2025fgs}.

\subsubsection{Second-order corrections to the charge-diffusion currents}

By substituting Eq.~\eqref{71} into Eq.~\eqref{76} and invoking Curie's theorem, we arrive at the following expression for the expectation value of the conserved current $\hat{\mathscr{J}}_{c\mu}$ to second order
\begin{eqnarray}
\begin{aligned}
	\langle\hat{\mathscr{J}}_{c\mu}\left(x\right)\rangle_{2}^{1}= & -\frac{1}{3}\Delta_{\mu\rho}\left(x\right)\sum_{a}\left[\partial_{\tau}\left(\nabla^{\rho}\alpha_{a}\right)-\beta\theta\left(\partial_{\tau}u^{\rho}\right)\delta_{a}\right]_{x}\int d^{4}x_{1}\left(\hat{\mathscr{J}}_{c\lambda}\left(x\right),\hat{\mathscr{J}}_{a}^{\lambda}\left(x_{1}\right)_{x}\right)\left(x_{1}-x\right)^{\tau}\\
	&+\frac{1}{3}\Delta_{\mu\rho}\left(x\right)\left[-2\beta\sigma^{\rho\sigma}\left(\partial_{\tau}u_{\sigma}\right)-y\beta\theta\left(\partial_{\tau}u^{\rho}\right)+\sum_{a}\partial_{\tau}\left(n_{a}h^{-1}\right)\nabla^{\rho}\alpha_{a}\right]_{x}\\
	& \times\int d^{4}x_{1}\left(\hat{\mathscr{J}}_{c\lambda}\left(x\right),\hat{h}^{\lambda}\left(x_{1}\right)_{x}\right)\left(x_{1}-x\right)^{\tau}\\
	& +\frac{1}{3}\Delta_{\mu\rho}\left(x\right)\left[\partial_{\tau}\left(\beta M^{\rho}\right)-2\beta\xi^{\rho\sigma}\left(\partial_{\tau}u_{\sigma}\right)\right]_{x}\int d^{4}x_{1}\left(\hat{\mathscr{J}}_{c\lambda}\left(x\right),\hat{q}^{\lambda}\left(x_{1}\right)_{x}\right)\left(x_{1}-x\right)^{\tau}.
\end{aligned}
\label{163}
\end{eqnarray}
This derivation relies on Eqs.~\eqref{85} and \eqref{86}, along with the analogous relation
\begin{eqnarray*}
	\left(\hat{\mathscr{J}}_{a}^{\mu}\left(x\right),\hat{h}^{\alpha}\left(x_{1}\right)_{x}\right)=\frac{1}{3}\Delta^{\mu\alpha}\left(x\right)\left(\hat{\mathscr{J}}_{a}^{\lambda}\left(x\right),\hat{q}_{\lambda}\left(x_{1}\right)_{x}\right).
\end{eqnarray*}
Furthermore, we define the quantity $y$ as
\begin{eqnarray}
y=\frac{2}{3}-2\Gamma-\sum_{a}\delta_{a}n_{a}h^{-1}.
\label{164}
\end{eqnarray}
From Appendix B of Ref.~\cite{Harutyunyan:2025fgs}, it can be obtained that
\begin{align}
 -\frac{1}{3}\int d^{4}x_{1}\left(\hat{\mathscr{J}}_{c\lambda}\left(x\right),\hat{\mathscr{J}}_{a}^{\lambda}\left(x_{1}\right)_{x}\right)\left(x_{1}-x\right)^{\tau}=&\widetilde{\chi}_{ac}u^{\tau},\label{165}\\
 \frac{1}{3}\int d^{4}x_{1}\left(\hat{\mathscr{J}}_{c\lambda}\left(x\right),\hat{h}^{\lambda}\left(x_{1}\right)_{x}\right)\left(x_{1}-x\right)^{\tau}=&\widetilde{\chi}_{\mathscr{J}_{c}h}u^{\tau},\label{166}\\
 \frac{1}{3}\beta\int d^{4}x_{1}\left(\hat{\mathscr{J}}_{c\lambda}\left(x\right),\hat{q}^{\lambda}\left(x_{1}\right)_{x}\right)\left(x_{1}-x\right)^{\tau}=&\widetilde{\chi}_{\mathscr{J}_{c}q}u^{\tau},\label{167}
\end{align}
where the transport coefficients $\widetilde{\chi}_{ac},\widetilde{\chi}_{\mathscr{J}_{c}h}$ and $\widetilde{\chi}_{\mathscr{J}_{c}q}$ are defined as
\begin{align}
\widetilde{\chi}_{ac}= & i\frac{d}{d\omega}\chi_{ac}\left(\omega\right)\bigg|_{\omega=0}=\frac{T}{6}\frac{d^{2}}{d\omega^{2}}\text{Re}G_{\hat{\mathscr{J}}_{a}^{\lambda}\hat{\mathscr{J}}_{c\lambda}}^{R}\left(\omega\right)\bigg|_{\omega=0},\label{168}\\
\widetilde{\chi}_{\mathscr{J}_{c}h}= & i\frac{d}{d\omega}\chi_{\mathscr{J}_{c}h}\left(\omega\right)\bigg|_{\omega=0}=-\frac{T}{6}\frac{d^{2}}{d\omega^{2}}\text{Re}G_{\hat{\mathscr{J}}_{c}^{\lambda}\hat{h}_{\lambda}}^{R}\left(\omega\right)\bigg|_{\omega=0},\label{169}\\
\widetilde{\chi}_{\mathscr{J}_{c}q}= & i\frac{d}{d\omega}\chi_{\mathscr{J}_{c}q}\left(\omega\right)\bigg|_{\omega=0}=-\frac{1}{6}\frac{d^{2}}{d\omega^{2}}\text{Re}G_{\hat{\mathscr{J}}_{c}^{\lambda}\hat{q}_{\lambda}}^{R}\left(\omega\right)\bigg|_{\omega=0},\label{170}
\end{align}
with the frequency-dependent transport coefficients $\chi_{ac}\left(\omega\right),\chi_{\mathscr{J}_{c}h}\left(\omega\right),\chi_{\mathscr{J}_{c}q}\left(\omega\right)$ are given by
\begin{align}
\chi_{ac}\left(\omega\right)= & -\frac{1}{3}\int d^{4}x_{1}\int_{-\infty}^{t}e^{i\omega\left(t-t_{1}\right)}\left(\hat{\mathscr{J}}_{c}^{\lambda}\left(\boldsymbol{x},t\right),\hat{\mathscr{J}}_{a\lambda}\left(\boldsymbol{x}_{1},t_{1}\right)\right),\label{171}\\
\chi_{\mathscr{J}_{c}h}\left(\omega\right)= & \frac{1}{3}\int d^{4}x_{1}\int_{-\infty}^{t}e^{i\omega\left(t-t_{1}\right)}\left(\hat{\mathscr{J}}_{c}^{\lambda}\left(\boldsymbol{x},t\right),\hat{h}_{\lambda}\left(\boldsymbol{x}_{1},t_{1}\right)\right),\label{172}\\
\chi_{\mathscr{J}_{c}q}\left(\omega\right)= & \frac{1}{3}\beta\int d^{4}x_{1}\int_{-\infty}^{t}e^{i\omega\left(t-t_{1}\right)}\left(\hat{\mathscr{J}}_{c}^{\lambda}\left(\boldsymbol{x},t\right),\hat{q}_{\lambda}\left(\boldsymbol{x}_{1},t_{1}\right)\right).\label{173}
\end{align}

The non-local corrections to the charge-diffusion currents arising from the two-point correlation function are derived by combining Eqs.~\eqref{163}-\eqref{167} and applying the approximation $D\beta\simeq\beta\theta\Gamma$. This yields
\begin{eqnarray}
\begin{aligned}
\langle\hat{\mathscr{J}}_{c\mu}\left(x\right)\rangle_{2}^{1}= & \sum_{a}\widetilde{\chi}_{ac}\Delta_{\mu\rho}D\left(\nabla^{\rho}\alpha_{a}\right)+\widetilde{\chi}_{\mathscr{J}_{c}h}\sum_{a}D\left(n_{a}h^{-1}\right)\nabla_{\mu}\alpha_{a}+\left(M_{\mu}\theta\Gamma+\Delta_{\mu\rho}DM^{\rho}\right)\widetilde{\chi}_{\mathscr{J}_{c}q}\\
&-\beta\theta\dot{u}_{\mu}\sum_{a}\delta_{a}\widetilde{\chi}_{ac}-\beta\widetilde{\chi}_{\mathscr{J}_{c}h}\left(2\sigma_{\mu\nu}\dot{u}^{\nu}+y\theta\dot{u}_{\mu}\right)-2\widetilde{\chi}_{\mathscr{J}_{c}q}\xi_{\mu\nu}\dot{u}^{\nu}.
\end{aligned}
\label{174}
\end{eqnarray}
Since Eq.~\eqref{174} is of second order, we can substitute $D\left(n_{a}h^{-1}\right)=-n_{a}h^{-2}Dp$ using relations from Eqs.~\eqref{32}-\eqref{34}. The expression for $Dp$ is obtained from Eqs.~\eqref{32}-\eqref{34} and \eqref{44}
\begin{eqnarray}
Dp =\Gamma D\epsilon+\sum_{a}\delta_{a}Dn_{a}+\mathcal{K}_{\alpha\beta}DS^{\alpha\beta}
 =-\left(\Gamma h\theta+\sum_{a}\delta_{a}n_{a}\theta\right).
\label{175}
\end{eqnarray}
Substituting this result for $D\left(n_{a}h^{-1}\right)$ into Eq.~\eqref{174} yields
\begin{eqnarray}
\begin{aligned}
\langle\hat{\mathscr{J}}_{c\mu}\left(x\right)\rangle_{2}^{1}= & \sum_{a}\widetilde{\chi}_{ac}\Delta_{\mu\rho}D\left(\nabla^{\rho}\alpha_{a}\right)+\widetilde{\chi}_{\mathscr{J}_{c}h}h^{-2}\left(\Gamma h+\sum_{b}\delta_{b}n_{b}\right)\theta\sum_{a}n_{a}\nabla_{\mu}\alpha_{a}+\widetilde{\chi}_{\mathscr{J}_{c}q}\theta\Gamma M_{\mu}+\widetilde{\chi}_{\mathscr{J}_{c}q}\Delta_{\mu\rho}DM^{\rho}\\
&-\beta\theta\dot{u}_{\mu}\sum_{a}\delta_{a}\widetilde{\chi}_{ac}-\beta\widetilde{\chi}_{\mathscr{J}_{c}h}\left(2\sigma_{\mu\nu}\dot{u}^{\nu}+y\theta\dot{u}_{\mu}\right)-2\widetilde{\chi}_{\mathscr{J}_{c}q}\xi_{\mu\nu}\dot{u}^{\nu}.
\end{aligned}
\label{176}
\end{eqnarray}
The second line of Eq.~\eqref{176} introduces novel contributions. Specifically, the term proportional to $\theta\dot{u}_{\mu}$ characterizes the non-local coupling between the charge diffusion currents and the bulk viscous pressure. The term proportional to $\sigma_{\mu\nu}\dot{u}^{\nu}$
describes the non-local interaction with the shear stress tensor, while the term proportional to $\xi_{\mu\nu}\dot{u}^{\nu}$ represents the non-local mixing with the rotational stress tensor.

In accordance with our previous work~\cite{She:2024rnx}, the second-order corrections to $\hat{\mathscr{J}}_{c\mu}$ that stem from generalized thermodynamic forces and three-point correlation functions are given by
\begin{align}
	\langle\hat{\mathscr{J}}_{c\mu}\rangle_{2}^{2}= & \beta\chi_{\mathscr{J}_{c}h}\mathcal{H}_{\mu}+\chi_{\mathscr{J}_{c}q}\mathcal{Q}_{\mu},\label{177}\\
	\langle\hat{\mathscr{J}}_{c\mu}\rangle_{2}^{3}= & 2\sum_{a}\chi_{\mathscr{J}_{c}p\mathscr{J}_{a}}\theta\nabla_{\mu}\alpha_{a}+2\chi_{\mathscr{J}_{c}pq}\theta M_{\mu}+2\sum_{a}\chi_{\mathscr{J}_{c}\mathscr{J}_{a}\pi}\nabla^{\nu}\alpha_{a}\sigma_{\mu\nu}\nonumber\\
	&+2\sum_{a}\chi_{\mathscr{J}_{c}\mathscr{J}_{a}\phi}\nabla^{\nu}\alpha_{a}\xi_{\mu\nu}+2\chi_{\mathscr{J}_{c}q\pi}M^{\nu}\sigma_{\mu\nu}+2\chi_{\mathscr{J}_{c}q\phi}M^{\nu}\xi_{\mu\nu},\label{178}
\end{align}
where the coefficients are defined as follows (see the appendix of Ref.~\cite{Harutyunyan:2025fgs})
\begin{align}
	\chi_{\mathscr{J}_{c}h}&=\frac{1}{3}\int d^{4}x_{1}\left(\hat{\mathscr{J}}_{c\alpha}\left(x\right),\hat{h}^{\alpha}\left(x_{1}\right)_{x}\right)=-\frac{T}{3}\frac{d}{d\omega}\text{Im}G_{\hat{\mathscr{J}}_{c\alpha}\hat{h}^{\alpha}}^{R}\left(\omega\right)\bigg|_{\omega=0},\label{179}\\
	\chi_{\mathscr{J}_{c}p\mathscr{J}_{a}} & =\frac{1}{3}\beta\int d^{4}x_{1}d^{4}x_{2}\left(\hat{\mathscr{J}}_{c\beta}\left(x\right),\hat{p}^{*}\left(x_{1}\right),\hat{\mathscr{J}}_{a}^{\beta}\left(x_{2}\right)\right)=-\frac{T}{6}\frac{\partial}{\partial\omega_{1}}\frac{\partial}{\partial\omega_{2}}\text{Re}G_{\hat{\mathscr{J}}_{c\beta}\hat{p}^{*}\hat{\mathscr{J}}_{a}^{\beta}}^{R}\left(\omega_{1},\omega_{2}\right)\bigg|_{\omega_{1,2}=0},\label{180}\\
	\chi_{\mathscr{J}_{c}pq} & =-\frac{1}{3}\beta^{2}\int d^{4}x_{1}d^{4}x_{2}\left(\hat{\mathscr{J}}_{c\beta}\left(x\right),\hat{p}^{*}\left(x_{1}\right),\hat{q}^{\beta}\left(x_{2}\right)\right)=\frac{1}{6}\frac{\partial}{\partial\omega_{1}}\frac{\partial}{\partial\omega_{2}}\text{Re}G_{\hat{\mathscr{J}}_{c\beta}\hat{p}^{*}\hat{q}^{\beta}}^{R}\left(\omega_{1},\omega_{2}\right)\bigg|_{\omega_{1,2}=0},\label{181}\\
	\chi_{\mathscr{J}_{c}\mathscr{J}_{a}\pi} & =-\frac{1}{5}\beta\int d^{4}x_{1}d^{4}x_{2}\left(\hat{\mathscr{J}}_{c\lambda}\left(x\right),\hat{\mathscr{J}}_{a\delta}\left(x_{1}\right),\hat{\pi}^{\lambda\delta}\left(x_{2}\right)\right)=\frac{T}{10}\frac{\partial}{\partial\omega_{1}}\frac{\partial}{\partial\omega_{2}}\text{Re}G_{\hat{\mathscr{J}}_{c\lambda}\hat{\mathscr{J}}_{a\delta}\hat{\pi}^{\lambda\delta}}^{R}\left(\omega_{1},\omega_{2}\right)\bigg|_{\omega_{1,2}=0},\label{182}\\
	\chi_{\mathscr{J}_{c}\mathscr{J}_{a}\phi} & =-\frac{1}{3}\beta\int d^{4}x_{1}d^{4}x_{2}\left(\hat{\mathscr{J}}_{c\lambda}\left(x\right),\hat{\mathscr{J}}_{a\delta}\left(x_{1}\right),\hat{\phi}^{\lambda\delta}\left(x_{2}\right)\right)=\frac{T}{6}\frac{\partial}{\partial\omega_{1}}\frac{\partial}{\partial\omega_{2}}\text{Re}G_{\hat{\mathscr{J}}_{c\lambda}\hat{\mathscr{J}}_{a\delta}\hat{\phi}^{\lambda\delta}}^{R}\left(\omega_{1},\omega_{2}\right)\bigg|_{\omega_{1,2}=0},\label{183}\\
	\chi_{\mathscr{J}_{c}q\pi} & =\frac{1}{5}\beta^{2}\int d^{4}x_{1}d^{4}x_{2}\left(\hat{\mathscr{J}}_{c\lambda}\left(x\right),\hat{q}_{\delta}\left(x_{1}\right),\hat{\pi}^{\lambda\delta}\left(x_{2}\right)\right)=-\frac{1}{10}\frac{\partial}{\partial\omega_{1}}\frac{\partial}{\partial\omega_{2}}\text{Re}G_{\hat{\mathscr{J}}_{c\lambda}\hat{q}_{\delta}\hat{\pi}^{\lambda\delta}}^{R}\left(\omega_{1},\omega_{2}\right)\bigg|_{\omega_{1,2}=0},\label{184}\\
	\chi_{\mathscr{J}_{c}q\phi} & =\frac{1}{3}\beta^{2}\int d^{4}x_{1}d^{4}x_{2}\left(\hat{\mathscr{J}}_{c\lambda}\left(x\right),\hat{q}_{\delta}\left(x_{1}\right),\hat{\phi}^{\lambda\delta}\left(x_{2}\right)\right)=-\frac{1}{6}\frac{\partial}{\partial\omega_{1}}\frac{\partial}{\partial\omega_{2}}\text{Re}G_{\hat{\mathscr{J}}_{c\lambda}\hat{q}_{\delta}\hat{\phi}^{\lambda\delta}}^{R}\left(\omega_{1},\omega_{2}\right)\bigg|_{\omega_{1,2}=0}.\label{185}
\end{align}
By synthesizing the results from Eqs.~\eqref{32}, \eqref{93}, and \eqref{176}-\eqref{178}, we obtain the comprehensive second-order expression for the charge-diffusion current $\mathscr{J}_{c\mu}$
\begin{equation}
\begin{aligned}
	\mathscr{J}_{c\mu}= & \sum_{b}\chi_{cb}\nabla_{\mu}\alpha_{b}+\chi_{\mathscr{J}_{c}q}M_{\mu}+\sum_{a}\widetilde{\chi}_{ac}\Delta_{\mu\rho}D\left(\nabla^{\rho}\alpha_{a}\right)+\widetilde{\chi}_{\mathscr{J}_{c}h}h^{-2}\left(\Gamma h+\sum_{b}\delta_{b}n_{b}\right)\theta\sum_{a}n_{a}\nabla_{\mu}\alpha_{a}\\
	& +\widetilde{\chi}_{\mathscr{J}_{c}q}\theta\Gamma M_{\mu}+\widetilde{\chi}_{\mathscr{J}_{c}q}\Delta_{\mu\rho}DM^{\rho}-\beta\theta\dot{u}_{\mu}\sum_{a}\delta_{a}\widetilde{\chi}_{ac}-\beta\widetilde{\chi}_{\mathscr{J}_{c}h}\left(2\sigma_{\mu\nu}\dot{u}^{\nu}+y\theta\dot{u}_{\mu}\right)-2\widetilde{\chi}_{\mathscr{J}_{c}q}\xi_{\mu\nu}\dot{u}^{\nu}\\
	&+\beta\chi_{\mathscr{J}_{c}h}\mathcal{H}_{\mu}+\chi_{\mathscr{J}_{c}q}\mathcal{Q}_{\mu}+2\sum_{a}\chi_{\mathscr{J}_{c}p\mathscr{J}_{a}}\theta\nabla_{\mu}\alpha_{a}+2\chi_{\mathscr{J}_{c}pq}\theta M_{\mu}+2\sum_{a}\chi_{\mathscr{J}_{c}\mathscr{J}_{a}\pi}\nabla^{\nu}\alpha_{a}\sigma_{\mu\nu}\\
	&+2\sum_{a}\chi_{\mathscr{J}_{c}\mathscr{J}_{a}\phi}\nabla^{\nu}\alpha_{a}\xi_{\mu\nu}+2\chi_{\mathscr{J}_{c}q\pi}M^{\nu}\sigma_{\mu\nu}+2\chi_{\mathscr{J}_{c}q\phi}M^{\nu}\xi_{\mu\nu}.
\end{aligned}
\label{186}
\end{equation}

To derive a relaxation-type equation for the charge-diffusion currents, we substitute the first-order approximation 
\begin{eqnarray}
	\nabla^{\beta}\alpha_{a}=\sum_{b}\left(\chi^{-1}\right)_{ab}\left(\mathscr{J}_{b}^{\beta}-\chi_{\mathscr{J}_{b}q}M^{\beta}\right),
	\label{187}
\end{eqnarray}
into the term $\sum_{a}\widetilde{\chi}_{ca}\Delta_{\mu\beta}D\left(\nabla^{\beta}\alpha_{a}\right)$ on the right-hand side of Eq.~\eqref{186}. This substitution yields
\begin{eqnarray}
	\begin{aligned}
		\sum_{a}\widetilde{\chi}_{ca}\Delta_{\mu\beta}D\left(\nabla^{\beta}\alpha_{a}\right)\simeq & \sum_{b}\left(\widetilde{\chi}\chi^{-1}\right)_{cb}\Delta_{\mu\beta}D\mathscr{J}_{b}^{\beta}-\sum_{b}\left(\widetilde{\chi}\chi^{-1}\right)_{cb}M_{\mu}\beta\theta\left(\frac{\partial\chi_{\mathscr{J}_{b}q}}{\partial\beta}\Gamma-\sum_{d}\frac{\partial\chi_{\mathscr{J}_{b}q}}{\partial\alpha_{d}}\delta_{d}\right)\\
		&-\sum_{b}\left(\widetilde{\chi}\chi^{-1}\right)_{cb}\Delta_{\mu\beta}\chi_{\mathscr{J}_{b}q}DM^{\beta}+\sum_{ab}\widetilde{\chi}_{ca}\mathscr{J}_{b\mu}\beta\theta\left(\frac{\partial\left(\chi^{-1}\right)_{ab}}{\partial\beta}\Gamma-\sum_{d}\frac{\partial\left(\chi^{-1}\right)_{ab}}{\partial\alpha_{d}}\delta_{d}\right)\\
		&-\sum_{ab}\widetilde{\chi}_{ca}\chi_{\mathscr{J}_{b}q}M_{\mu}\beta\theta\left(\frac{\partial\left(\chi^{-1}\right)_{ab}}{\partial\beta}\Gamma-\sum_{d}\frac{\partial\left(\chi^{-1}\right)_{ab}}{\partial\alpha_{d}}\delta_{d}\right).
	\end{aligned}
	\label{188}
\end{eqnarray}
By combining Eqs.~\eqref{186} and~\eqref{188} and introducing the following coefficients
\begin{align}
	\tau_{\mathscr{J}}^{cb}= & -\left(\widetilde{\chi}\chi^{-1}\right)_{cb}=-\sum_{a}\widetilde{\chi}_{ca}\left(\chi^{-1}\right)_{ab},\label{189}\\
	\widetilde{\chi}_{\mathscr{J}}^{cb}= & \beta\sum_{a}\widetilde{\chi}_{ca}\left(\Gamma\frac{\partial\left(\chi^{-1}\right)_{ab}}{\partial\beta}-\sum_{d}\delta_{d}\frac{\partial\left(\chi^{-1}\right)_{ab}}{\partial\alpha_{d}}\right),\label{190}\\
	\overline{\chi}^{c}= & \beta\sum_{b}\tau_{\mathscr{J}}^{cb}\left(\Gamma\frac{\partial\chi_{\mathscr{J}_{b}q}}{\partial\beta}-\sum_{d}\delta_{d}\frac{\partial\chi_{\mathscr{J}_{b}q}}{\partial\alpha_{d}}\right),\label{191}
\end{align}
we obtain the relaxation equation for the charge-diffusion currents
\begin{eqnarray}
\begin{aligned}
	\mathscr{J}_{c\mu}+\sum_{b}\tau_{\mathscr{J}}^{cb}\dot{\mathscr{J}}_{b\mu}= & \sum_{b}\chi_{cb}\nabla_{\mu}\alpha_{b}+\chi_{\mathscr{J}_{c}q}M_{\mu}+\widetilde{\chi}_{\mathscr{J}_{c}h}h^{-2}\left(\Gamma h+\sum_{b}\delta_{b}n_{b}\right)\theta\sum_{a}n_{a}\nabla_{\mu}\alpha_{a}+\overline{\chi}^{c}\theta M_{\mu}\\
	&+\sum_{b}\tau_{\mathscr{J}}^{cb}\chi_{\mathscr{J}_{b}q}\Delta_{\mu\beta}DM^{\beta}+\sum_{b}\widetilde{\chi}_{\mathscr{J}}^{cb}\theta\mathscr{J}_{b\mu}-\sum_{b}\widetilde{\chi}_{\mathscr{J}}^{cb}\chi_{\mathscr{J}_{b}q}\theta M_{\mu}+\widetilde{\chi}_{\mathscr{J}_{c}q}\theta\Gamma M_{\mu}\\
	&+\widetilde{\chi}_{\mathscr{J}_{c}q}\Delta_{\mu\beta}DM^{\beta}-\beta\theta\dot{u}_{\mu}\sum_{a}\delta_{a}\widetilde{\chi}_{ac}-\beta\widetilde{\chi}_{\mathscr{J}_{c}h}\left(2\sigma_{\mu\nu}\dot{u}^{\nu}+y\theta\dot{u}_{\mu}\right)-2\widetilde{\chi}_{\mathscr{J}_{c}q}\xi_{\mu\nu}\dot{u}^{\nu}\\
	&+\beta\chi_{\mathscr{J}_{c}h}\mathcal{H}_{\mu}+\chi_{\mathscr{J}_{c}q}\mathcal{Q}_{\mu}+2\sum_{a}\chi_{\mathscr{J}_{c}p\mathscr{J}_{a}}\theta\nabla_{\mu}\alpha_{a}+2\chi_{\mathscr{J}_{c}pq}\theta M_{\mu}\\
	&+2\sum_{a}\chi_{\mathscr{J}_{c}\mathscr{J}_{a}\pi}\nabla^{\nu}\alpha_{a}\sigma_{\mu\nu}+2\sum_{a}\chi_{\mathscr{J}_{c}\mathscr{J}_{a}\phi}\nabla^{\nu}\alpha_{a}\xi_{\mu\nu}+2\chi_{\mathscr{J}_{c}q\pi}M^{\nu}\sigma_{\mu\nu}+2\chi_{\mathscr{J}_{c}q\phi}M^{\nu}\xi_{\mu\nu}.
\end{aligned}
\label{192}
\end{eqnarray}
Here, $\dot{\mathscr{J}}_{b\mu}$ is defined as $\dot{\mathscr{J}}_{b\mu}=\Delta_{\mu\nu}D\mathscr{J}_{b}^{\nu}$.In contrast to our previous work~\cite{She:2024rnx}, the derived relaxation equation contains additional terms proportional to $\dot{u}^\mu$ (specifically, those in the third line on the right-hand side). These terms represent new physical contributions to the dynamics of charge-diffusion currents. It should be noted that while the terms $-\beta\theta\dot{u}_{\mu}\sum_{a}\delta_{a}\widetilde{\chi}_{ac}-\beta\widetilde{\chi}_{\mathscr{J}_{c}h}\left(2\sigma_{\mu\nu}\dot{u}^{\nu}+y\theta\dot{u}_{\mu}\right)$ are also present in viscous fluids~\cite{Harutyunyan:2025fgs}, the term $-2\widetilde{\chi}_{\mathscr{J}_{c}q}\xi_{\mu\nu}\dot{u}^{\nu}$ is unique to spin fluids.

\subsubsection{Second-order corrections to the rotational stress tensor}

By substituting Eqs.~\eqref{71} and \eqref{87} into Eq.~\eqref{76} and invoking Curie's theorem, we derive the expression for $\langle\hat{\phi}_{\mu\nu}\rangle_{2}^{1}$, given by
\begin{eqnarray}
\langle\hat{\phi}_{\mu\nu}\rangle_{2}^{1}=\frac{1}{3}\mydelta_{\mu\nu\rho\sigma}\left(x\right)\left[\partial_{\tau}\left(\beta\xi^{\rho\sigma}\right)+\left(\partial_{\tau}u^{\rho}\right)\sum_{a}\frac{n_{a}}{h}\nabla^{\sigma}\alpha_{a}\right]_{x}\int d^{4}x_{1}\left(\hat{\phi}_{\gamma\delta}\left(x\right),\hat{\phi}^{\gamma\delta}\left(x_{1}\right)_{x}\right)\left(x_{1}-x\right)^{\tau},
\label{193}
\end{eqnarray}
Based on the appendix of Ref.~\cite{Harutyunyan:2025fgs}, it can be derived that
\begin{eqnarray}
	\frac{\beta}{6}\int d^{4}x_{1}\left(\hat{\phi}_{\gamma\delta}\left(x\right),\hat{\phi}^{\gamma\delta}\left(x_{1}\right)_{x}\right)\left(x_{1}-x\right)^{\tau}=\widetilde{\gamma}u^{\tau},
	\label{194}
\end{eqnarray}
where the frequency-dependent transport coefficient $\gamma\left(\omega\right)$ is defined as
\begin{eqnarray}
	\gamma\left(\omega\right)=\frac{\beta}{6}\int d^{4}x_{1}\int_{-\infty}^{t}e^{i\omega\left(t-t_{1}\right)}\left(\hat{\phi}_{\mu\nu}\left(\boldsymbol{x},t\right),\hat{\phi}^{\mu\nu}\left(\boldsymbol{x}_{1},t_{1}\right)\right).
	\label{195}
\end{eqnarray}
The coefficient $\widetilde{\gamma}$ is defined as the derivative of $\gamma\left(\omega\right)$ with respect to frequency evaluated at $\omega=0$
\begin{eqnarray}
	\widetilde{\gamma}=i\frac{d}{d\omega}\gamma\left(\omega\right)\bigg|_{\omega=0}=-\frac{1}{12}\frac{d^{2}}{d\omega^{2}}\text{Re}G_{\hat{\phi}_{\mu\nu}\hat{\phi}^{\mu\nu}}^{R}\left(\omega\right)\bigg|_{\omega=0}.
	\label{196}
\end{eqnarray}
Combining Eqs.~\eqref{193} and \eqref{194} and applying the approximation $D\beta\simeq\beta\theta\Gamma$, we obtain the non-local corrections to the rotational stress tensor originating from the two-point correlation function
\begin{eqnarray}	\langle\hat{\phi}_{\mu\nu}\rangle_{2}^{1}=2\widetilde{\gamma}\xi_{\mu\nu}\theta\Gamma+2\widetilde{\gamma}\mydelta_{\mu\nu\rho\sigma}D\xi^{\rho\sigma}+2\widetilde{\gamma}Th^{-1}\sum_{a}n_{a}\dot{u}_{[\mu}\nabla_{\nu]}\alpha_{a}.
\label{197}
\end{eqnarray}
The final term on the right-hand side of Eq.~\eqref{197} quantifies the non-local coupling between the rotational stress tensor and charge-diffusion currents. It is worth noting that this term is unique to spin fluids.

In accordance with our prior work~\cite{She:2024rnx}, the second-order corrections to the expectation value of the rotational stress tensor $\langle\hat{\phi}_{\mu\nu}\rangle$ arising from generalized thermodynamic forces and three-point correlation functions are given by
\begin{align}
\langle\hat{\phi}_{\mu\nu}\rangle_{2}^{2}= & \gamma_{\phi S}\mathcal{R}_{\langle\mu\rangle\langle\nu\rangle},\label{198}\\
\langle\hat{\phi}_{\mu\nu}\rangle_{2}^{3}= & 2\gamma_{\phi p\phi}\theta\xi_{\mu\nu}+\sum_{ab}\gamma_{\phi\mathscr{J}_{a}\mathscr{J}_{b}}\nabla_{[\mu}\alpha_{a}\nabla_{\nu]}\alpha_{b}+2\sum_{a}\gamma_{\phi\mathscr{J}_{a}q}\nabla_{[\mu}\alpha_{a}M_{\nu]}\nonumber\\
& +\gamma_{\phi qq}M_{[\mu}M_{\nu]}+\gamma_{\phi\pi\pi}\sigma_{\alpha[\mu}\sigma_{\nu]}^{\,\,\,\,\alpha}+2\gamma_{\phi\pi\phi}\sigma_{\alpha[\mu}\xi_{\nu]}^{\,\,\,\,\alpha}+\gamma_{\phi\phi\phi}\xi_{\alpha[\mu}\xi_{\nu]}^{\,\,\,\,\alpha},\label{199}
\end{align}
where the coefficients are defined as (see the appendix of Ref.~\cite{Harutyunyan:2025fgs})
\begin{align}
	\gamma_{\phi S}= & \frac{1}{3}\beta\int d^{4}x_{1}\left(\hat{\phi}_{\lambda\eta}\left(x\right),\hat{S}^{\lambda\eta}\left(x_{1}\right)_{x}\right)=-\frac{1}{3}\frac{d}{d\omega}\text{Im}G_{\hat{\phi}_{\lambda\eta}\hat{S}^{\lambda\eta}}^{R}\left(\omega\right)\bigg|_{\omega=0},\label{200}\\
	\gamma_{\phi p\phi} = &-\frac{1}{3}\beta^{2}\int d^{4}x_{1}d^{4}x_{2}\left(\hat{\phi}_{\gamma\delta}\left(x\right),\hat{p}^{*}\left(x_{1}\right),\hat{\phi}^{\gamma\delta}\left(x_{2}\right)\right)=\frac{1}{6}\frac{\partial}{\partial\omega_{1}}\frac{\partial}{\partial\omega_{2}}\text{Re}G_{\hat{\phi}_{\gamma\delta}\hat{p}^{*}\hat{\phi}^{\gamma\delta}}^{R}\left(\omega_{1},\omega_{2}\right)\bigg|_{\omega_{1,2}=0},\label{201}\\
	\gamma_{\phi\mathscr{J}_{a}\mathscr{J}_{b}} = &\frac{1}{3}\int d^{4}x_{1}d^{4}x_{2}\left(\hat{\phi}_{\gamma\delta}\left(x\right),\hat{\mathscr{J}}_{a}^{\gamma}\left(x_{1}\right),\hat{\mathscr{J}}_{b}^{\delta}\left(x_{2}\right)\right)=-\frac{T^{2}}{6}\frac{\partial}{\partial\omega_{1}}\frac{\partial}{\partial\omega_{2}}\text{Re}G_{\hat{\phi}_{\gamma\delta}\hat{\mathscr{J}}_{a}^{\gamma}\hat{\mathscr{J}}_{b}^{\delta}}^{R}\left(\omega_{1},\omega_{2}\right)\bigg|_{\omega_{1,2}=0},\label{202}\\
	\gamma_{\phi\mathscr{J}_{a}q} = &-\frac{1}{3}\beta\int d^{4}x_{1}d^{4}x_{2}\left(\hat{\phi}_{\gamma\delta}\left(x\right),\hat{\mathscr{J}}_{a}^{\gamma}\left(x_{1}\right),\hat{q}^{\delta}\left(x_{2}\right)\right)=\frac{T}{6}\frac{\partial}{\partial\omega_{1}}\frac{\partial}{\partial\omega_{2}}\text{Re}G_{\hat{\phi}_{\gamma\delta}\hat{\mathscr{J}}_{a}^{\gamma}\hat{q}^{\delta}}^{R}\left(\omega_{1},\omega_{2}\right)\bigg|_{\omega_{1,2}=0},\label{203}\\
	\gamma_{\phi qq} = &\frac{1}{3}\beta^{2}\int d^{4}x_{1}d^{4}x_{2}\left(\hat{\phi}_{\gamma\delta}\left(x\right),\hat{q}^{\gamma}\left(x_{1}\right),\hat{q}^{\delta}\left(x_{2}\right)\right)=-\frac{1}{6}\frac{\partial}{\partial\omega_{1}}\frac{\partial}{\partial\omega_{2}}\text{Re}G_{\hat{\phi}_{\gamma\delta}\hat{q}^{\gamma}\hat{q}^{\delta}}^{R}\left(\omega_{1},\omega_{2}\right)\bigg|_{\omega_{1,2}=0},\label{204}\\
	\gamma_{\phi\pi\pi} = &-\frac{4}{15}\beta^{2}\int d^{4}x_{1}d^{4}x_{2}\left(\hat{\phi}_{\lambda}^{\,\,\,\,\delta}\left(x\right),\hat{\pi}_{\delta}^{\,\,\,\,\eta}\left(x_{1}\right),\hat{\pi}_{\eta}^{\,\,\,\,\lambda}\left(x_{2}\right)\right)=\frac{2}{15}\frac{\partial}{\partial\omega_{1}}\frac{\partial}{\partial\omega_{2}}\text{Re}G_{\hat{\phi}_{\lambda}^{\,\,\,\,\delta}\hat{\pi}_{\delta}^{\,\,\,\,\eta}\hat{\pi}_{\eta}^{\,\,\,\,\lambda}}^{R}\left(\omega_{1},\omega_{2}\right)\bigg|_{\omega_{1,2}=0},\label{205}\\
	\gamma_{\phi\pi\phi} = &\frac{4}{5}\beta^{2}\int d^{4}x_{1}d^{4}x_{2}\left(\hat{\phi}_{\lambda}^{\,\,\,\,\delta}\left(x\right),\hat{\pi}_{\delta}^{\,\,\,\,\eta}\left(x_{1}\right),\hat{\phi}_{\eta}^{\,\,\,\,\lambda}\left(x_{2}\right)\right)=-\frac{2}{5}\frac{\partial}{\partial\omega_{1}}\frac{\partial}{\partial\omega_{2}}\text{Re}G_{\hat{\phi}_{\lambda}^{\,\,\,\,\delta}\hat{\pi}_{\delta}^{\,\,\,\,\eta}\hat{\phi}_{\eta}^{\,\,\,\,\lambda}}^{R}\left(\omega_{1},\omega_{2}\right)\bigg|_{\omega_{1,2}=0},\label{206}\\
	\gamma_{\phi\phi\phi} = &-\frac{4}{3}\beta^{2}\int d^{4}x_{1}d^{4}x_{2}\left(\hat{\phi}_{\lambda}^{\,\,\,\,\delta}\left(x\right),\hat{\phi}_{\delta}^{\,\,\,\,\eta}\left(x_{1}\right),\hat{\phi}_{\eta}^{\,\,\,\,\lambda}\left(x_{2}\right)\right)=\frac{2}{3}\frac{\partial}{\partial\omega_{1}}\frac{\partial}{\partial\omega_{2}}\text{Re}G_{\hat{\phi}_{\lambda}^{\,\,\,\,\delta}\hat{\phi}_{\delta}^{\,\,\,\,\eta}\hat{\phi}_{\eta}^{\,\,\,\,\lambda}}^{R}\left(\omega_{1},\omega_{2}\right)\bigg|_{\omega_{1,2}=0}.\label{207}
\end{align}
Synthesizing the first-order expression for $\phi_{\mu\nu}$ (derived from Eqs.~\eqref{36} and \eqref{92}) with these second-order corrections (\eqref{197}-\eqref{199}), we obtain the comprehensive second-order expression for the rotational stress tensor
\begin{eqnarray}
\begin{aligned}
\phi_{\mu\nu}= & 2\gamma\xi_{\mu\nu}+2\widetilde{\gamma}\xi_{\mu\nu}\theta\Gamma+2\widetilde{\gamma}\mydelta_{\mu\nu\rho\sigma}D\xi^{\rho\sigma}+2\widetilde{\gamma}Th^{-1}\sum_{a}n_{a}\dot{u}_{[\mu}\nabla_{\nu]}\alpha_{a}\\
 & +\gamma_{\phi S}\mathcal{R}_{\langle\mu\rangle\langle\nu\rangle}+2\gamma_{\phi p\phi}\theta\xi_{\mu\nu}+\sum_{ab}\gamma_{\phi\mathscr{J}_{a}\mathscr{J}_{b}}\nabla_{[\mu}\alpha_{a}\nabla_{\nu]}\alpha_{b}+2\sum_{a}\gamma_{\phi\mathscr{J}_{a}q}\nabla_{[\mu}\alpha_{a}M_{\nu]}\\
 & +\gamma_{\phi qq}M_{[\mu}M_{\nu]}+\gamma_{\phi\pi\pi}\sigma_{\alpha[\mu}\sigma_{\nu]}^{\,\,\,\,\alpha}+2\gamma_{\phi\pi\phi}\sigma_{\alpha[\mu}\xi_{\nu]}^{\,\,\,\,\alpha}+\gamma_{\phi\phi\phi}\xi_{\alpha[\mu}\xi_{\nu]}^{\,\,\,\,\alpha}.
\end{aligned}
\label{208}
\end{eqnarray}
The evolution equation for the rotational stress tensor is derived by substituting the relation $2\xi^{\rho\sigma}\sim\gamma^{-1}\phi^{\rho\sigma}$ into the term $D\xi^{\rho\sigma}$, yielding
\begin{eqnarray}
2\widetilde{\gamma}\mydelta_{\mu\nu\rho\sigma}D\xi^{\rho\sigma}\simeq\widetilde{\gamma}\gamma^{-1}\mydelta_{\mu\nu\rho\sigma}D\phi^{\rho\sigma}-\widetilde{\gamma}\gamma^{-2}\beta\left(\frac{\partial\gamma}{\partial\beta}\Gamma-\sum_{a}\frac{\partial\gamma}{\partial\alpha_{a}}\delta_{a}\right)\theta\phi_{\mu\nu}.\label{209}
\end{eqnarray}
By combining Eqs.~\eqref{208} and \eqref{209} and introducing the following coefficients
\begin{align}
\dot{\phi}_{\mu\nu}&=\mydelta_{\mu\nu\rho\sigma}D\phi^{\rho\sigma},\label{210}\\
\tau_{\phi}&=-\widetilde{\gamma}\gamma^{-1},\label{211}\\
\widetilde{\gamma}_{\phi}	&=\tau_{\phi}\gamma^{-1}\beta\left(\frac{\partial\gamma}{\partial\beta}\Gamma-\sum_{a}\frac{\partial\gamma}{\partial\alpha_{a}}\delta_{a}\right),\label{212}
\end{align}
we obtain the relaxation equation for the rotational stress tensor
\begin{eqnarray}
\begin{aligned}
\phi_{\mu\nu}+\tau_{\phi}\dot{\phi}_{\mu\nu}= & 2\gamma\xi_{\mu\nu}+\widetilde{\gamma}_{\phi}\theta\phi_{\mu\nu}+2\widetilde{\gamma}\theta\Gamma\xi_{\mu\nu}+2\widetilde{\gamma}Th^{-1}\sum_{a}n_{a}\dot{u}_{[\mu}\nabla_{\nu]}\alpha_{a}+\gamma_{\phi S}\mathcal{R}_{\langle\mu\rangle\langle\nu\rangle}\\
 & +2\gamma_{\phi p\phi}\theta\xi_{\mu\nu}+\sum_{ab}\gamma_{\phi\mathscr{J}_{a}\mathscr{J}_{b}}\nabla_{[\mu}\alpha_{a}\nabla_{\nu]}\alpha_{b}+2\sum_{a}\gamma_{\phi\mathscr{J}_{a}q}\nabla_{[\mu}\alpha_{a}M_{\nu]}\\
 & +\gamma_{\phi qq}M_{[\mu}M_{\nu]}+\gamma_{\phi\pi\pi}\sigma_{\alpha[\mu}\sigma_{\nu]}^{\,\,\,\,\alpha}+2\gamma_{\phi\pi\phi}\sigma_{\alpha[\mu}\xi_{\nu]}^{\,\,\,\,\alpha}+\gamma_{\phi\phi\phi}\xi_{\alpha[\mu}\xi_{\nu]}^{\,\,\,\,\alpha}.
\end{aligned}
\label{213}
\end{eqnarray}
Symmetry considerations allow the omission of the term $\gamma_{\phi qq}M_{[\mu}M_{\nu]}$. Notably, the term $2\widetilde{\gamma}Th^{-1}\sum_{a}n_{a}\dot{u}_{[\mu}\nabla_{\nu]}\alpha_{a}$ on the first line of the right-hand side represents a new contribution compared to our previous work~\cite{She:2024rnx}, underscoring an additional physical effect in the rotational stress tensor dynamics.

\subsubsection{Second-order corrections to the boost heat vector}

By substituting Eq.~\eqref{71} into Eq.~\eqref{76} and applying Curie's theorem, we derive the following expression for $\langle\hat{q}_{\mu}\left(x\right)\rangle_{2}^{1}$
\begin{equation}
	\begin{aligned}
		\langle\hat{q}_{\mu}\left(x\right)\rangle_{2}^{1}= & \frac{1}{3}\Delta_{\mu\rho}\left(x\right)\left[-2\beta\sigma^{\rho\sigma}\left(\partial_{\tau}u_{\sigma}\right)-2\left(\frac{1}{3}-\Gamma\right)\beta\theta\left(\partial_{\tau}u^{\rho}\right)+\sum_{a}\partial_{\tau}\left(n_{a}h^{-1}\right)\nabla^{\rho}\alpha_{a}+\beta\theta\left(\partial_{\tau}u^{\rho}\right)h^{-1}\sum_{a}n_{a}\delta_{a}\right]_{x}\\
		& \times\int d^{4}x_{1}\left(\hat{q}_{\lambda}\left(x\right),\hat{h}^{\lambda}\left(x_{1}\right)_{x}\right)\left(x_{1}-x\right)^{\tau}+\frac{1}{3}\Delta_{\mu\rho}\left(x\right)\left[\partial_{\tau}\left(\beta M^{\rho}\right)-2\beta\xi^{\rho\sigma}\left(\partial_{\tau}u_{\sigma}\right)\right]_{x}\int d^{4}x_{1}\left(\hat{q}_{\lambda}\left(x\right),\hat{q}^{\lambda}\left(x_{1}\right)_{x}\right)\left(x_{1}-x\right)^{\tau}\\
		&+\frac{1}{3}\Delta_{\mu\rho}\left(x\right)\sum_{a}\left[\beta\theta\left(\partial_{\tau}u^{\rho}\right)\delta_{a}-\partial_{\tau}\left(\nabla^{\rho}\alpha_{a}\right)\right]_{x}\int d^{4}x_{1}\left(\hat{q}_{\lambda}\left(x\right),\hat{\mathscr{J}}_{a}^{\lambda}\left(x_{1}\right)_{x}\right)\left(x_{1}-x\right)^{\tau}.
	\end{aligned}
	\label{214}
\end{equation}
This derivation relies on Eqs.~\eqref{88} and \eqref{89}, along with the relation
\begin{eqnarray*}
	\left(\hat{q}^{\mu}\left(x\right),\hat{h}^{\alpha}\left(x_{1}\right)_{x}\right)=\frac{1}{3}\Delta^{\mu\alpha}\left(x\right)\left(\hat{q}^{\lambda}\left(x\right),\hat{h}_{\lambda}\left(x_{1}\right)_{x}\right).
\end{eqnarray*}
The following equation can be obtained from the appendix of Ref.~\cite{Harutyunyan:2025fgs}
\begin{align}
	\frac{1}{3}\int d^{4}x_{1}\left(\hat{q}^{\lambda}\left(x\right),\hat{h}_{\lambda}\left(x_{1}\right)_{x}\right)\left(x_{1}-x\right)^{\tau}= & \widetilde{\lambda}_{qh}u^{\tau},\label{215}\\
	-\frac{1}{3}\int d^{4}x_{1}\left(\hat{q}^{\lambda}\left(x\right),\hat{\mathscr{J}}_{a\lambda}\left(x_{1}\right)_{x}\right)\left(x_{1}-x\right)^{\tau}= & \widetilde{\lambda}_{q\mathscr{J}_{a}}u^{\tau},\label{216}\\
	-\frac{1}{3}\beta\int d^{4}x_{1}\left(\hat{q}^{\lambda}\left(x\right),\hat{q}_{\lambda}\left(x_{1}\right)_{x}\right)\left(x_{1}-x\right)^{\tau}= & \widetilde{\lambda}u^{\tau},\label{217}
\end{align}
where the coefficients $\widetilde{\lambda}qh,\widetilde{\lambda}_{q\mathscr{J}_{a}}$, and $\widetilde{\lambda}$ are defined as
\begin{align}
		\widetilde{\lambda}_{qh}= & i\frac{d}{d\omega}\lambda_{qh}\left(\omega\right)\bigg|_{\omega=0}=-\frac{T}{6}\frac{d^{2}}{d\omega^{2}}\text{Re}G_{\hat{q}^{\lambda}\hat{h}_{\lambda}}^{R}\left(\omega\right)\bigg|_{\omega=0},\label{218}\\
		\widetilde{\lambda}_{q\mathscr{J}_{a}}= & i\frac{d}{d\omega}\lambda_{q\mathscr{J}_{a}}\left(\omega\right)\bigg|_{\omega=0}=\frac{T}{6}\frac{d^{2}}{d\omega^{2}}\text{Re}G_{\hat{q}^{\lambda}\hat{\mathscr{J}}_{a\lambda}}^{R}\left(\omega\right)\bigg|_{\omega=0},\label{219}\\
		\widetilde{\lambda}= & i\frac{d}{d\omega}\lambda\left(\omega\right)\bigg|_{\omega=0}=\frac{1}{6}\frac{d^{2}}{d\omega^{2}}\text{Re}G_{\hat{q}^{\lambda}\hat{q}_{\lambda}}^{R}\left(\omega\right)\bigg|_{\omega=0}.\label{220}
\end{align}
The frequency-dependent transport coefficients $\lambda_{qh}\left(\omega\right),\lambda_{q\mathscr{J}_{a}}\left(\omega\right),\lambda\left(\omega\right)$ are given by
\begin{align}
	\lambda_{qh}\left(\omega\right) & =\frac{1}{3}\int d^{4}x_{1}\int_{-\infty}^{t}e^{i\omega\left(t-t_{1}\right)}\left(\hat{q}^{\lambda}\left(\boldsymbol{x},t\right),\hat{h}_{\lambda}\left(\boldsymbol{x}_{1},t_{1}\right)\right),\label{221}\\
	\lambda_{q\mathscr{J}_{a}}\left(\omega\right) & =-\frac{1}{3}\int d^{4}x_{1}\int_{-\infty}^{t}e^{i\omega\left(t-t_{1}\right)}\left(\hat{q}^{\lambda}\left(\boldsymbol{x},t\right),\hat{\mathscr{J}}_{a\lambda}\left(\boldsymbol{x}_{1},t_{1}\right)\right),\label{222}\\
	\lambda\left(\omega\right) & =-\frac{1}{3}\beta\int d^{4}x_{1}\int_{-\infty}^{t}e^{i\omega\left(t-t_{1}\right)}\left(\hat{q}^{\lambda}\left(\boldsymbol{x},t\right),\hat{q}_{\lambda}\left(\boldsymbol{x}_{1},t_{1}\right)\right).\label{223}
\end{align}
Combining Eqs.~\eqref{214}-\eqref{217} and using the approximation $D\beta\simeq\beta\theta\Gamma$, we obtain the non-local corrections to the boost heat vector from the two-point correlation function
\begin{eqnarray}
	\begin{aligned}
		\langle\hat{q}_{\mu}\rangle_{2}^{1}= & \widetilde{\lambda}_{qh}\sum_{a}D\left(n_{a}h^{-1}\right)\nabla_{\mu}\alpha_{a}+\sum_{a}\widetilde{\lambda}_{q\mathscr{J}_{a}}\Delta_{\mu\gamma}D\left(\nabla^{\gamma}\alpha_{a}\right)-\widetilde{\lambda}\theta\Gamma M_{\mu}-\widetilde{\lambda}\Delta_{\mu\rho}DM^{\rho}\\
		&-2\widetilde{\lambda}_{qh}\beta\sigma_{\mu\nu}\dot{u}^{\nu}-\widetilde{\lambda}_{qh}y\beta\theta\dot{u}_{\mu}+2\widetilde{\lambda}\xi_{\mu\nu}\dot{u}^{\nu}-\sum_{a}\widetilde{\lambda}_{q\mathscr{J}_{a}}\beta\theta\dot{u}_{\mu}\delta_{a}.
	\end{aligned}
	\label{224}
\end{eqnarray}
Following the methodology for charge-diffusion currents and employing Eq.~\eqref{175}, we reformulate the above expression as
\begin{eqnarray}
	\begin{aligned}
		\langle\hat{q}_{\mu}\rangle_{2}^{1}= & \widetilde{\lambda}_{qh}h^{-2}\left(\Gamma h+\sum_{c}\delta_{c}n_{c}\right)\theta\sum_{a}n_{a}\nabla_{\mu}\alpha_{a}+\sum_{a}\widetilde{\lambda}_{q\mathscr{J}_{a}}\Delta_{\mu\gamma}D\left(\nabla^{\gamma}\alpha_{a}\right)-\widetilde{\lambda}\theta\Gamma M_{\mu}-\widetilde{\lambda}\Delta_{\mu\rho}DM^{\rho}\\
		&-2\widetilde{\lambda}_{qh}\beta\sigma_{\mu\nu}\dot{u}^{\nu}-\widetilde{\lambda}_{qh}y\beta\theta\dot{u}_{\mu}+2\widetilde{\lambda}\xi_{\mu\nu}\dot{u}^{\nu}-\sum_{a}\widetilde{\lambda}_{q\mathscr{J}_{a}}\beta\theta\dot{u}_{\mu}\delta_{a},
	\end{aligned}
	\label{225}
\end{eqnarray}
where the terms on the second line represent novel contributions unique to spin fluids.

As demonstrated in our previous work~\cite{She:2024rnx}, the second-order corrections to the boost heat vector $\hat{q}_\mu$ arising from generalized thermodynamic forces and three-point correlation functions are given by
\begin{align}
	\langle\hat{q}_{\mu}\rangle_{2}^{2}= & \lambda_{qh}\mathcal{H}_{\mu}-\lambda\mathcal{Q}_{\mu},\label{226}\\
	\langle\hat{q}_{\mu}\rangle_{2}^{3}= & 2\sum_{a}\lambda_{qp\mathscr{J}_{a}}\theta\nabla_{\mu}\alpha_{a}+2\lambda_{qpq}\theta M_{\mu}+2\sum_{a}\lambda_{q\mathscr{J}_{a}\pi}\nabla^{\nu}\alpha_{a}\sigma_{\mu\nu}\nonumber\\
	&+2\sum_{a}\lambda_{q\mathscr{J}_{a}\phi}\nabla^{\nu}\alpha_{a}\xi_{\mu\nu}+2\lambda_{qq\pi}M^{\nu}\sigma_{\mu\nu}+2\lambda_{qq\phi}M^{\nu}\xi_{\mu\nu},\label{227}
\end{align}
where the transport coefficients are defined as (see appendix of Ref.~\cite{Harutyunyan:2025fgs})
\begin{align}
	\lambda_{qh} & =\frac{1}{3}\beta\int d^{4}x_{1}\left(\hat{q}^{\lambda}\left(x\right),\hat{h}_{\lambda}\left(x_{1}\right)\right)=-\frac{1}{3}\frac{d}{d\omega}\text{Im}G_{\hat{q}^{\lambda}\hat{h}_{\lambda}}^{R}\left(\omega\right)\bigg|_{\omega=0},\label{228}\\
	\lambda_{qp\mathscr{J}_{a}} & =\frac{1}{3}\beta\int d^{4}x_{1}d^{4}x_{2}\left(\hat{q}_{\beta}\left(x\right),\hat{p}^{*}\left(x_{1}\right),\hat{\mathscr{J}}_{a}^{\beta}\left(x_{2}\right)\right)=-\frac{T}{6}\frac{\partial}{\partial\omega_{1}}\frac{\partial}{\partial\omega_{2}}\text{Re}G_{\hat{q}_{\beta}\hat{p}^{*}\hat{\mathscr{J}}_{a}^{\beta}}^{R}\left(\omega_{1},\omega_{2}\right)\bigg|_{\omega_{1,2}=0}\label{229},\\
	\lambda_{qpq} & =-\frac{1}{3}\beta^{2}\int d^{4}x_{1}d^{4}x_{2}\left(\hat{q}_{\beta}\left(x\right),\hat{p}^{*}\left(x_{1}\right),\hat{q}^{\beta}\left(x_{2}\right)\right)=\frac{1}{6}\frac{\partial}{\partial\omega_{1}}\frac{\partial}{\partial\omega_{2}}\text{Re}G_{\hat{q}_{\beta}\hat{p}^{*}\hat{q}^{\beta}}^{R}\left(\omega_{1},\omega_{2}\right)\bigg|_{\omega_{1,2}=0},\label{230}\\
	\lambda_{q\mathscr{J}_{a}\pi} & =-\frac{1}{5}\beta\int d^{4}x_{1}d^{4}x_{2}\left(\hat{q}_{\lambda}\left(x\right),\hat{\mathscr{J}}_{a\delta}\left(x_{1}\right),\hat{\pi}^{\lambda\delta}\left(x_{2}\right)\right)=\frac{T}{10}\frac{\partial}{\partial\omega_{1}}\frac{\partial}{\partial\omega_{2}}\text{Re}G_{\hat{q}_{\lambda}\hat{\mathscr{J}}_{a\delta}\hat{\pi}^{\lambda\delta}}^{R}\left(\omega_{1},\omega_{2}\right)\bigg|_{\omega_{1,2}=0},\label{231}\\
	\lambda_{q\mathscr{J}_{a}\phi} & =-\frac{1}{3}\beta\int d^{4}x_{1}d^{4}x_{2}\left(\hat{q}_{\lambda}\left(x\right),\hat{\mathscr{J}}_{a\delta}\left(x_{1}\right),\hat{\phi}^{\lambda\delta}\left(x_{2}\right)\right)=\frac{T}{6}\frac{\partial}{\partial\omega_{1}}\frac{\partial}{\partial\omega_{2}}\text{Re}G_{\hat{q}_{\lambda}\hat{\mathscr{J}}_{a\delta}\hat{\phi}^{\lambda\delta}}^{R}\left(\omega_{1},\omega_{2}\right)\bigg|_{\omega_{1,2}=0},\label{232}\\
	\lambda_{qq\pi} & =\frac{1}{5}\beta^{2}\int d^{4}x_{1}d^{4}x_{2}\left(\hat{q}_{\lambda}\left(x\right),\hat{q}_{\delta}\left(x_{1}\right),\hat{\pi}^{\lambda\delta}\left(x_{2}\right)\right)=-\frac{1}{10}\frac{\partial}{\partial\omega_{1}}\frac{\partial}{\partial\omega_{2}}\text{Re}G_{\hat{q}_{\lambda}\hat{q}_{\delta}\hat{\pi}^{\lambda\delta}}^{R}\left(\omega_{1},\omega_{2}\right)\bigg|_{\omega_{1,2}=0},\label{233}\\
	\lambda_{qq\phi} & =\frac{1}{3}\beta^{2}\int d^{4}x_{1}d^{4}x_{2}\left(\hat{q}_{\lambda}\left(x\right),\hat{q}_{\delta}\left(x_{1}\right),\hat{\phi}^{\lambda\delta}\left(x_{2}\right)\right)=-\frac{1}{6}\frac{\partial}{\partial\omega_{1}}\frac{\partial}{\partial\omega_{2}}\text{Re}G_{\hat{q}_{\lambda}\hat{q}_{\delta}\hat{\phi}^{\lambda\delta}}^{R}\left(\omega_{1},\omega_{2}\right)\bigg|_{\omega_{1,2}=0}.\label{234}
\end{align}
The complete second-order expression for the boost heat vector combines these contributions
\begin{eqnarray}
\begin{aligned}
	q_{\mu}= & \sum_{a}\lambda_{q\mathscr{J}_{a}}\nabla_{\mu}\alpha_{a}-\lambda M_{\mu}+\widetilde{\lambda}_{qh}h^{-2}\left(\Gamma h+\sum_{c}\delta_{c}n_{c}\right)\theta\sum_{a}n_{a}\nabla_{\mu}\alpha_{a}+\sum_{a}\widetilde{\lambda}_{q\mathscr{J}_{a}}\Delta_{\mu\gamma}D\left(\nabla^{\gamma}\alpha_{a}\right)\\
	&-\widetilde{\lambda}\theta\Gamma M_{\mu}-\widetilde{\lambda}\Delta_{\mu\rho}DM^{\rho}-2\widetilde{\lambda}_{qh}\beta\sigma_{\mu\nu}\dot{u}^{\nu}-\widetilde{\lambda}_{qh}y\beta\theta\dot{u}_{\mu}+2\widetilde{\lambda}\xi_{\mu\nu}\dot{u}^{\nu}-\sum_{a}\widetilde{\lambda}_{q\mathscr{J}_{a}}\beta\theta\dot{u}_{\mu}\delta_{a}\\
	&+\lambda_{qh}\mathcal{H}_{\mu}-\lambda\mathcal{Q}_{\mu}+2\sum_{a}\lambda_{qp\mathscr{J}_{a}}\theta\nabla_{\mu}\alpha_{a}+2\lambda_{qpq}\theta M_{\mu}+2\sum_{a}\lambda_{q\mathscr{J}_{a}\pi}\nabla^{\nu}\alpha_{a}\sigma_{\mu\nu}\\
	&+2\sum_{a}\lambda_{q\mathscr{J}_{a}\phi}\nabla^{\nu}\alpha_{a}\xi_{\mu\nu}+2\lambda_{qq\pi}M^{\nu}\sigma_{\mu\nu}+2\lambda_{qq\phi}M^{\nu}\xi_{\mu\nu},
\end{aligned}
\label{235}
\end{eqnarray}

To derive a relaxation-type equation for the boost heat vector $q_\mu$, we begin by substituting the first-order approximation
\begin{eqnarray}
M^{\gamma}=-\lambda^{-1}\left(q^{\gamma}-\sum_{a}\lambda_{q\mathscr{J}_{a}}\nabla^{\gamma}\alpha_{a}\right),
	\label{236}
\end{eqnarray}
into the term $-\widetilde{\lambda}\Delta_{\mu\gamma}DM^{\gamma}$ in Eq.~\eqref{235}. This substitution yields
\begin{eqnarray}
	\begin{aligned}
		-\widetilde{\lambda}\Delta_{\mu\gamma}DM^{\gamma}\simeq & \widetilde{\lambda}\lambda^{-1}\Delta_{\mu\gamma}Dq^{\gamma}-\widetilde{\lambda}\lambda^{-1}\Delta_{\mu\gamma}\sum_{a}\lambda_{q\mathscr{J}_{a}}D\left(\nabla^{\gamma}\alpha_{a}\right)-\widetilde{\lambda}\lambda^{-1}\sum_{a}\nabla_{\mu}\alpha_{a}\beta\theta\left(\frac{\partial\lambda_{q\mathscr{J}_{a}}}{\partial\beta}\Gamma-\sum_{d}\frac{\partial\lambda_{q\mathscr{J}_{a}}}{\partial\alpha_{d}}\delta_{d}\right)\\
		&-\widetilde{\lambda}\lambda^{-2}q_{\mu}\beta\theta\left(\frac{\partial\lambda}{\partial\beta}\Gamma-\sum_{d}\frac{\partial\lambda}{\partial\alpha_{d}}\delta_{d}\right)+\widetilde{\lambda}\lambda^{-2}\sum_{a}\lambda_{q\mathscr{J}_{a}}\nabla_{\mu}\alpha_{a}\beta\theta\left(\frac{\partial\lambda}{\partial\beta}\Gamma-\sum_{d}\frac{\partial\lambda}{\partial\alpha_{d}}\delta_{d}\right).
	\end{aligned}
	\label{237}
\end{eqnarray}
Combining Eqs.~\eqref{235} and \eqref{237} and introducing the following coefficients
\begin{align}
	\tau_{q}= & -\widetilde{\lambda}\lambda^{-1},\label{238}\\
	\widetilde{\lambda}_{q}= & \beta\tau_{q}\lambda^{-1}\left(\frac{\partial\lambda}{\partial\beta}\Gamma-\sum_{d}\frac{\partial\lambda}{\partial\alpha_{d}}\delta_{d}\right),\label{239}\\
	\overline{\lambda}^{a}= & \beta\tau_{q}\left(\frac{\partial\lambda_{q\mathscr{J}_{a}}}{\partial\beta}\Gamma-\sum_{d}\frac{\partial\lambda_{q\mathscr{J}_{a}}}{\partial\alpha_{d}}\delta_{d}\right),\label{240}
\end{align}
we obtain the relaxation equation for the boost heat vector $q_\mu$
\begin{eqnarray}
	\begin{aligned}
		q_{\mu}+\tau_{q}\dot{q}_{\mu}= & \sum_{a}\lambda_{q\mathscr{J}_{a}}\nabla_{\mu}\alpha_{a}-\lambda M_{\mu}+\widetilde{\lambda}_{qh}h^{-2}\left(\Gamma h+\sum_{c}\delta_{c}n_{c}\right)\theta\sum_{a}n_{a}\nabla_{\mu}\alpha_{a}+\sum_{a}\widetilde{\lambda}_{q\mathscr{J}_{a}}\Delta_{\mu\gamma}D\left(\nabla^{\gamma}\alpha_{a}\right)\\
		& -\widetilde{\lambda}\theta\Gamma M_{\mu}+\tau_{q}\Delta_{\mu\gamma}\sum_{a}\lambda_{q\mathscr{J}_{a}}D\left(\nabla^{\gamma}\alpha_{a}\right)+\sum_{a}\overline{\lambda}^{a}\theta\nabla_{\mu}\alpha_{a}+\widetilde{\lambda}_{q}\theta q_{\mu}-\sum_{a}\widetilde{\lambda}_{q}\lambda_{q\mathscr{J}_{a}}\theta\nabla_{\mu}\alpha_{a}\\
		& -2\widetilde{\lambda}_{qh}\beta\sigma_{\mu\nu}\dot{u}^{\nu}-\widetilde{\lambda}_{qh}y\beta\theta\dot{u}_{\mu}+2\widetilde{\lambda}\xi_{\mu\nu}\dot{u}^{\nu}-\sum_{a}\widetilde{\lambda}_{q\mathscr{J}_{a}}\beta\theta\dot{u}_{\mu}\delta_{a}+\lambda_{qh}\mathcal{H}_{\mu}-\lambda\mathcal{Q}_{\mu}\\
		& +2\sum_{a}\lambda_{qp\mathscr{J}_{a}}\theta\nabla_{\mu}\alpha_{a}+2\lambda_{qpq}\theta M_{\mu}+2\sum_{a}\lambda_{q\mathscr{J}_{a}\pi}\nabla^{\nu}\alpha_{a}\sigma_{\mu\nu}+2\sum_{a}\lambda_{q\mathscr{J}_{a}\phi}\nabla^{\nu}\alpha_{a}\xi_{\mu\nu}\\
		& +2\lambda_{qq\pi}M^{\nu}\sigma_{\mu\nu}+2\lambda_{qq\phi}M^{\nu}\xi_{\mu\nu}.
	\end{aligned}
	\label{241}
\end{eqnarray}
Here, $\dot{q}_{\mu}$ is defined as $\dot{q}_{\mu}=\Delta_{\mu\nu}Dq^{\nu}$. Compared to our previous work~\cite{She:2024rnx}, four new terms related to $\dot{u}^\mu$ emerge as distinctive features of spin hydrodynamics.

\subsubsection{Second-order corrections to the spin-related dissipative flux $\varpi^{\lambda\mu\nu}$}

Since Eq.~\eqref{71} contains no third-rank tensor component, Curie's theorem dictates that the non-local correction to $\varpi^{\lambda\mu\nu}$ from two-point correlation functions must vanish
\begin{eqnarray}
\langle\hat{\varpi}^{\lambda\mu\nu}\rangle_{2}^{1}=0.
\label{242}
\end{eqnarray}
As shown in our previous work~\cite{She:2024rnx}, the second-order corrections to $\hat{\varpi}^{\lambda\mu\nu}$ arising from generalized thermodynamic forces and three-point correlation functions take the form
\begin{align}
	\langle\hat{\varpi}^{\lambda\mu\nu}\rangle_{2}^{2}	=&\varphi\varXi^{\lambda\mu\nu},\label{243}\\
	\langle\hat{\varpi}^{\lambda\mu\nu}\rangle_{2}^{3}	=&2\sum_{a}\varphi_{\varpi\mathscr{J}_{a}\phi}\myDelta^{\lambda\mu\nu\rho\sigma\delta}\nabla_{\rho}\alpha_{a}\xi_{\sigma\delta}+2\varphi_{\varpi q\phi}\myDelta^{\lambda\mu\nu\rho\sigma\delta}M_{\rho}\xi_{\sigma\delta},\label{244}
\end{align}
where the transport coefficients are defined as (see appendix of Ref.~\cite{Harutyunyan:2025fgs})
\begin{align}
	\varphi	=&\int d^{4}x_{1}\left(\hat{\varpi}_{\gamma\varepsilon\eta}\left(x\right),\hat{\varpi}^{\gamma\varepsilon\eta}\left(x_{1}\right)_{x}\right)=-T\frac{d}{d\omega}\text{Im}G_{\hat{\varpi}_{\gamma\varepsilon\eta}\hat{\varpi}^{\gamma\varepsilon\eta}}^{R}\left(\omega\right)\bigg|_{\omega=0},\label{245}\\
	\varphi_{\varpi\mathscr{J}_{a}\phi}	=&-\beta\int d^{4}x_{1}d^{4}x_{2}\left(\hat{\varpi}^{\gamma\varepsilon\zeta}\left(x\right),\hat{\mathscr{J}}_{a\gamma}\left(x_{1}\right),\hat{\phi}_{\varepsilon\zeta}\left(x_{2}\right)\right)=\frac{T}{2}\frac{\partial}{\partial\omega_{1}}\frac{\partial}{\partial\omega_{2}}\text{Re}G_{\hat{\varpi}^{\gamma\varepsilon\zeta}\hat{\mathscr{J}}_{a\gamma}\hat{\phi}_{\varepsilon\zeta}}^{R}\left(\omega_{1},\omega_{2}\right)\bigg|_{\omega_{1,2}=0},\label{246}\\
	\varphi_{\varpi q\phi}	=&\beta^{2}\int d^{4}x_{1}d^{4}x_{2}\left(\hat{\varpi}^{\gamma\varepsilon\zeta}\left(x\right),\hat{q}_{\gamma}\left(x_{1}\right),\hat{\phi}_{\varepsilon\zeta}\left(x_{2}\right)\right)=-\frac{1}{2}\frac{\partial}{\partial\omega_{1}}\frac{\partial}{\partial\omega_{2}}\text{Re}G_{\hat{\varpi}^{\gamma\varepsilon\zeta}\hat{q}_{\gamma}\hat{\phi}_{\varepsilon\zeta}}^{R}\left(\omega_{1},\omega_{2}\right)\bigg|_{\omega_{1,2}=0}.\label{247}
\end{align}
The complete second-order expression for $\varpi^{\lambda\mu\nu}$ is
\begin{eqnarray}
\varpi^{\lambda\mu\nu}=\varphi\varXi^{\lambda\mu\nu}+2\sum_{a}\varphi_{\varpi\mathscr{J}_{a}\phi}\myDelta^{\lambda\mu\nu\rho\sigma\delta}\xi_{\sigma\delta}\nabla_{\rho}\alpha_{a}+2\varphi_{\varpi q\phi}\myDelta^{\lambda\mu\nu\rho\sigma\delta}M_{\rho}\xi_{\sigma\delta}.
\label{248}
\end{eqnarray}

\section{Conclusions}
\label{section4}

Recent studies~\cite{Harutyunyan:2025fgs} have revealed that two-point correlation functions involving tensors of different ranks at distinct spacetime points generate additional second-order nonlocal corrections in the hydrodynamic constitutive relations. Motivated by this finding, we extend our previous framework of relativistic second-order spin hydrodynamics~\cite{She:2024rnx}, developed through the nonequilibrium statistical operator method, by systematically incorporating these corrections. Specifically, we generalize the second-order viscous hydrodynamic formulation of Ref.~\cite{Harutyunyan:2025fgs} to include spin degrees of freedom.

Following the methodology of Ref.~\cite{She:2024rnx}, we consider a quantum system described by its energy-momentum tensor, multiple conserved charge currents, and spin tensor. Through a second-order expansion of the statistical operator, we derive a complete set of relativistic, canonical-like second-order spin hydrodynamic equations for the shear-stress tensor, bulk-viscous pressure, flavor-diffusion currents, rotational stress tensor, boost heat vector, and Spin tensor-related dissipative fluxes.

Our analysis not only reproduces the previously identified second-order modifications in viscous fluid dynamics~\cite{Harutyunyan:2025fgs}, but also reveals novel second-order terms in the charge diffusion currents, rotational stress tensor, and boost heat vector. These new terms originate from nonlocal corrections associated with two-point correlations between tensors of different ranks. As emphasized in Ref.~\cite{Harutyunyan:2025fgs}, all such corrections couple to the comoving derivative of the flow velocity, $\dot{u}^\mu$. Furthermore, we provide explicit expressions for the transport coefficients in terms of retarded Green's functions, which are determined by two-point or three-point correlation functions.

A key finding is that these new second-order terms emerge from memory effects in the statistical operator—a crucial aspect of relativistic spin hydrodynamics that was not accounted for in our previous work~\cite{She:2024rnx}. These terms characterize nonlocal interactions between different dissipative processes through two-point correlation functions. Notably, while appearing as quadratic terms in thermodynamic gradients, they fundamentally originate from first-order terms in the statistical operator's Taylor expansion. Their explicit form requires proper treatment of memory effects and nonlocality in the correlation functions.

Future research efforts should prioritize three key directions: (1) precise determination of transport coefficients through thermal field theoretic approaches, (2) systematic analysis of stability and causality properties in spin hydrodynamic systems across both linear and nonlinear regimes, and (3) development of robust numerical frameworks for spin hydrodynamic simulations. These comprehensive studies will yield fundamental insights into the non-equilibrium dynamics of spin-polarized systems while enabling quantitatively reliable predictions that can be directly confronted with experimental measurements.

\section*{Acknowledgments}

We are grateful to Armen Sedrakian for his important and constructive comments, which significantly enhanced this manuscript. This research was partly funded by the Startup Research Fund of Henan Academy of Sciences (No. 231820058), the 2024 Henan Province International Science and Technology Cooperation Projects (No. 242102521068), and High-level Achievements Reward and Cultivation Projects (No. 20252320001). 
Ze-Fang Jiang's research is funded by the National Natural Science Foundation of China (NSFC) under Grant Nos. 12305138. The research of Defu Hou is support in part by the National Key Research and Development Program of China under Contract No. 2022YFA1604900. Additionally, he received partial support from the National Natural Science Foundation of China (NSFC) under Grants No.12435009 and No. 12275104. 

\bibliography{rotation.bib}

\end{document}